\documentclass[numberedappendix]{emulateapj}

\usepackage{amsmath,bm}
\usepackage{epstopdf}

\begin{document}

\title{Type Ia Supernova Colors and Ejecta Velocities:\\ Hierarchical Bayesian Regression with Non-Gaussian Distributions}
\author{Kaisey S. Mandel\altaffilmark{1,4}, Ryan J. Foley\altaffilmark{2,3} and Robert P. Kirshner\altaffilmark{1}}
\altaffiltext{1}{Harvard-Smithsonian Center for Astrophysics, 60 Garden St., Cambridge, MA 02138 USA}
\altaffiltext{2}{
Astronomy Department,
University of Illinois at Urbana-Champaign,
1002 W.\ Green Street,
Urbana, IL 61801 USA}
\altaffiltext{3}{
Department of Physics,
University of Illinois Urbana-Champaign,
1110 W.\ Green Street,
Urbana, IL 61801 USA}
\altaffiltext{4}{kmandel@cfa.harvard.edu}

\slugcomment{\today; Accepted for publication in ApJ}

 \begin{abstract} 
 We investigate the statistical dependence of the peak intrinsic colors of Type Ia supernovae (SN Ia) on their expansion velocities at maximum light, measured from the Si II $\lambda 6355$ spectral feature. We construct a new hierarchical Bayesian regression model, accounting for the random effects of intrinsic scatter, measurement error, and reddening by host galaxy dust, and implement a Gibbs sampler and deviance information criteria to estimate the correlation. The method is applied to the apparent colors from BVRI light curves and Si II velocity data for 79 nearby SNe Ia.  The apparent color distributions of high (HV) and normal velocity (NV) supernovae exhibit significant discrepancies for $B-V$ and $B-R$, but not other colors. Hence, they are likely due to intrinsic color differences originating in the $B$-band, rather than dust reddening. The mean intrinsic $B-V$ and $B-R$ color differences  between HV and NV groups are $0.06 \pm 0.02$ and $0.09 \pm 0.02$ mag, respectively.  A linear model finds significant slopes of $-0.021 \pm 0.006$ and $-0.030 \pm 0.009$ mag $(10^3 \text{ km s}^{-1})^{-1}$ for intrinsic $B-V$ and $B-R$ colors versus velocity, respectively. Since the ejecta velocity distribution is skewed towards high velocities, these effects imply non-Gaussian intrinsic color distributions with skewness up to $+0.3$. Accounting for the intrinsic color-velocity correlation results in corrections to $A_V$ extinction estimates as large as $-0.12$ mag for HV SNe Ia and $+0.06$ mag for NV events. Velocity measurements from SN Ia spectra have potential to diminish systematic errors from the confounding of intrinsic colors and dust reddening affecting supernova distances.
 \end{abstract}
   
\section{Introduction}

Type Ia supernova (SN Ia) light curves have been used as cosmological distance indicators to trace the history of cosmic expansion, detect cosmic acceleration \citep{riess98,perlmutter99}, and to constrain the equation-of-state parameter $w$ of dark energy  \citep{garnavich98b,wood-vasey07, astier06, kowalski08, hicken09b, kessler09, freedman09, amanullah10,conley11,sullivan11,rest14,scolnic14b}.   Determining supernova distances with high precision and small systematic error is essential to accurate  constraints on the cosmic  expansion history and the properties of dark energy.  
However, the confounding of extrinsic host galaxy dust reddening with the intrinsic color variations of SNe Ia  presents a systematic limitation to their precision and accuracy in cosmological applications \citep{conley07}.  In this paper, we investigate how measurements of the expansion velocity of the supernova atmosphere can be used to learn more about the intrinsic color distribution of SNe Ia and improve inferences of host galaxy dust extinction.

Dust along the line of sight in the host galaxy reddens SN Ia colors and dims their magnitudes, and can lead to systematic errors in distance estimates if not properly accounted for.  The ratio of total to selective dust extinction, $R_V = A_V/(A_B - A_V)$, typically parameterizes the wavelength dependence of dust absorption and scattering, and has an average value of 3.1 for interstellar dust in the Milky Way (MW) Galaxy, although it can vary between 2.1 and 5.8 \citep{draine03}.   Studies of external galaxies have found similar extinction curves with $R_V \approx 2.8$ \citep{finkelman08, finkelman10}.  However, several analyses of both individual SNe Ia and large samples have found anomalously low effective values of $R_V \lesssim 1.8$ \citep{branch92, tripp98,trippbranch99,conley07,nobili08,hicken09b,krisciunas07, elias-rosa06, elias-rosa07, wangx08}.    Interestingly, recent analyses of multi-wavelength light curve and color data, including near-infrared (NIR) observations, have indicated that SNe Ia with low reddening are subject to dust extinction $A_V$ with a reddening law closer to $R_V \approx 3$, whereas highly reddened objects appear extinguished by dust with $R_V \approx 1.7$ \citep{folatelli10, mandel11, burns14}.  Systematic uncertainties in the treatment of dust and color of SNe Ia have important implications for cosmological inference.  For example, \citet{scolnic14} find that attributing the Hubble diagram residual scatter to SN Ia color dispersion rather than luminosity dispersion yields an effective $R_V$ consistent with MW dust, and yields a 4\% shift in the inferred value of $w$.  Resolving the confusion between the intrinsic color variation in the SN Ia population and extrinsic host galaxy dust extinction is needed for the proper analysis of SN Ia observables.

One promising strategy is to observe SNe Ia at rest-frame NIR wavelengths, where host galaxy dust extinction is diminished, and SNe Ia have been shown to be excellent standard candles using a burgeoning nearby sample \citep{elias85, meikle00, krisciunas04a, krisciunas04c, wood-vasey08, mandel09, contreras10, folatelli10, stritzinger11, barone-nugent12, weyant13, friedman14}.  At rest-frame optical wavelengths, where the bulk of nearby and high-$z$ observations of SNe Ia have been made, we must find better ways to decompose the observed apparent magnitudes and colors into the components intrinsic to the SN Ia and those extrinsic and due to dust reddening and extinction.   Current SN Ia analyses using the SALT2 method \citep{guy07} do not distinguish between intrinsic SN Ia variations and dust effects.  One way to identify intrinsic color variations is to find the portion of SN Ia color correlated with a measurable intrinsic property of SNe Ia, unaffected by dust.  

Narrow features in SN Ia spectra are intrinsic to the supernovae and can be correlated with photometric observables, and used as additional parameters to try to improve the estimation of SN Ia luminosities.  Even among ``normal'' SNe Ia there exists diversity in spectral features, such as the widths, strengths, and expansion velocities of specific spectral lines, and their evolution with the phase of the SN Ia.   For example, \citet{benetti05} subdivided normal SNe Ia into two classes, one with high velocity gradients and one with low velocity gradients.  \citet{branch06} classified optical spectra of SNe Ia into ``core-normal'', ``broad-line'', ``cool'', and ``shallow-silicon.''   \citet{blondin12} examined the diversity of SNe Ia in the context of these classification schemes using the large spectroscopic dataset of the CfA Supernova Program.  \citet{chotard11} modeled the components of the  apparent SN Ia spectroscopic variations depending upon Si II and Ca II H\&K equivalent widths, finding that the remainder is well described  by a CCM \citep{ccm89} dust reddening law with $R_V = 2.8 \pm 0.3$, consistent with the MW value.

The use of spectroscopic variations and their correlations with SN Ia luminosity was explored by \citet{nugent95,bongard06, foley08} and \citet{hachinger08}.   \citet{bailey09} investigated ratios of fluxes at different wavelengths using spectra from the Nearby Supernova Factory and found that they could be used to standardize SN Ia magnitudes and estimate distances with a lower scatter ($\sim 0.12$ mag) than with light curve information alone.   \citet*{blondin11} examined the utility of flux ratios and other spectroscopic indicators for predicting SN Ia luminosities and distances using a large spectroscopic data set collected by the CfA Supernova Group.  They used a cross-validation procedure to avoid overfitting the SN Ia sample and to robustly gauge the impact on SN Ia distance estimates in the Hubble diagram.  They found that spectral flux ratios led to modest improvements in the Hubble diagram scatter, but at low statistical significance given the sample size, whereas spectral line profile measurements did not improve luminosity estimates and distance predictions beyond the usual optical light curve width and color standardization.

\citet{wangx09b}  divided SNe Ia into normal velocity (NV) and high velocity (HV) subclasses based upon the photospheric expansion velocity of the explosion ejecta near maximum light, as measured from the prominent Si II $\lambda$6355 absorption feature.  Measured from the blueshift of the Si II line, the ejecta velocities are conventionally negative (towards the observer), so HV objects  have more negative velocities ($v < -11,800 \text{ km s}^{-1}$), and larger absolute velocities $|v|$, whereas NV objects have less negative velocities ($v > -11,800 \text{ km s}^{-1}$), and smaller absolute velocities $|v|$.  Applying this classification scheme to 156 SNe Ia observed by the Lick Observatory Supernova Search \citep{ganeshalingam10,silverman12}  and the CfA Supernova Program  \citep{matheson08}, they analyzed the $B$- and $V$-band magnitudes, light curve shapes and colors of each class.  They found that the Hubble diagram scatter could be reduced from 0.178 mag for the full sample to  $0.12-0.13$ mag by treating each class separately.  They estimated the dust extinction law slope parameter $R_V = 1.9$ for the full sample, but $R_V = 2.4$ for the normal velocity sample alone, and $R_V = 1.6$ for the high velocity sample alone.   Finding that high velocity SNe Ia had redder apparent colors overall, they suggested that treating these subsamples separately with different $R_V$s for host galaxy dust would lead to improved SN Ia distance measurements.   

\citet{foleykasen11} re-analyzed the data set from \citet{wangx09b} and found a reddening law slope $R_V \approx 2.5$ fit each subsample separately, when the reddest SNe Ia were excluded ($E(B-V) < 0.35$ mag).  They examined the relation between the $B_\text{max} - V_\text{max}$ pseudocolors and the velocity classes, finding that the cumulative distribution function  of the apparent $B_\text{max} - V_\text{max}$ pseudocolors of high-velocity SNe Ia were offset by $\sim 0.07$ mag to the red relative to the CDF of normal SNe Ia, and that the apparently bluest SNe Ia in the normal set were $\sim 0.10$ mag bluer than the bluest of the high-velocity SNe Ia.  They argued that the offset in the distributions of apparent colors between the two classes is due to an intrinsic color difference between normal-velocity and high-velocity SNe Ia, and found that the rms Hubble residual could be reduced to 0.11 mag by excluding high ejecta velocity SNe Ia.   Hence, accounting for this velocity-color relation (VCR) could lead to better SN Ia distance estimates.

\citet*{foleysanderskirshner11} examined spectra and color measurements of low redshift SNe Ia observed by the CfA Supernova Program \citep{blondin12}, and found correlations between the maximum light Si II $\lambda$6355 velocity and pseudo-equivalent width and the scatter about the mean relation between the peak absolute magnitude in $V$, controlling for light curve shape, and the apparent $B_\text{max}-V_\text{max}$ color at peak, which they interpret as due to variations of the intrinsic $B_\text{max}-V_\text{max}$ colors.   Using 65 SNe Ia, they estimate a slope of this quantity versus velocity of $-0.033$ mag $(10^3 \text{ km s}^{-1})^{-1}$, finding that lower velocity SNe Ia tended to have bluer intrinsic colors.  In a similar analysis, \citet{foley12a} studied spectra and colors of high-$z$ SNe Ia from the SDSS-II and SNLS surveys,  finding similar correlations of ejecta velocity and intrinsic colors.

\citet{blondin12} presented the spectroscopy from the CfA Supernova Program and compared the maximum light Si II velocities with the intrinsic $B-V$ colors inferred from the \textsc{BayeSN} statistical model for optical and NIR SN Ia light curves \citep{mandel11}.   This comprehensive statistical model analyzes the apparent light curve data, incorporating uncertainties due to peculiar velocities, measurement error, host galaxy reddening and extinction, and the intrinsic population distribution to infer intrinsic colors, luminosities and distances.  Regressing the estimates of intrinsic color (in units of mag) inferred from the model, with their uncertainties, against the Si II ejecta velocities (measured in units of $\text{ km s}^{-1}$) at maximum light, they found a linear slope of $-0.013 \pm 0.005$ mag $(10^3 \text{ km s}^{-1})^{-1}$ at 2.9$\sigma$ significance using 73 SNe Ia.  Analyzing the Carnegie Supernova Project data, \citet{folatelli13} estimated a similar slope between intrinsic $B_\text{max} - V_\text{max}$ and Si II velocity of $-0.012 \pm 0.016$ mag $(10^3 \text{ km s}^{-1})^{-1}$, but at lower statistical significance.  The lower significance is not surprising as they used only a small ``low-reddening'' sample   of 9 NV and 4 HV SNe Ia.

The physical origin of the heterogeneity of SN Ia ejecta velocities is still somewhat uncertain.  \citet{foleykasen11} offer a simplistic, heuristic physical explanation for the correlation between peak intrinsic $B-V$ color and Si II velocity.   Higher ejecta velocity is associated with broader Fe-group absorption lines and thus greater line opacity at wavelengths shorter than $\sim 4300$ \AA, in the $B$-band.  The $V$-band is less affected by line opacity and thus by higher ejecta velocities.  Hence, higher ejecta velocity could be correlated with redder intrinsic $B-V$ colors.  Such a trend can be seen in the asymmetric, detonating failed deflagration explosion models of \citet{kasenplewa07}.  Both the Si II ejecta velocity and the $B-V$ color at maximum are functions of viewing angle into the asymmetric explosion, and there is a roughly linear relation between the two \citep{foleykasen11}.  \citet{maeda10} found an association between the early phase velocity gradients, which are correlated with ejecta velocities, and late phase nebular velocity shifts, which probe the inner ejecta of SNe Ia.  The trend supports the hypothesis that their spectroscopic diversity is caused by the effects of viewing angle in observations of asymmetric explosions \citep{krw09}.  \citet{maeda11} found correlations between the peak $B_\text{max} - V_\text{max}$ pseudocolor, controlling for light curve decline rate, and the nebular emission line shift, further supporting a connection between SN Ia color and viewing angle into an asymmetric explosion.  Correlations between early phase colors and nebular line shift were also found by \citet{cartier11}.
However, \citet{blondin11b} examined the 2D delayed detonation models of \citet{krw09} and found that a strong correlation between the $B-V$ intrinsic color and ejecta velocity was not a generic feature of these models.    A recent study by \citet{wangx13} found that HV SNe Ia tended to occur in brighter regions, closer to the host galaxy center, and within brighter and larger host galaxies as compared to NV SNe Ia.  They suggest that these differences indicate two distinct populations of SNe Ia, possibly associated with different progenitor populations.

Accounting for a significant correlation between intrinsic colors of SNe Ia and their ejecta velocities has important consequences for estimating their dust extinction.   Estimates of dust extinction and reddening in the host galaxies of SNe Ia depend on the statistical properties of the intrinsic colors of SNe Ia, as they are determined from the difference between the observed apparent colors and the inferred intrinsic colors.    Significant correlations of ejecta velocity with intrinsic colors mean that measurement of the ejecta velocity from the SN Ia spectrum would provide additional information to estimate the intrinsic colors of individual SNe Ia more precisely.  This in turn would lead to more accurate inferences of the host galaxy dust extinction that should improve luminosity distance estimates.  Since this effect is not incorporated into current schemes for SN Ia light curve analysis, it has the potential to improve distance estimates.

Although \citet{wangx09b, wangx13} and \citet{foleykasen11} split the SNe Ia into two categories based on their Si II velocities, there is no definitive boundary separating the two since the velocities form a continuous distribution.  Since the models of \citet{kasenplewa07} predict a linear relation between intrinsic color and velocity as a function of viewing angle, it makes sense to construct a statistical model for SN Ia colors with potential linear, or even nonlinear, dependence upon ejecta velocity, treated as a continuous parameter.  In this work, we estimate the functional dependence of multiple intrinsic colors on ejecta velocity by hierarchically modeling the conditional distribution of the observed, apparent optical colors given the measured Si II velocities.  Whereas previous studies have focused on the $B-V$ color, in this work, we estimate the effect in multiple optical colors in the $BVRI$ bands simultaneously, while consistently accounting for dust reddening across these wavelengths.

In \S \ref{sec:data}, we describe our dataset of apparent colors from $BVRI$ optical light curves and spectroscopic velocity measurements for a sample of 79 nearby SNe Ia consisting of the compilation of \citet*{foleysanderskirshner11} plus more recent additional supernovae.   Comparing the apparent color distributions between HV and NV groups, we find statistically significant discrepancies in $B-V$ and $B-R$, but not in other colors.  We demonstrate that these velocity-dependent discrepancies are likely to be caused by intrinsic color differences rather than host galaxy dust extinction.

In \S \ref{sec:statmodel}, we construct a new hierarchical Bayesian regression model describing the SN Ia apparent colors and ejecta velocity data as a combination of the mean intrinsic colors-velocity relation, random intrinsic scatter, measurement error, and dust reddening.     The marginal likelihood function, the probability model for the distribution of the observed data depending on the parameters of the population, is described in \S \ref{sec:marginal_lkhd} and derived in Appendix \ref{sec:math_lkhd}.   The empirical distribution of ejecta velocities has a long tail towards high velocities.  Significant correlations of intrinsic colors with velocity generically imply a non-Gaussian marginal population distribution of intrinsic colors.
In \S \ref{sec:nongaussian}, we demonstrate the capability of our model to capture the resulting non-Gaussian population distributions of the intrinsic SN Ia colors.  In \S \ref{sec:globalpost}, the global posterior probability of the unknowns, given the observed data, is derived and depicted as a probabilistic graphical model.     Our model can be used with various hypotheses about the functional form of the relations between the intrinsic colors and ejecta velocity (\S \ref{sec:nonlinear}).  We employ  the deviance information criterion \citep[DIC;][]{spiegelhalter02} to gauge whether the more complex hypotheses are justified by their improved representation of the data.  As it has not been used previously in supernova analyses, we provide a brief introduction to the DIC in Appendix \ref{sec:modelcomp} for the unfamiliar reader.

We implement a new Gibbs sampling algorithm (Markov Chain for Regressing Colors, \textsc{MCRC}, Appendix \ref{sec:mcrc}) to compute the posterior distribution of the unknown intrinsic colors and dust extinctions for individual SNe Ia as well as the population hyperparameters, including those governing the mean relation between intrinsic colors and ejecta velocity.  In \S \ref{sec:simulations}, we validate our method on simulated data generated from  non-Gaussian distributions with different underlying intrinsic color-velocity trends to show that the true model is recovered accurately.   The simulations also show that DIC effectively discriminates between competing models for the mean intrinsic color-velocity function.

In \S \ref{sec:application}, we apply our new statistical method to our observed data and we find significant trends of intrinsic $B-V$ and $B-R$ colors versus Si II velocity. In \S \ref{sec:apply_dic}, using the DIC to evaluate the model fits,  we find that the information criteria significantly favor models with simple non-constant trends over the basic model that assumes a constant Gaussian intrinsic color distribution with no trend with ejecta velocity.  Higher order polynomial fits are disfavored.  In \S \ref{sec:implied_color_distr}, we compute the non-Gaussian shape of the population distribution of intrinsic colors implied by the fitted model relations and the ejecta velocity distribution, and estimate its skewness.  In \S \ref{sec:effect}, we show that our model capturing the intrinsic colors-velocity trend leads to significant velocity-dependent corrections to intrinsic color and dust extinction estimates for individual SNe Ia.  The accuracy and precision of these estimates are best when using both apparent colors and velocity information.  But even with only the apparent color measurements of an individual SN Ia, one can still  obtain better inferences by using the population distribution of intrinsic colors implied by the model capturing these trends.  This distribution
accounts for the skewed probability of intrinsically red events, which is underestimated by the basic Gaussian model that ignores these trends.
In \S \ref{sec:dust_impact}, we demonstrate the significant velocity-dependent corrections to the dust extinction estimates for the full sample.  We conclude in \S \ref{sec:conclusion}.

\section{Apparent Colors and Velocity Data}\label{sec:data}

In this section, we describe our dataset consisting of a large sample of low redshift $(z < 0.06)$ SNe Ia with observed spectra and $BVRI$ light curves.   The bulk of our sample was compiled by \citet*{foleysanderskirshner11}.  They gathered photometric measurements of nearby SNe Ia drawn from the data compilations of \citet{hamuy96_29sne, jha07}, CfA1-3, \citep{riess99, jha06,  hicken09a} and LOSS \citep{ganeshalingam10}.  The spectroscopic measurements were drawn from the dataset from the CfA Supernova Program \citep{matheson08,blondin12} as well as from the literature.   \citet{foleysanderskirshner11} presented an empirical model for the velocity evolution of SNe Ia that we used to interpolate the spectroscopic  measurements  from the time of spectroscopic observation to the time of maximum light in $B$.  This method is only accurate for SNe Ia with mid-range decline rates between $1 \le \Delta m_{15}(B) < 1.5$ and requires a spectrum observed at a rest-frame phase of $-6 \leq t \leq 10$ days relative to maximum light.  

\citet{foleykasen11} found that high- and normal-velocity SNe Ia seemed to be subject to different dust reddening laws (parameterized by $R_V$) only when highly reddened SNe Ia were included.    Such highly reddened SNe Ia are not seen at high redshift and typically not used in cosmological analyses.
The set retaining objects with low to moderate reddening (within the apparent color range $B_\text{max} - V_\text{max} \le 0.32$ or $(B-V)_\text{max} \le 0.36$), was adequately fit with a single reddening law for both velocity groups (with $R_V \approx 2.5$).  The cuts on decline rate $\Delta m_{15}(B)$, peak apparent color, and the requirement of a spectrum near maximum light exclude about one-third of the nearby, normal SNe Ia in the photometric light curve sample compiled by  \citet*{foleysanderskirshner11}.   We adopt the 65 SNe Ia that remain after imposing these cuts, with both $BV$ photometry and Si II velocity measurements, as selected by \citet{foleysanderskirshner11}, with the following modifications and additions.

In contrast to previous studies that investigated only the $B-V$ color dependence on ejecta velocity, we analyze multiple colors from $BVRI$ light curves.  Hence, we require measurements in all four filters.  This removed two SNe Ia (SN 1992ag and SN 1981B) from the sample.  We added 17 more recent SNe Ia,  from the LOSS \citep{ganeshalingam10}, CfA4 \citep{hicken12} and CSP \citep{stritzinger11} samples, that have light curves consistent with the same color, decline rate cuts, and spectral phase cuts above.    The Si II spectroscopic measurements for these were presented in \citet{foley12b} or will be presented in a forthcoming paper (R. Foley, 2014, in prep.).   

We omitted the highest velocity object in the sample, SN 2004dt, with $|v| = 15,928 \text{ km s}^{-1}$, which has unusual spectroscopic characteristics.  Its Si II $\lambda 6355$ absorption feature likely contains two components at different velocities, and is the SN Ia with the highest measured polarization \citep{lwang06-04dt,altavilla07,leonard05}.  It is a significant outlier in the relation between the late-phase nebular line shift and early-phase Si II velocity or velocity gradient \citep{maeda10, blondin12}.  \citet{maeda10} note that the late-phase spectrum of SN 2004dt resembles that of the peculiar SN 1991bg, which defines a faint SN Ia subclass \citep{filippenko92,leibundgut93}.  Hence, SN 2004dt appears to be spectroscopically distinct from most normal SNe Ia, even those in the HV class \citep*{foleysanderskirshner11}.   It is also well-separated in velocity from the rest of the  sample.  Conservatively, we omit SN 2004dt and restrict our conclusions to the densely sampled Si II velocity range  $9,300 \text{ km s}^{-1} < |v| < 14,700 \text{ km s}^{-1}$.

For the final sample of 79 SNe Ia, we estimated the peak apparent colors at the date of $B$ maximum from \textsc{BayeSN} fits \citep{mandel11} to their multi-band optical  light curve data.  These fits include corrections for Milky Way dust as well as redshift-dependent $K$-corrections between the observer-frame filters and rest-frame $BVRI$ filters, including cross-filter corrections between e.g. observer $ri$ or $r'i'$ and rest-frame $RI$.  We account for 0.02 mag of error in these corrections for each light curve point, in addition to its photometric uncertainty, which was typically a few hundredths of a magnitude. Since these SNe Ia are all at roughly the same (and at low) redshifts, we expect that any systematic errors in $K$-corrections to be minimal.  The estimated error on the fitted peak apparent color depended on the light curve for each SN Ia, but had a median value of 0.04 mag.

\begin{figure}[t]
\centering
\includegraphics[angle=0,scale=0.35]{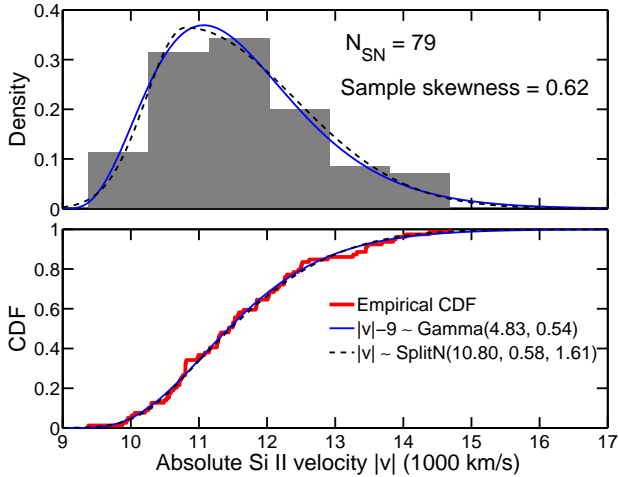}
\caption{\label{fig:siII_vel_distr_n79} Empirical distribution Si II velocity measurements for the sample of 79 nearby SNe Ia. (Top) The histogram of velocities (grey), the best-fitting Gamma distribution, with shape parameter $a=4.83$ and scale parameter $b=0.54$ (blue), and the best-fitting split-normal distribution (Eq. \ref{eqn:splitgaussian}) with mode $\tilde{\mu} = 10.8$, left-width $\sigma_{-} = 0.58$, and right-width $\sigma_{+} = 1.61$, in units of $10^3 \text{ km s}^{-1}$ (black). (Bottom) The empirical cumulative distribution function of the velocity measurements (red), and the CDFs of the best-fitting distributions.}
\end{figure}

The empirical distribution of this sample of Si II velocities is shown in Fig. \ref{fig:siII_vel_distr_n79}.   The maximum-light absolute ejecta velocities $|v|$ range from $9,377$ to $14,685 \text{ km s}^{-1}$.  The typical velocity measurement error is $250 \text{ km s}^{-1}$ \citep{foleysanderskirshner11}.  Since the velocity error is much smaller than the range of velocities, we can neglect it in our regression.   The empirical distribution of velocities is non-Gaussian and skewed, with a long tail towards higher velocities.   We  show the best-fitting Gamma distribution, with maximum likelihood estimates of its shape and scale parameters. The Gamma distribution is an excellent approximation to the skewed distribution of velocities.  (A Gamma random variable $\Gamma_{a,b}$ with shape parameter $a$ and scale parameter $b$ has probability density proportional to  $p(x) \propto x^{a-1} \exp(-x/b)$ on the domain $x>0$.) We will use this to simulate velocity data in \S \ref{sec:simulations}.

The sample skewness of the empirical distribution of the absolute Si II velocities is $0.62 \pm 0.19$, where the uncertainty was estimated using bootstrap resampling.  For comparison, a Gaussian distribution has zero skewness, since its tails are symmetric.  We  quantify the asymmetry in the tails by fitting a split-Normal distribution $\mathcal{SN}(\tilde{\mu}, \sigma_{-}, \sigma_{+})$ with probability density
\begin{equation}\label{eqn:splitgaussian}
\begin{split}
P( x &| \, \tilde{\mu}, \sigma_{-}, \sigma_{+} ) \\ &= \frac{2}{\sigma_{-} + \sigma_{+}}  \begin{cases} (\sigma_{-}) \, N( x | \, \tilde{\mu}, \sigma^2_{-}), & x \le \tilde{\mu} \\ (\sigma_{+}) \, N( x | \, \tilde{\mu}, \sigma^2_{+}), & x > \tilde{\mu} \end{cases} 
\end{split}
\end{equation}
to the velocity data.  In this asymmetric model distribution, the mode is $\tilde{\mu}$, and the widths of the half-Gaussians to the left and right of the mode are $\sigma_{-}$  and $\sigma_{+}$, respectively.  A Gaussian probability density in $x$ with mean $\mu$ and variance $\sigma^2$ is denoted by $N( x | \, \mu, \sigma^2)$.   A maximum likelihood fit to the absolute velocities $|v|$ (in units of $10^3 \text{ km s}^{-1}$) yields $\tilde{\mu} = 10.8 \pm 0.20$, $\sigma_{-} = 0.58 \pm 0.13$, and $\sigma_{+} = 1.61 \pm 0.18$ (also shown in Fig. \ref{fig:siII_vel_distr_n79}).  The high velocity tail is thus almost three times longer than the low velocity tail.  Adopting the \citet{wangx09b} division between HV and NV at $v_0 = -11,800 \text{ km s}^{-1}$, there are 47 SNe Ia with normal velocities $|v| < |v_0|$ and 32 SNe Ia with high velocities $|v| > |v_0|$.

\begin{figure}[t]
\centering
\includegraphics[angle=0,scale=0.28]{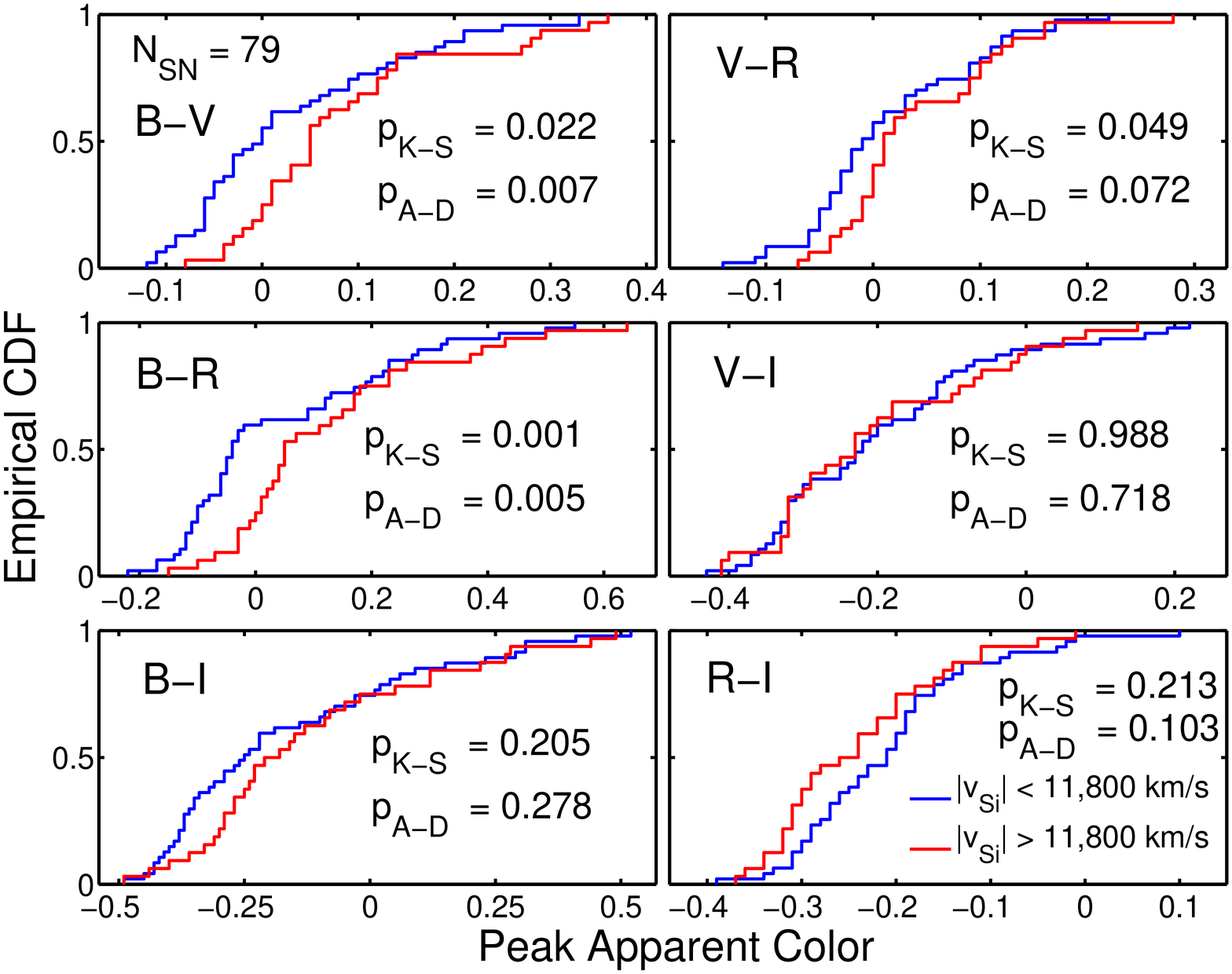}
\caption{\label{fig:plot_appcol_ecdfs_all_n79} Empirical cumulative distribution functions of the apparent SN Ia optical colors at maximum light.   The red CDFs include only high-velocity SNe Ia, and the blue CDFs include only normal-velocity SNe Ia.   We indicate the $p$-values resulting from two-sample Kolmogorov-Smirnov and Anderson-Darling tests comparing the high-velocity versus normal-velocity samples for each color.  The tests show a significant discrepancy in color distribution for $B-V$ and $B-R$.  In particular, the ``blue edge'' (left tail) of these distributions is redder (more positive) for high-velocity SNe Ia than for normal velocity SNe Ia.  The discrepancies in the apparent color distributions are not statistically significantly in the $B-I$, $V-R$, $V-I$, and $R-I$ colors.}
\end{figure}

In Figure \ref{fig:plot_appcol_ecdfs_all_n79}, we illustrate the differences in the distribution of the \emph{apparent} colors between the high velocity subgroup and the normal velocity subgroup.  The apparent color distributions of HV and NV objects have roughly similar shapes, but appear to be offset in some colors, notably $B-V$ and $B-R$.  We perform the two-sample Kolmogorov-Smirnov (K-S) and Anderson-Darling (A-D) tests \citep{scholzstephens87} to evaluate the statistical significance of discrepancies  between the empirical cumulative distribution functions (CDFs) of the apparent colors of each velocity group, treating each color separately. The K-S test uses the maximum absolute difference between the empirical CDFs to compare the samples and is most sensitive to differences in the middle of the distributions, while the A-D test uses a weighted squared difference integrated over the whole distribution, and is more sensitive to deviations in the tails.  The differences between the velocity groups are statistically significant in $B-V$ ($p_{KS} = 0.022$) and $B-R$  ($p_{KS} = 0.001$).  The ``blue edges'' (left tails) of these color distributions are redder (more positive) for the high-velocity SNe Ia.    The differences are not statistically significant in the other colors.  The same conclusions are reached using either the K-S test or A-D test.

We estimate the relative apparent color shifts of the observed data, $\Delta O_\text{obs}$, for each color, by finding the constant that, when subtracted from the apparent colors of the HV sample, minimizes the K-S distance between the NV and HV samples.  These estimated color shifts are shown in Table \ref{table:ecdfs}.  To estimate the uncertainty of each color shift, we bootstrap resampled $10^3$ pairs of mock NV and HV color datasets from the observed NV and HV datasets, respectively.  For each pair of NV and HV mock datasets, we estimated the color shift.  From the resulting distribution of $10^3$ color shift estimates, we computed the standard deviation.

If HV and NV SNe Ia had the same distributions of intrinsic colors and were reddened by the same distributions of line-of-sight extrinsic host galaxy dust, then we would expect their apparent color distributions to be  consistent.  It is unlikely that the discrepancies at the blue edges are caused by an overall larger extrinsic dust reddening of only the HV population; it is more plausible that the intrinsic color of the SN is correlated with the ejecta velocity, which is also an intrinsic physical property of the SN.   If the mean intrinsic colors of HV and NV objects were the same, and the difference in the blue edges were  caused only by dust reddening, then one would need an extra overall host galaxy dust extinction of $\Delta A_V \approx 0.15$ mag  exclusively for the HV objects to explain the significant $0.06$ and $0.085$ mag offsets in the apparent $B-V$ and $B-R$ distributions between the two velocity groups.

However, this extra dust extinction of $\Delta A_V \approx 0.15$ mag would also cause relative color excesses $\Delta E_\text{dust}$ in $B-I$, $V-R$, $V-I$, and $R-I$ between the two velocity groups, as shown in Table \ref{table:ecdfs} (assuming a CCM dust law with $R_V = 2.5$).   While the $V-R$ distributions are consistent with a small color excess due to dust, the $B-I$, $V-I$, and $R-I$ distributions are not.  The apparent $V-I$ color distributions do not exhibit any relative color excess, and are completely consistent, between the NV and HV groups $\Delta O_\text{obs} = (0.00 \pm 0.03)$.  The estimated color shift in $R-I$ appears to be negative, whereas the dust would cause a positive shift.

We ran a set of simulations to assess the discrepancy between the observed color shifts and the expected dust reddening, while accounting for  the variance and finite sampling of the empirical color distributions.   In each simulation, we generated mock NV and HV apparent color data sets of the same size as the observed data sets, under the hypothesis that the two groups have the same intrinsic color distribution, but the HV colors are reddened by an overall extra $\Delta A_V \approx 0.15$ mag of dust extinction relative to NV objects.  We bootstrapped a mock NV apparent color sample by resampling from the observed NV distribution.  We then bootstrapped a mock HV apparent color sample by resampling from the observed NV color distribution, and then adding the expected dust reddening in each color for $\Delta A_V \approx 0.15$ mag and $R_V = 2.5$.  For each simulation, we estimated the relative color shift $\Delta O_\text{sim}$ between the mock NV and HV apparent color data by minimizing the K-S distance between their empirical CDFs.  This procedure was repeated for $10^3$ simulations to compute the distribution of estimated color shifts $\Delta O_\text{sim}$ for these mock data sets.  We then compared these distributions of $\Delta O_\text{sim}$ to the estimated $\Delta O_\text{obs}$ for the observed data set.   Specifically, we computed the fraction of simulations with estimated shift equal to or less than the observed shift, $p_\text{sim} \equiv P(\Delta O_\text{sim} \le \Delta O_\text{obs})$, shown in Table \ref{table:ecdfs}.   Values of $p_\text{sim}$ close to 0.5 indicate that the observed shift is consistent with expected distribution under the dust hypothesis; whereas small values of $p_\text{sim}$ indicate that the observed shift is in the far left tail of the expected distribution.  The small estimated $V-R$ shift was consistent with the simulations, but  $\Delta O_\text{obs}$ for $(B-I, V-I, R-I)$ were much smaller than the mean of the $\Delta O_\text{sim}$ distributions, with tail probabilities $p_\text{sim} = (0.05, 0.01, 0)$, respectively.     The observed multicolor distributions of NV and HV SNe Ia are inconsistent with the hypothesis that the significant $B-V$ and $B-R$ discrepancies are caused by $\Delta A_V \approx 0.15$ mag dust extinction for HV objects.

\begin{deluxetable}{lrrrr}
\tabletypesize{\small}
\tablecaption{Statistics of HV and NV peak Apparent Color Distributions}
\tablewidth{0pt}
\tablehead{ \colhead{Color}  & \colhead{std\tablenotemark{a}} & \colhead{$\Delta O_\text{obs}$ (mag)\tablenotemark{b}} & \colhead{$\Delta E_\text{dust}$\tablenotemark{c}} & \colhead{$p_\text{sim}$\tablenotemark{d}}  }
\startdata
$B-V$  & 0.11 & $0.061 \pm 0.021$ & 0.061 & 0.49 \\ 
 $B-R$ & 0.18 & $0.085 \pm 0.031$ & 0.085 & 0.54\\
 $B-I$ & 0.25 & $0.066 \pm 0.043$ & 0.125 & 0.05 \\
 $V-R$ & 0.08 & $0.025 \pm 0.015$  & 0.028 & 0.51 \\ 
 $V-I$ & 0.15 & $0.000 \pm 0.031$ & 0.067 & 0.01 \\
 $R-I$ & 0.09 & $-0.037 \pm 0.021$ & 0.039 & 0.00
 \enddata
\tablecomments{\label{table:ecdfs} See \S \ref{sec:data} for details.}
\tablenotetext{a}{Sample standard deviation of the observed color distribution (in mag).  The values for the NV and HV subsets are consistent within 0.01 mag with the value for the full dataset.}
\tablenotetext{b}{Estimated apparent color shift obtained by minimizing the K-S distance between the NV and shifted HV color distributions.}
\tablenotetext{c}{Relative color excess (in mag) due to dust reddening if there were an overall $\Delta A_V \approx 0.15$ mag for the HV SN Ia relative to the NV SN Ia, assuming $R_V = 2.5$.  This value was chosen to match the significant observed color shifts in $B-V$ and $B-R$.}
\tablenotetext{d}{We bootstrapped 1000 mock data sets simulated with dust reddening $\Delta E_\text{dust}$ for HV objects relative to NV objects.  $p_\text{sim} \equiv P(\Delta O_\text{sim} \le \Delta O_\text{obs})$ is the fraction of simulations with estimated color shifts $\Delta O_\text{sim}$ less than or equal to that of the observed data $\Delta O_\text{obs}$.}
\end{deluxetable}

This indicates that the statistically significant discrepancies in the apparent $B-V$ and $B-R$ color distributions are not caused by extrinsic host galaxy dust, but are intrinsic to the SNe Ia.  The apparent color distributions in the $V-R$, $V-I$, and $R-I$ are not significantly discrepant between the velocity groups. This suggests that the color-velocity effects originate mainly in the SN Ia spectra at $B$-band wavelengths.

The empirical CDFs of the apparent $B-I$ colors also exhibit a ``blue edge'' discrepancy between the HV and NV velocity groups, as one might expect if the color-velocity effects originate in the $B$-band.  However, this difference is not statistically significant under a K-S test, with $p_{KS} = 0.205$.  This may seem surprising, since the NV and HV apparent $B-V$ distributions are discrepant, their apparent $V-I$ distributions are consistent, and $(B-I) = (B-V) + (V-I)$.  The detectability of a relative shift between two distributions depends on the size of the shift relative to the width of each distribution.  The sample standard deviations of the observed color distributions  (Table \ref{table:ecdfs}) are consistent between the NV and HV samples.   The standard deviations of the observed $B-V$ and $B-R$ distributions are $0.11$ and $0.18$ mag, respectively.  Their estimated relative color shifts between NV and HV SNe Ia is roughly 50\% of the widths of their distributions ($0.06/0.11$ and $0.085/0.18$) and thus relatively easy to detect.  In contrast, the sample standard deviation of the apparent $B-I$ distribution is a much larger $0.25$ mag, while the estimated relative shift is only $0.066$ mag, or about 25\% of the width of the distribution.  Even if the true relative $B-I$ intrinsic color shift between NV and HV SNe Ia was actually 0.066 mag, it would be much harder to detect than in $B-V$ and $B-R$.

We ran another set of $10^3$ simulations to assess the detectability of an intrinsic color shift, while accounting for the variance of the color distributions and finite sampling.  We bootstrapped  pairs of mock NV and HV color data sets from the observed NV color data, and then added to the HV colors a relative color shift equal to the estimated color shift $\Delta O_\text{obs}$ in the actual observed data (Table \ref{table:ecdfs}).   For each pair, we computed the K-S statistic to compare the empirical color distributions of the NV and HV samples.  For $B-V$ and $B-R$, a statistically significant discrepancy ($p_\text{KS} < 0.05$) was found in greater than 90\% of the simulations.  In $B-I$, however, $p_\text{KS} < 0.05$ was only found in 40\% of the simulations.  Thus, even if the true $B-I$ intrinsic color shift were 0.066 mag, it would be unlikely to be consistently detected in similar datasets. We reached the same conclusions using the A-D test rather than the K-S test.  A larger sample may help determine the reality of the discrepancy in apparent $B-I$ distributions between HV and NV groups.

Although splitting the sample into HV and NV groups is convenient to illustrate these color differences, the division between the two is arbitrary, and the ejecta velocity is a continuous parameter and its empirical distribution (Fig. \ref{fig:siII_vel_distr_n79}) does not strongly suggest distinct velocity groups.  Furthermore, analyzing each apparent color separately in this way ignores cross-color information in the data:  SNe Ia that are redder (more positive) or bluer (more negative) in one color are also likely to be redder or bluer in other colors.   In the following sections, we develop and apply a hierarchical Bayesian regression model for the dependence of multiple intrinsic colors of a SN Ia on the continuously distributed ejecta velocity, using the observed, apparent data.

\section{The Statistical Model}\label{sec:statmodel}

We adopt a hierarchical Bayesian, or multi-level modeling, framework  to build a structured probability model describing the multiple random effects that produce the observed data. This principled strategy enables us to coherently model and make inferences at both the level of an ensemble or population of objects as well as at the level of individuals from the ensemble \citep{gelman_bda, loredohendry10, loredo12,mandel_scmav}.   The hierarchical Bayesian approach was first applied to SNe Ia by \citet{mandel09,mandel11} to model  optical and the near-infrared light curves, and to improve inferences on host galaxy dust extinction and the precision of distance predictions.   \citet{march11} describe a hierarchical Bayesian model for fitting the SN Ia Hubble diagram using SALT2 parameters \citep{guy07}.   Hierarchical Bayesian statistical models for SN Ia colors were developed by \citet{mandelthesis} and recently by \citet{burns14}.  Other recent astrophysical applications of hierarchical Bayesian modeling are described by \citet{hogg10, kelly12, shetty13, foster13, brewer14}; and \citet{sanders14}.

In this paper, our primary statistical task is that of regression: modeling and estimating the relation between the dependent variables (colors) and independent covariates (velocities) based on observed data.  Hierarchical linear regression in which the observables are affected by Gaussian intrinsic scatter around the mean relation and measurement error has been discussed elsewhere (e.g. \citealt{bkelly07, march11}).  Here, we consider regression of observables that have measurement error and intrinsic scatter about the mean relation, but are also affected by non-Gaussian, asymmetric deviations caused by positive dust reddening.

We build a hierarchical model for the multiple random and uncertain effects underlying the SN Ia data: measurement error, reddening of SN Ia colors due to host galaxy dust, and the variation and correlation of intrinsic SN Ia colors and their dependence upon spectroscopic variables.   This statistical model is used to perform coherent probabilistic inference of the populations and individuals underlying the ensemble of SN Ia data.  The unknowns we want to estimate are the parameters of individual SNe Ia (their intrinsic colors and dust extinctions), and the hyperparameters describing the intrinsic SN Ia population and the extrinsic host galaxy dust distribution. Inference with the hierarchical model may be thought of as a probabilistic deconvolution of the observed SN data into the multiple, unobserved, latent random effects generating it.

Bayesian models for SN Ia apparent color distributions typically assume that the shape of the intrinsic color distribution is Gaussian \citep{jha07, mandel09, mandel11}.  Convolving this with an asymmetric (e.g. exponential) distribution for (positive) dust reddening yields the likelihood function for the apparent color distribution.    The intrinsic colors of SNe Ia may also be correlated with other observable covariates, e.g. spectroscopic line velocities and equivalent widths \citep{foleykasen11, foleysanderskirshner11, blondin12, mandelthesis}.  However, these covariates have non-Gaussian distributions.  In particular, the empirical distribution of Si II velocities has a positive skew (long tail) towards higher absolute velocities.  In the simple case that the relation between intrinsic colors and velocity is linear, one should expect  the shape of the intrinsic color distribution to be similarly skewed, as we demonstrate in \S \ref{sec:nongaussian}.   An incorrect assumption of the Gaussianity of the intrinsic color distribution will then tend to discount (underestimate the probability of) very intrinsically red events at high Si II ejecta velocities, leading to biased estimates of dust extinction, and may reduce the inferred correlation and its estimated statistical significance.   

We formulate a statistical model for SN Ia colors and velocities that allows for their non-Gaussianity.  It enables the non-Gaussianity in the velocity distribution to be reflected in the implied intrinsic color distribution.   We do this by modeling the conditional probability of the colors given the velocity, rather than the joint distribution of colors and velocity.  This can be done because the spectroscopic velocities are well measured, so we do not need to assume a model for their distribution.   In the simplest non-trivial case, we assume that the mean intrinsic colors are a linear function of velocity (\S \ref{sec:linear}),  but the model is also easily extended to nonlinear functions of velocity (e.g. polynomial and step functions, \S \ref{sec:nonlinear}).  Hence, the method can be used with any arbitrary distribution of spectroscopic velocities, and a flexible family of nonlinear relations between intrinsic colors and velocity.  The model is general and could be used to estimate correlations between intrinsic colors and any well-measured independent variable using apparent color data.

In the following subsections, we lay out the modeling assumptions  relating the apparent color data to the intrinsic colors and dust reddening of individual supernovae, as well as the population models for the dust extinction and the mean trends of intrinsic colors versus ejecta velocity.   Together, these assumptions describe the marginal likelihood, or the probability distribution of observed color data of the SN Ia ensemble.  The global posterior probability density, derived from the modeling assumptions and Bayes' Theorem, provides a unified measure of the joint uncertainties in the unknowns given the observed data and a clear objective function for the analysis. It quantifies the trade-offs and degeneracies in inference between competing latent effects, e.g. the intrinsic color and  dust reddening, underlying the data.

\subsection{Model Assumptions: Linear Intrinsic Color-Velocity Correlation}\label{sec:linear}

We have a vector of measurements $\bm{O}_s$ of the $n_C$ apparent colors at the time of maximum light, $T_{B\text{max}}$ (e.g. apparent $B-V$, $B-R$, $B-I$) for each supernova $s$ in a set of $s = 1\ldots N_\text{SN}$ objects.  We also have well-measured estimates of their ejecta velocities $v_s$, from the Si II absorption line, so that their error may be ignored.   The observed, apparent colors of SN $s$ are the combinations of the intrinsic colors $\bm{C}_s$, the dust reddening, and measurement error:
\begin{equation}\label{eqn:structeqn}
\bm{O}_s = \bm{C}_s  + A_V^s \bm{\gamma}(R_V) + \bm{\epsilon}_s.
\end{equation}
We assume that the color measurement error $\bm{\epsilon}_s$ is a zero-mean Gaussian random variable with known covariance: $\bm{\epsilon}_s \sim N(\bm{0}, \bm{W}_s)$.   The measurement covariance matrix $\bm{W}_s$ will generically contain non-zero off-diagonal terms encoding the correlations between the color measurements.  For example, if the apparent magnitudes in $B, V, R$, and $I$ (at time of maximum light in $B$) are estimated independently with the same measurement variance, each pair of resulting colors in the set $(B-V, B-R, B-I)$ will have a 50\% correlation in the covariance matrix $\bm{W}_s$.  Our analysis accounts for these correlations to encode the fact that the color measurements are not independent.

(Note that, in this work, the intrinsic colors $\bm{C}_s$ are latent variables referring to the part of the total observed colors attributed to the SN Ia without any dust reddening or measurement error.   In other supernova contexts, $c$ refers to the color parameter in the SALT2 model \citep{guy07}, which is a proxy for the peak \emph{apparent} $B-V$ color, inclusive of dust reddening.  In those contexts, the ``intrinsic color'' may refer to the latent ``true'' apparent color unaffected by measurement error.)

The second term on the right side of Eq. \ref{eqn:structeqn} describes the reddening effect of host galaxy dust extinction $A_V$ on each color through the assumed reddening law \citep[CCM;][]{ccm89}: $\bm{\gamma}(R_V) = ( \Delta \bm{\alpha} + \Delta \bm{\beta}/R_V )$.  The coefficients of this reddening law were obtained from \citet{jha07}, who examined the effect of dust reddening on SNe Ia spectra within each filter.   The host galaxy dust extinction $A_V$ is assumed to be drawn from an exponential distribution with average $\tau$: $A_V^s \sim \text{Expon}(\tau)$ \citep{jha07}.  This has a probability density of $P(A_V | \tau) = \tau^{-1} \exp(A_V /\tau)$ for $A_V > 0$ and zero otherwise, as dust only causes dimming and reddening.  \citet{mandel11} found that this model describes well the distribution of peak apparent $B-V$ colors of nearby SNe Ia up to $B-V \lesssim 1$.  

We model the mean relation $\bm{\mu}_C(v; \bm{\theta})$ between the vector of intrinsic colors $\bm{C}_s$ and the velocity, with some intrinsic scatter about the average trend:
\begin{equation}\label{eqn:intcol_dependence}
\bm{C}_s = \bm{\mu}_C(v_s; \bm{\theta})+ \bm{\epsilon}^C_s,
\end{equation}
where $\bm{\theta}$ are hyperparameters governing the regression relation.
If the mean intrinsic colors are linear functions of velocity, then  
\begin{equation}\label{eqn:intcol_linear}
\bm{\mu}_C(v; \bm{\theta}) = \bm{c}_0 + \bm{b}(v - v_0),
\end{equation}
and $\bm{\theta} = (\bm{c}_0, \bm{b})$.
This function models the \emph{conditional mean} of the intrinsic colors given the known covariate $v$.  A characteristic Si II velocity is $v_0 = -11,800 \text{ km s}^{-1}$.  The expected intrinsic colors at $v_s = v_0$ are given by the offsets $\bm{c_0}$, and the slopes of intrinsic colors versus velocity are $\bm{b}$.   Since the response variables $\bm{C}_s$ are vectorial, this is equivalent to a multiple-outcome linear regression model, with trends for each scalar component.  The trends may not be exact, and we expect some intrinsic random scatter about the mean trends that is uncorrelated with ejecta velocity.
We assume that the scatter term is Gaussian distributed about the linear trend: $\bm{\epsilon}_s^C \sim N(\bm{0}, \bm{\Sigma}_C)$.  The residual scatter covariance matrix $\bm{\Sigma}_C$ allows for the scatter about the linear trend to be correlated between different colors.  This covariance matrix is composed of the standard deviations of the residual color scatter $\bm{\sigma}_C$ and the correlation matrix $\bm{R}_C$: $\bm{\Sigma}_C = \text{diag}(\bm{\sigma}_C) \bm{R}_C \, \text{diag}(\bm{\sigma}_C)$.  For example, for a given Si II velocity, the deviation of the true intrinsic $B-V$ from the mean trend may be correlated with the deviation of the true intrinsic $B-R$ from the trend.  These residual intrinsic correlations would be captured in the off-diagonal elements of $\bm{R}_C$.

\subsection{Generalizations}\label{sec:nonlinear}

If the mean intrinsic color-velocity relation $\bm{\mu}_C(v; \bm{\theta})$ in Eq. \ref{eqn:intcol_dependence} is linear in the hyperparameters $\bm{\theta}$, then one can construct a $n_C \times \dim(\bm{\theta})$ matrix function of the covariate $v_s$, $\bm{M}_s = \bm{M}(v_s)$, such that $\bm{\mu}_C(v; \bm{\theta}) = \bm{M}_s \bm{\theta}$.  Then we can write the likelihood function for a set of intrinsic colors as
\begin{equation}\label{eqn:intcol_lkhd}
P(\bm{C}_s | \, v_s; \bm{\theta},  \bm{\Sigma}_C) = N( \bm{C}_s | \, \bm{M}_s \bm{\theta}, \bm{\Sigma}_C), 
\end{equation}
where $N( \bm{x} | \,\bm{\mu}, \bm{\Sigma})$ denotes a multivariate Gaussian probability density for the random vector $\bm{x}$ with mean $\bm{\mu}$ and covariance $\bm{\Sigma}$. In the linear case of Eq. \ref{eqn:intcol_linear}, this matrix is
\begin{equation}
\bm{M}(v_s) = \left[ \bm{I}_{n_C}, \bm{I}_{n_C} (v_s-v_0) \right],
\end{equation}
a horizontal concatenation of the identity matrix of dimension $n_C$, and the identity matrix times the covariate.  In this case, $\bm{\theta} = (\bm{c_0}, \bm{b})$ is a column vector of the hyperparameters (the intercepts and slopes) describing the mean intrinsic color-velocity function.  For the linear model, $\bm{\theta}$ contains $2 n_C$ scalar parameters.

We can easily extend the formalism to non-linear functions of the velocity, but it is computationally convenient to choose such nonlinear functions of velocity that retain a linear dependence on the hyperparameters $\bm{\theta}$.  We need only define the conditional mean function $\mathbb{E}[\bm{C} | \, v]  = \bm{\mu}_C(v \,; \bm{\theta})$, the covariate matrix $\bm{M}_s$, and the hyperparameters $\bm{\theta}$ in a way such that the intrinsic color likelihood can be written in the form of Eq. \ref{eqn:intcol_lkhd}.

\subsubsection{Polynomial dependence}\label{sec:poly}
To model a nonlinear polynomial dependence of order $p$ of the conditional mean intrinsic colors on the scalar covariate $v_s$, we write:
\begin{equation}\label{eqn:poly}
\bm{\mu}_C(v; \bm{\theta}) = \bm{c}_0 + \sum_{j=1}^p \bm{b}_i (v - v_0)^j
\end{equation}
and $\bm{\theta} = (\bm{c}_0, \bm{b}_1, \ldots \bm{b}_m)$. 
The linear case is obtained with $p=1$.  The $p=0$ case assumes that the mean intrinsic color is a constant with respect to ejecta velocity.

\subsubsection{Step Function dependence}\label{sec:step}

A step-function dependence of the conditional mean intrinsic colors on the scalar covariate $v$, with a discontinuous step at $v = v_0$, is written as
\begin{equation}\label{eqn:stepfcn}
\bm{\mu}_C(v; \bm{\theta}) = \begin{cases} \bm{\theta}_{HV}, & |v| > v_0 \\ \bm{\theta}_{NV}, & |v| \le v_0 \end{cases},
\end{equation}
with $\bm{\theta} = ( \bm{\theta}_{HV}, \bm{\theta}_{NV} )$ denoting the mean intrinsic color in high velocity and normal velocity groups.

\subsubsection{Multiple covariates}\label{sec:mult_covariates}

Suppose we have vectors $\bm{v}_s$, with $m$ covariates for each supernova $s$: $v_s^i$, $i = 1,\ldots, m$.  A multi-linear dependence of the intrinsic colors on these vectors is written as
\begin{equation}
\bm{\mu}_C(\bm{v}; \bm{\theta}) = \bm{c}_0 + \sum_{i=1}^m \bm{b}_i (v^i - v_0^i) 
\end{equation}
where $\bm{\theta} = (\bm{c}_0, \bm{b}_1, \ldots \bm{b}_m)$, and $v_0^i$ is some characteristic value for the $i$th  covariate. For example, \citet{foleysanderskirshner11} compiled measurements of velocities and pseudo-equivalent widths of Si II and Ca H\&K lines.  This model could be used to examine the dependence of SN Ia colors on these multiple spectroscopic measurements simultaneously.

\subsection{Non-Gaussian Population Distributions of Intrinsic Color}\label{sec:nongaussian}

Critically, we have not assumed a specific shape for the population distribution of intrinsic colors $\{\bm{C}_s\}$, nor for the distribution of velocities $\{v_s\}$. Rather, the shape of the intrinsic color distribution reflects that of the velocity distribution when there is a significant trend between the two quantities.   Specifically, if the intrinsic scatter term were always zero, then for any arbitrary distribution of velocities, $P_V(v)$, Eq. \ref{eqn:intcol_linear} implies a distribution for the intrinsic colors that is a scaled and shifted version of $P_V(v)$.   If the residual intrinsic scatter is significant, then the implied intrinsic color distribution results from a scaled and shifted version of $P_V(v)$ convolved with a Gaussian distribution with width and shape given by $\bm{\Sigma}_C$.  In the case where the slopes $\bm{b}$ are all zero, then the model automatically reverts to the assumption that the intrinsic color distribution is jointly Gaussian with mean $\bm{c}_0$ and covariance matrix $\bm{\Sigma}_C$.

\begin{figure}[t]
\centering
\includegraphics[angle=0,scale=0.353]{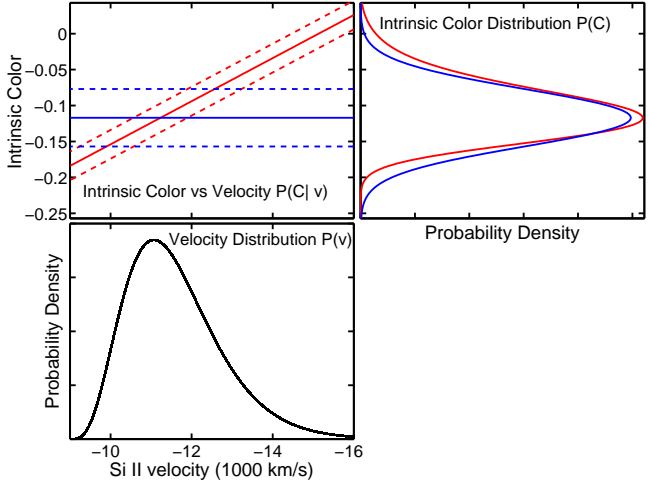}
\caption{\label{fig:pv_pcv_pc_linear} Illustration of the intrinsic color distribution $P(C)$ implied by the Si II velocity distribution $P_V(v)$ and an assumed linear relation between the mean intrinsic color and velocity.  (bottom left)  The Si II velocity distribution is a gamma distribution with a long tail towards high velocities.   (top left)  The red solid line is a mean linear relation with slope $b =-0.03$ mag $(10^3 \text{ km s}^{-1})^{-1}$ and intercept at $v = -11,800 \text{ km s}^{-1}$ of $c_0 = -0.10$ mag, with a residual intrinsic scatter of $\sigma_C = 0.02$ mag (red dashed).  The blue solid line is a mean constant relation with zero slope and some intrinsic scatter (blue dashed).  (top right)  With zero slope (blue line), the implied marginal intrinsic color distribution is Gaussian.  With nonzero slope (red line) the intrinsic color distribution has a skewed distribution with a tail towards redder intrinsic color.  The intrinsic mean and variance for the blue-line model were chosen so that the resulting red-curve and blue-curve $P(C)$ distributions  would match in mode and variance.  The model with the strong linear trend implies a non-Gaussian intrinsic color distribution $P(C)$  with a longer red (positive) tail and a shorter blue (negative) tail than the model with no trend.}
\end{figure}

For fixed hyperparameters of the SN Ia population, $\bm{\theta}, \bm{\Sigma}_C$, and a conditional mean function $\bm{\mu}_C(v; \bm{\theta})$, the implied marginal intrinsic color population distribution is found by integrating the joint distribution $P( \bm{C}_s, v_s | \bm{\theta}, \bm{\Sigma}_C)$ over the  velocity distribution $P_V(v)$:
\begin{equation}\label{eqn:implied_intrcolor_distr}
P( \bm{C}_s | \, \bm{\theta}, \bm{\Sigma}_C ) = \int dv \, P( \bm{C}_s | \, v;  \bm{\theta}, \bm{\Sigma}_C) \, P_V(v).
\end{equation}
In Figure \ref{fig:pv_pcv_pc_linear}, we illustrate the implied intrinsic color distribution for a single color ($n_C = 1$).  We assume a non-Gaussian gamma distribution for the velocity in units of $1000 \text{ km s}^{-1}$, $-(v+9) \sim \text{Gamma}(a,b)$, with the shape  $a = 4.83$ and the scale $b = 0.54$ parameters chosen to fit the actual data distribution (Fig. \ref{fig:siII_vel_distr_n79}).   For simplicity,  measurement errors are set to zero, $\bm{W} = \bm{0}$.  For a constant mean intrinsic color (blue line), the implied intrinsic color distribution is Gaussian.  However, for a linear trend with non-zero slope ($b =-0.03$ mag $(10^3 \text{ km s}^{-1})^{-1}$, red line), the implied intrinsic color distribution is non-Gaussian and skewed, with a long tail towards redder colors.

\begin{figure}[t]
\centering
\includegraphics[angle=0,scale=0.36]{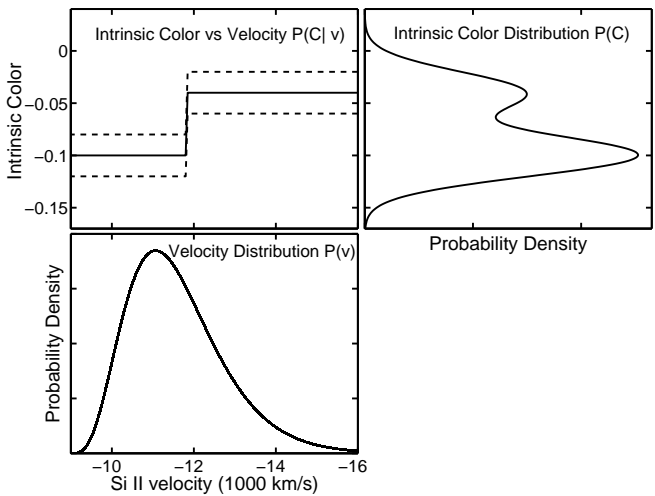}
\caption{\label{fig:pv_pcv_pc_step}  Illustration of the implied color distribution $P(C)$ implied by the Si II velocity distribution $P_V(v)$ and an assumed step function for the mean intrinsic color vs. velocity.  (bottom left)  The Si II velocity distribution is a gamma distribution with a long tail towards high velocities.   (top left) The high velocity group has a mean intrinsic color of $-0.04$ mag while the normal velocity group has a mean intrinsic color of $-0.10$ mag.  The residual intrinsic scatter in both cases is $\sigma_C = 0.02$ mag (dashed lines).    (top right)  The marginal intrinsic color distribution has a double peaked structure, with a small hump at redder colors associated with the high velocity subgroup.}
\end{figure}

In Figure \ref{fig:pv_pcv_pc_step}, we show the implied intrinsic color distribution implied by the same skewed Gamma distribution, but with a step function for the mean intrinsic color-velocity relation.  The difference between the mean intrinsic color for the high and normal velocity groups is set to $\bm{\theta}_\text{HV} - \bm{\theta}_\text{LV} = 0.06$ mag, with a residual scatter about the mean of $\sigma_C = 0.02$ mag.   The implied marginal distribution of the intrinsic color is then bimodal, the sum of two Gaussians, with the relative heights of the peaks determined by the proportion of SNe Ia in the high velocity vs. normal velocity groups.

\subsection{The Marginal Likelihood}\label{sec:marginal_lkhd}

The marginal likelihood for a single SN $s$ with a given Si II velocity $v_s$ is the probability density of its apparent color data, $\bm{O}_s$, under a set of population hyperparameters.   Given the preceding model assumptions,  $P(\bm{O}_s | \, v_s ; \bm{\theta}, \bm{\Sigma_C}, \tau)$ can be derived analytically by integrating over the latent variables  $\bm{C}_s, A_V^s$ of the individual SN (Appendix \ref{sec:math_lkhd}).   The mathematical form is given in Eq. \ref{eqn:marginal_lkhd} for general $n_C$ and arbitrary mean intrinsic color functions $\bm{\mu}_C(v; \bm{\theta})$ and simplifies to Eq. \ref{eqn:marginal_lkhd_1c} in the case of $n_C = 1$ color.  The marginal likelihood for the full sample is the product of $N_{SN}$ individual marginal likelihood functions.  This marginal likelihood can be maximized to estimate the hyperparameters ($ \bm{\theta}, \bm{\Sigma_C}, \tau$).  It is also needed to compute the deviance information criterion used for model comparison (Appendix \ref{sec:modelcomp}).  To gain some intuition, we illustrate some salient aspects of the marginal likelihood function graphically.

\begin{figure}[t]
\centering
\includegraphics[angle=0,scale=0.4]{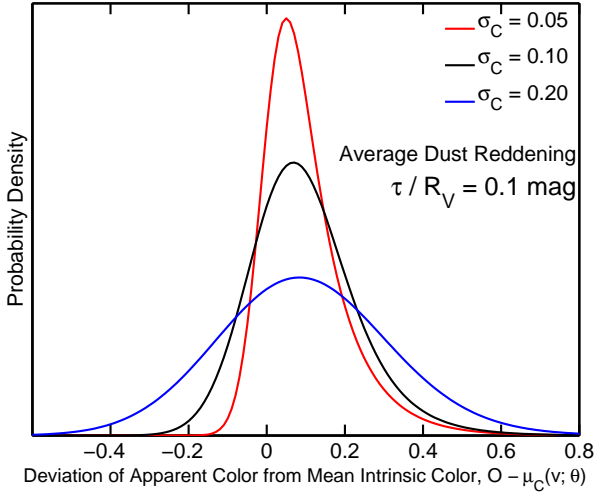}
\caption{\label{fig:marglkhd_1c} Conditional probability density (Eq. \ref{eqn:marginal_lkhd}) of the deviation of a single ($n_C = 1$) apparent color from the intrinsic mean color for a given velocity $v$, for an average dust reddening $\tau / R_V = 0.1$ mag and a range of values of intrinsic color scatter $\sigma_C$.   Measurement errors $\bm{W}$ have been set to zero.  When the average dust reddening dominates (red curve), the apparent distribution is positively skewed.  When the intrinsic scatter dominates (blue curve), it is less asymmetric.}
\end{figure}

In Figure \ref{fig:marglkhd_1c}, we show the conditional probability density of the apparent $B-V$ color relative to the mean intrinsic color for a given velocity, $O - \mu_C(v; \bm{\theta})$.  This shows the expected distribution of the apparent color measurements about the mean intrinsic color-velocity relation.  For simplicity, we have assumed  measurements errors are negligible, $\bm{W} = 0$. We show this for a fixed $\tau / R_V = 0.1$ mag (e.g. $\tau = 0.31$ mag for $R_V = 3.1$), and several values of $\sigma_C$.  For values of $\sigma_C < \tau/ R_V$, the asymmetry in the apparent distribution (conditional on a specific velocity) is dominant, with a skewed tail towards redder (positive) color.  For values of $\sigma_C > \tau / R_V$, the spread in intrinsic color is greater than the effects of positive dust reddening, and the apparent distribution for a given velocity is less asymmetric.

\begin{figure}[t]
\centering
\includegraphics[angle=0,scale=0.4]{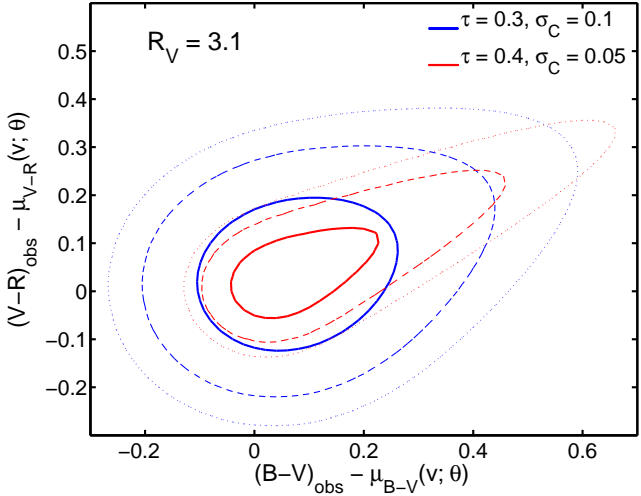}
\caption{\label{fig:marglkhd_2c} Conditional probability density (Eq. \ref{eqn:marginal_lkhd}) of  the deviation of a pair ($n_C = 2$) of apparent colors from the mean intrinsic colors for a given velocity $v$.   The solid, dashed, and dotted contours enclose approximately 68\%, 95\%, and 99\% of the two-dimensional probability.   For an average dust extinction of $\tau = 0.4$ mag and $\sigma_C = 0.05$ mag (for both $B-V$ and $V-R$), the dust reddening is dominant over the intrinsic color variance, leading to a sharp tail towards redder colors (red contours).  For an average dust extinction of $\tau = 0.3$ mag and $\sigma_C = 0.1$ mag, the effects of intrinsic variance and dust reddening are about equal, so the shape of the contours is more rounded and the dust reddened tail less pronounced (blue contours).  The tilt of the dust-reddened tail is set by the value of $R_V$, which is fixed to 3.1 here.}
\end{figure}

Figure \ref{fig:marglkhd_2c} shows the two-dimensional conditional probability density for a pair $n_C = 2$ of colors ($B-V$ and $V-R$).  This shows the expected shape of the joint distribution of observed color measurements around the mean intrinsic color-velocity relation.   If the average dust reddening is large relative to the intrinsic color scatter, then there is a narrow tail towards redder colors.  The tilt of the tail is set by $R_V$.  If the intrinsic color scatter and dust reddening are comparable in value, then the contours are more rounded and egg-shaped.

\begin{figure}[h]
\centering
\includegraphics[angle=0,scale=0.39]{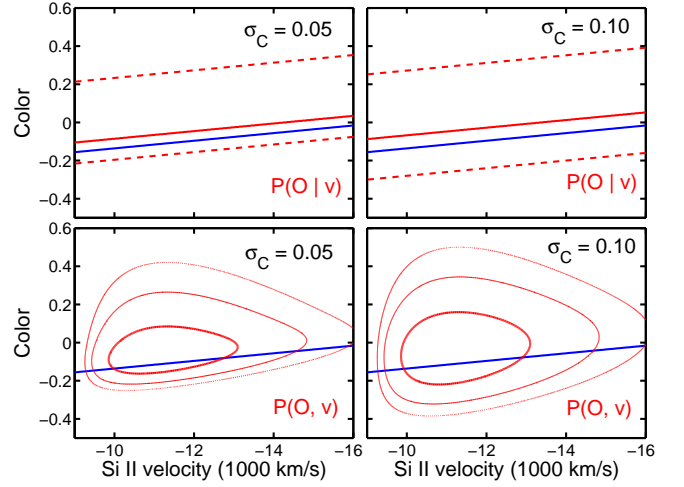}
\caption{\label{fig:pov_linear} (top) Conditional probability density (Eq. \ref{eqn:marginal_lkhd}) of the apparent color at each value of the velocity for an assumed linear intrinsic color-velocity trend $\mu_C(v)$ with slope of $b = -0.02$ mag  $(10^3 \text{ km s}^{-1})^{-1}$ (blue line) and an average dust reddening of $\tau / R_V =  0.1$ mag.  The solid red line is the color of the peak of $P(O | v)$ and the dashed red lines are the 2.5\% and 97.5\% quantiles of $P( O | v)$.  This is shown for the residual intrinsic scatter of $\sigma_C = 0.05$ mag (left) and $\sigma_C = 0.1$ mag (right).  (bottom) The joint probability density $P(O, v) = P( O | v) P_V(v)$ assuming a Gamma distribution for the marginal velocity distribution $P_V(v)$.  The equiprobability density contours approximately enclose 68\%, 95\% and 99\% of the joint probability.  For small values of the residual intrinsic scatter $\sigma_C$, the conditional and joint distribution of the color and velocity is more asymmetric around the apparent mode (solid red line) and the intrinsic mean (solid blue line).  Furthermore, the ``blue edge'' of the apparent color distribution is more squashed.}
\end{figure}

Figure \ref{fig:pov_linear} shows the conditional and joint probability densities of the apparent color ($n_C = 1$) and velocities for a linear trend with an intercept $c_0 = -0.1$ mag at $v = -11,800 \text{ km s}^{-1}$ and a slope of $b = -0.02$ mag per $10^3 \text{ km s}^{-1}$, for an average dust reddening $\tau / R_V = 0.1$ mag, and a residual intrinsic scatter of $\sigma_C = 0.05$ mag (left) and $\sigma_C = 0.1$ mag (right).  We depict the conditional probability density of the apparent color for each value of the velocity, $P(O | v)$ (Eq. \ref{eqn:marginal_lkhd}), as well as the joint probability density $P( O, v)  = P(O | v) P_V(v)$, assuming the same Gamma distribution for $P_V(v)$, as fitted for the data distribution in Fig. \ref{fig:siII_vel_distr_n79}.    For smaller values of the residual intrinsic scatter $\sigma_C$, the asymmetry in both the conditional and joint distributions is more pronounced.   The ``blue edge'' of the apparent color distribution (depicted here by the region between the 2.5\% quantile and the mode), is much narrower than the ``red tail'' (between the mode and the 97.5\% quantile).  For larger values of $\sigma_C$, the difference between the blue edge and the red tail is smaller, as the distributions are less asymmetric about the mode.

\subsection{Hyperpriors}

We generally use non-informative or standard, diffuse hyperpriors on the hyperparameters.  We use flat priors on  $\bm{\theta}$, $\log \tau_A$: $P(\bm{\theta}, \log \tau_A) \propto 1$.  For the residual color covariance matrix $\bm{\Sigma}_C$, we use a standard inverse Wishart density
\begin{equation}
P(\bm{\Sigma}_C) = \text{Inv-Wishart}_{\nu_0}( \bm{\Sigma}_C | \, \bm{\Lambda}_0^{-1})
\end{equation}
with prior degrees of freedom $\nu_0 = n_C + 1$ which ensures flat marginal prior densities on the correlation coefficients of the residuals \citep{barnard00}. Covariance matrices are required to be positive semidefinite, and this prior density assigns positive probabilities to only such matrices.  The prior scale matrix is chosen as $\bm{\Lambda}_0 = \epsilon_0^2 \, \bm{I}_{n_C}$, where $\epsilon_0$ the expected order of magnitude of the intrinsic color residuals (typically $\sim 0.05$ mag).  We have checked that the inferences are not strongly sensitive to the choice of $\epsilon_0$ over a reasonable range of values.  Hyperparameter estimates from the posterior, using these hyperpriors, are consistent with those
obtained by maximizing the marginal likelihood, Eq. \ref{eqn:marginal_lkhd}, independently from these hyperpriors (or equivalently, assuming flat priors on all hyperparameters).
\newpage
\subsection{Global Posterior Probability Density}\label{sec:globalpost}

The unknown parameters for each individual SN $s$ are $\bm{C}_s, A_V^s$, and the hyperparameters of the populations of intrinsic SN colors and dust  are $\bm{\theta}, \bm{\Sigma}_C$, and $\tau$.  The data for SN $s$ are the measured peak apparent colors $\bm{O}_s$ and the spectral line velocity $v_s$.   If we have estimates for the population hyperparameters, then the conditional posterior probability of the intrinsic colors and dust extinction for a single SN $s$, given these estimates and the data, is proportional to the product of observed color likelihood, the population distribution of intrinsic colors given the velocity, and the population distribution of dust extinction:
\begin{equation}\label{eqn:posterior_CAv}
\begin{split}
P&( \bm{C}_s, A_V^s  | \, \bm{\theta}, \bm{\Sigma}_C, \tau; \bm{O}_s, v_s  ) \\ &\propto  N(\bm{O}_s | \, \bm{C}_s + A_V^s \bm{\gamma}(R_V), \bm{W}_s )\, N( \bm{C}_s | \, \bm{\mu}_C(v_s; \bm{\theta}), \bm{\Sigma}_C ) \\ &\times \text{Expon}( A_V^s |\, \tau).
\end{split}
\end{equation}
To jointly estimate the intrinsic colors and dust extinctions of the individual SNe Ia in the full sample, together with the population hyperparameters, we use the global posterior probability density.  The full posterior probability is proportional to the product of $N_\text{SN}$ likelihoods times the hyperpriors. 
\begin{equation}\label{eqn:globalposterior}
\begin{split}
P&( \{\bm{C}_s, A_V^s\}; \bm{\theta}, \bm{\Sigma}_C, \tau | \, \{ \bm{O}_s, v_s \} ) \\ &\propto P(\tau, \bm{\Sigma}_C, \bm{\theta} ) \prod_{s=1}^{N_\text{SN}} \Big[ N(\bm{O}_s | \, \bm{C}_s + A_V^s \bm{\gamma}(R_V), \bm{W}_s )\, \\ &\times N( \bm{C}_s | \, \bm{\mu}_C(v_s; \bm{\theta}), \bm{\Sigma}_C ) \, \text{Expon}( A_V^s | \, \tau)  \Big] 
\end{split}
\end{equation}
This is the objective function from which all probabilistic inferences with the hierarchical model are computed.  In Appendix \ref{sec:mcrc}, we present a Gibbs sampling algorithm to generate an MCMC chain of samples from this posterior probability density.  These chains are then used to compute posterior estimates of all parameters and hyperparameters.   The samples are also used to compute the deviance information criterion (DIC), as described in Appendix \ref{sec:modelcomp}, to compare different models for the intrinsic color-velocity function.

\begin{figure}[t]
\centering
\includegraphics[angle=0,scale=0.35]{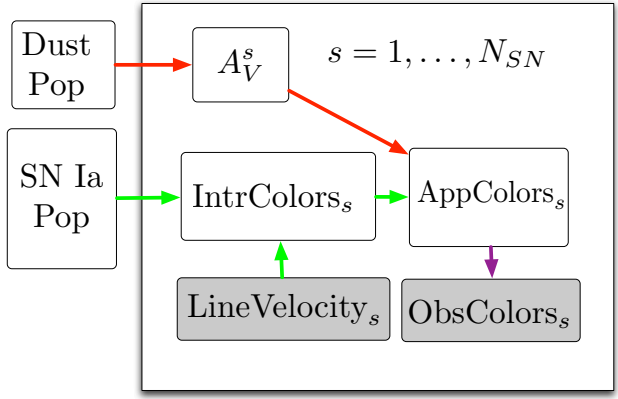}
\caption{\label{fig:dag_color_regress} A directed acyclic graph describing the hierarchical Bayesian model for colors and spectroscopic velocities.  The open boxes represent the unknown parameters of individual supenovae (the dust extinction $A_V^s$ and intrinsic colors $\bm{C}_s$) and the unknown hyperparameters of the SN Ia population ($\bm{\theta}, \bm{\Sigma}_C$) and the dust population ($\tau$).  The shaded boxes represent the measured data: the spectroscopic velocity measurement $v_s$ and the observed colors $\bm{O}_s$.  The arrows represent links of conditional probability relating the hyperparameters, parameters and the data (green: intrinsic effects; red: dust effects; purple: measurement error).}
\end{figure}

Our hierarchical Bayesian model can be expressed visually using a probabilistic graphical model known as a directed acyclic graph (DAG).  Graphical models were first used to express hierarchical Bayesian inference with SNe Ia by \citet{mandel09,mandel11}.  A DAG for a conceptually similar hierarchical model for stellar colors and dust extinction was recently presented by \citet{foster13}.   Figure \ref{fig:dag_color_regress} describes how the unknown parameters of individual SNe Ia (labelled by index $s$) and the hyperparameters of the dust and SN Ia populations are related to the measured supernova data.  

\section{Simulations}\label{sec:simulations}

In this section, we demonstrate and validate our method with simulated data, for which the true hyperparameters are selected and known.   We begin each simulation by sampling Si II velocities from a non-Gaussian distribution, $\{ v_s\} \sim P_V(v)$, and then use a chosen intrinsic color-velocity model, together with an exponential dust distribution and assumed measurement errors to generate observed colors $\{ \bm{O}_s \}$, using Eqs. \ref{eqn:structeqn} and \ref{eqn:intcol_dependence}. Conceptually, we are sampling forward through the graphical model in Figure \ref{fig:dag_color_regress} to generate the data.  For each SN $s$ in a sample of $N_{SN}$ objects,  $n_C = 3$ colors are observed ($B-V, B-R, B-I$).  For the true residual correlation matrix and the residual variances, we assumed $\bm{\sigma}_C = (0.02, 0.03, 0.03)$ mag and
\begin{equation}
\bm{R}_C = \begin{pmatrix} 1 & 0.5 & -0.6 \\ 0.5 & 1 & -0.6 \\  -0.6 & -0.6 & 1 \end{pmatrix}
\end{equation}
and constructed the true residual covariance matrix as $\bm{\Sigma}_C = \text{diag}(\bm{\sigma}_C)\, \bm{R}_C \, \text{diag}(\bm{\sigma}_C) $.  For the average extinction of the exponential dust distribution, we assumed $\langle A_V \rangle = \tau = 0.3$ mag, and $R_V = 2.5$ for a CCM dust law.  The measurement error standard deviation for each observed color was assumed to be 0.04 mag, which was typical for our actual color data.  The measurement covariance matrix $\bm{W}_s$ encoded a 50\% measurement error correlation between each pair of colors.  These values were chosen to be similar to those we ultimately found from fitting the real data.  The size of the simulated samples ($N_{SN} = 79$) is the size of our real SN Ia sample.

For the simulations, we know the true form of the intrinsic colors-velocity function $\bm{\mu}_C(v; \bm{\theta})$ in the model that generated the data.  When applying the method to real data, we will not know which relation to use; indeed the model can be fit for any chosen form.   We proceed by fitting a small set of simple functional forms, and then use the DIC (Appendix \ref{sec:modelcomp}) to choose the model that best negotiates the trade-off between model fit and complexity.    Differences in DIC greater than 2 represent positive support for the model with the lower numerical value, and differences greater than 6 represent strong support for that model.  We test this approach on the simulated data, trying a few models, as if we did not know the true model that generated the data.  For each simulation, we run the \textsc{MCRC} sampler (Appendix \ref{sec:mcrc}) to fit the simulated $\{ \bm{O}_s, v_s \}$ data, assuming these different models for the mean intrinsic colors-velocity function:  
\begin{enumerate}
\item \textsc{Constant} (Gaussian).  This assumes there is no mean trend of the intrinsic colors versus ejecta velocity, i.e. the mean intrinsic colors are constant $\bm{\mu}_C(v) = \bm{c}_0$.   Hence, the marginal intrinsic color distributions are Gaussian.
\item \textsc{Linear}, as in Eq. \ref{eqn:intcol_linear}.  The mean intrinsic colors are a linear function of the velocity.
\item \textsc{Step} function, as in Eq. \ref{eqn:stepfcn}, with the division between high and normal ejecta velocities set \emph{a priori} at $v_0 = -11,800 \text{ km s}^{-1}$.
\item \textsc{Quadratic}, a polynomial of order $p = 2$ (Eq. \ref{eqn:poly}).
\end{enumerate}

For each model, we sample the global posterior density of all the parameters and hyperparameters conditional on the simulated dataset.  We use the MCMC samples to compute the DIC (Appendix \ref{sec:modelcomp}) using the marginal likelihood (Appendix \ref{sec:math_lkhd}) for each model.   Selecting the model with the lowest DIC, we examine the posterior estimates of the hyperparameters.  We check the selected model and the estimates against the true model and true hyperparameters originally used to generated the data to validate our method.

In each of the following three scenarios, we illustrate our methodology using a single random realization of simulated data.  For each scenario, we have additionally simulated 9 other datasets (not shown) with the same parameters and computed DIC for the various models applied to each simulation in the same manner.  We tabulate $\overline{\Delta\text{DIC}}$, the mean $\Delta\text{DIC}$ averaged over the 10 simulations, to show that, over different random realizations of the data, the information criterion consistently selects the correct model underlying the simulated data.

\subsection{Bimodal ejecta velocity distribution with step function}\label{sec:bimodal_step}

We simulate a scenario in which SNe Ia are comprised of two populations with distinct expansion velocities and intrinsic colors.  We generate a sample of velocities $v$ from a bivariate Gaussian distribution
\begin{equation}
P(v) = 0.5 \,N( v | \, \mu_1, \sigma_v^2 ) + 0.5 \, N( v | \, \mu_2, \sigma_v^2 )
\end{equation}
truncated to $ -14,900 \text{ km s}^{-1} > v > -9,100 \text{ km s}^{-1}$.  This is an equal-weighted mixture of two Gaussians, one centered at $\mu_1 = -13,000 \text{ km s}^{-1}$ and one centered at $\mu_2 = -11,000 \text{ km s}^{-1}$, both with standard deviation of $\sigma_v = 500 \text{ km s}^{-1}$.  This distribution does not reflect the actual velocity data (Fig. \ref{fig:siII_vel_distr_n79}), which is unimodal. Our model can be used for any distribution of velocities; this example is for illustrative purposes. 

\begin{figure}[t]
\centering
\includegraphics[angle=0,scale=0.395]{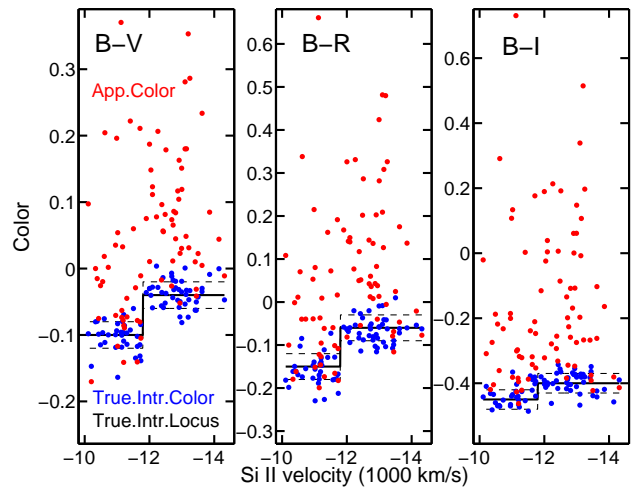}
\caption{\label{fig:sim_bm_step_BVBRBI} The joint distribution of intrinsic colors, observed colors and velocities along with the intrinsic color locus for each color for the bimodal-step simulation (\S \ref{sec:bimodal_step}).  This simulation assumes a bimodal distribution of velocities (a mixture of two Gaussians), with a step function dependence of the mean intrinsic colors on velocities (black solid line) with some intrinsic scatter (black dashed lines). The observed colors (red points) are generated from the true intrinsic colors (blue points) by adding reddening due to dust extinction randomly drawn from an exponential distribution, $A_V \sim \text{Expon}( \tau = 0.3 \text{ mag})$, and by adding random measurement error.}
\end{figure}

We assumed a step function for the mean intrinsic color-velocity function $\bm{\mu}_C(v; \bm{\theta}_\text{HV}, \bm{\theta}_\text{NV})$ with $\bm{\theta}_\text{HV} = ( -0.04, -0.06, -0.40)$ mag for the mean intrinsic colors for high velocity SNe Ia ($|v| > 11,800 \text{ km s}^{-1}$), and  $\bm{\theta}_\text{HV} = ( -0.10, -0.15, -0.45)$ mag for the normal velocity SNe Ia  ($|v | < 11,800 \text{ km s}^{-1}$).  These color differences are quantitatively the same as those we will find when fitting the step model to the actual data (\S \ref{sec:app_step}).  The observed colors were generated by adding random dust reddening and measurement error. The joint distribution of intrinsic colors, observed colors and velocities of the simulated sample is shown in Fig. \ref{fig:sim_bm_step_BVBRBI} along with the intrinsic color locus for each color.

\begin{figure}[t]
\centering
\includegraphics[angle=0,scale=0.41]{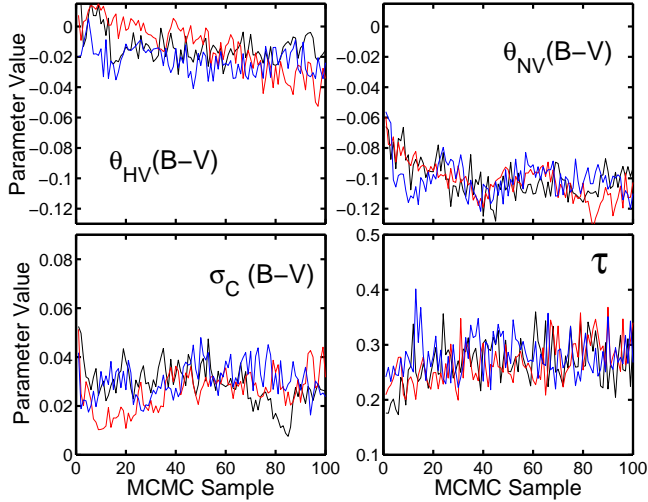}
\caption{\label{fig:sim_bm_step_mcmc_trace_044448} Trace plot for a run of \textsc{MCRC} for 1000 cycles with three independent chains.  The current values of all parameters were recorded every 10 cycles.  Each color represents a different chain, and each panel represents a different dimension of each chain corresponding to a particular scalar parameter.  The parameters are: (top left) The mean intrinsic $B-V$ color for HV SNe Ia, (top right) the mean intrinsic $B-V$ color for NV SNe Ia,  (bottom left) the residual standard deviation of the intrinsic $B-V$ around the mean trend, and (bottom right) the average extinction of the exponential dust distribution.  The chains converge rapidly to the posterior distribution, and the initial portions of each chain are discarded as ``burn-in'' before analysis.}
\end{figure}

We ran the \textsc{MCRC} sampler to estimate the unknown parameters, and to compute the DIC, inputting only the observable data $\{ \bm{O}_s, v_s\}$ of the SN Ia sample, and assuming $R_V = 2.5$.  Trace plots of the Markov chain projected along particular parameters (Fig. \ref{fig:sim_bm_step_mcmc_trace_044448}) show that it converges quickly to the posterior distribution.   

Table \ref{table:dic_sim_bm_step} shows the information criteria calculations.  The models that allow intrinsic-velocity trends are clearly favored and the constant-Gaussian model is strongly disfavored ($\Delta \text{DIC} < -13$).  The model with the lowest value of DIC is the step function model with a difference in DIC (relative to constant) of $-23.5$.  The DIC of this model is also much better than those of the linear model and the more complex quadratic model.  Hence, we find the reassuring result that the model with the lowest value of DIC coincides with the true model that generated the simulated data.  

\begin{deluxetable}{lrrrrrr}
\tabletypesize{\small}
\tablecaption{Information Criteria for Bimodal-Step Simulation}
\tablewidth{0pt}
\tablehead{ \colhead{Model} & \colhead{$\hat{D}$} & \colhead{$\langle D \rangle$} &\colhead{$p_D$} & \colhead{DIC} & \colhead{$\Delta$DIC\tablenotemark{a}} & \colhead{$\overline{\Delta\text{DIC}}$\tablenotemark{b}}}
\startdata
Constant & -462.0 &	 -453.9 & 8.1 & -445.8  & 0.0 & 0.0 \\ 
Linear & -481.2 & -470.5 & 10.7 & -459.8  & -14.0 & -17.6\\ 
Step & -490.9 & -480.1 & 10.8 & -469.3 & -23.5  &   -22.3\\
Quadratic & -486.6 	& -472.9 	& 13.8 & -459.1 & -13.3 & -13.1  
 \enddata
\tablenotetext{a}{Difference in DIC relative to constant-Gaussian.}
\tablenotetext{b}{$\Delta$DIC averaged over 10 simulations generated from the same model.}
\tablecomments{\label{table:dic_sim_bm_step} For a single simulation, $\hat{D}$ is the deviance at the posterior mean, $\langle D \rangle$ is the posterior mean of the deviance,  $p_D$ is the effective number of hyperparameters, and  DIC is the deviance information criterion.  See \S \ref{sec:modelcomp} for details.  The first five numerical columns refer to the simulated dataset in Fig. \ref{fig:sim_bm_step_BVBRBI}, described in \S \ref{sec:bimodal_step}. }
\end{deluxetable}

Within the step function model, we check that the true values of the hyperparameters $\tau, \bm{\theta} = ( \bm{\theta}_\text{HV}, \bm{\theta}_\text{NV})$, and $\bm{\Sigma}_C$ are recovered within the uncertainties of the posterior.    In particular, the mean intrinsic colors $\bm{\theta}$ for each velocity group are recovered.  We computed the posterior mean and standard deviations for the mean intrinsic colors using the Markov chains.  For the simulation shown in Fig. \ref{fig:sim_bm_step_BVBRBI}, they were $\bm{\hat{\theta}}_{HV} =( -0.024 \pm 0.011, -0.053 \pm 0.015, -0.368 \pm 0.020)$ mag for the HV group and $\bm{\hat{\theta}}_{NV} =( -0.105 \pm 0.011, -0.160 \pm 0.015, -0.431 \pm 0.020)$ mag for the NV group.  The estimated average dust extinction of the population was $\hat{\tau}  = 0.29 \pm 0.04$ mag.

\subsection{Gamma velocity distribution with Constant-Gaussian Intrinsic Colors Model}\label{sec:gamma_const}

\begin{figure}[t]
\centering
\includegraphics[angle=0,scale=0.41]{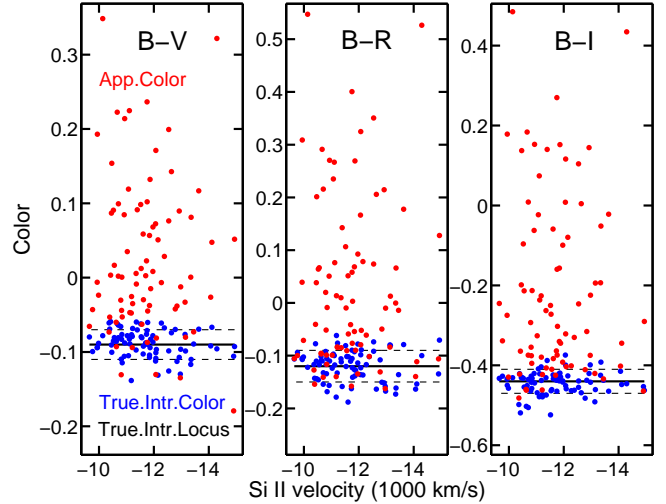}
\caption{\label{fig:sim_gamma_const_BVBRBI} The joint distribution of intrinsic colors, observed colors and velocities along with the intrinsic color locus for each color for the gamma-constant simulation (\S \ref{sec:gamma_const}).  This simulation assumes a gamma distribution of velocities, and constant mean intrinsic colors independent from velocities (black solid line), with some intrinsic scatter (black dashed lines). The observed colors (red points) are generated from the intrinsic colors (blue points) by adding reddening due to a random dust extinction drawn from an exponential distribution, $A_V \sim \text{Expon}( \tau = 0.3\text{ mag})$, and by adding random measurement error.}
\end{figure}

In this simulation, we generate a sample of ejecta velocities $v$ from the distribution of
\begin{equation}\label{eqn:gamma_distr}
v / (1000 \text{ km s}^{-1}) = -9 - \Gamma_{4.83, 0.54}
\end{equation}
truncated to $ -14,900 \text{ km s}^{-1} < v < -9,100 \text{ km s}^{-1}$.   This is the same gamma distribution that best fits the actual velocity data (Fig. \ref{fig:siII_vel_distr_n79}).   We assumed the constant-Gaussian model in which the population of intrinsic colors has a joint Gaussian distribution with zero trend with ejecta velocity.  For the mean intrinsic colors, we assumed $\bm{c}_0 = (-0.09, -0.12, -0.44)$ mag for $(B-V, B-R, B-I)$, respectively.  These values are those that we will find when fitting the constant-Gaussian model to the actual color-velocity data (\S \ref{sec:app_gaussian}). The joint distribution of intrinsic colors, observed colors and velocities of the simulated sample is shown in Fig. \ref{fig:sim_gamma_const_BVBRBI} along with the intrinsic color locus for each color.

Using only the simulated data $\{ \bm{O}_s, v_s\}$ of the SN Ia sample, and assuming $R_V = 2.5$, we ran the \textsc{MCRC} sampler for each model to estimate the unknown parameters and compute the information criteria.  Table \ref{table:dic_sim_gam_const} shows the resulting DIC for the various models.    The estimate of the deviance ($\hat{D}$) decreases with model complexity, indicating that the data appear more likely under the more complex models.  However, the DIC penalizes the deviance by the effective number of parameters, a measure of the model complexity.
The model with the lowest DIC is the constant-Gaussian, which is the true model that generated the data.   The other models that allow for trends with velocity are disfavored with $\Delta \text{DIC} \gtrsim 3$ relative to the simplest model.  This indicates that the fits achieved with the more complex models are not significantly better compared to the added complexity.   Within the constant-Gaussian model, we checked that the inferred mean intrinsic colors were consistent with the true values used to generated the data.  The posterior means and standard deviations computed from the Markov chains were $\bm{\hat{c}}_0 = (-0.079 \pm 0.008, -0.123 \pm 0.011, -0.439 \pm 0.014)$ mag.  The estimated average dust extinction was $\hat{\tau} = 0.28 \pm 0.04$ mag.

\begin{deluxetable}{lrrrrrr}
\tabletypesize{\small}
\tablecaption{Information Criteria for Gamma-Constant Simulation}
\tablewidth{0pt}
\tablehead{ \colhead{Model} & \colhead{$\hat{D}$} & \colhead{$\langle D \rangle$} &\colhead{$p_D$} & \colhead{DIC} & \colhead{$\Delta$DIC\tablenotemark{a}} & \colhead{$\overline{\Delta\text{DIC}}$\tablenotemark{b}}}
\startdata
Constant & -544.2 & -536.0 & 8.2 & -527.8 & 0.0 & 0.0 \\ 
Linear &  -545.8 & -534.8  & 11.0 & -523.9  & +3.9 & +2.6\\ 
Step & -545.9  & -535.0 & 11.0  & -524.0 & +3.8  &  +2.9\\
Quadratic & -548.7 & -534.6 & 14.0 & -520.6 & +7.2 & +5.1  
 \enddata
\tablenotetext{a}{Difference in DIC relative to constant-Gaussian.}
\tablenotetext{b}{$\Delta$DIC averaged over 10 simulations generated from the same model.}
\tablecomments{\label{table:dic_sim_gam_const} See \S \ref{sec:modelcomp} for details. The first five numerical columns refer to the simulated dataset in Fig. \ref{fig:sim_gamma_const_BVBRBI}, described in \S \ref{sec:gamma_const}. }
\end{deluxetable}

\subsection{Gamma velocity distribution with a linear model}\label{sec:gamma_linear}

\begin{figure}[t]
\centering
\includegraphics[angle=0,scale=0.41]{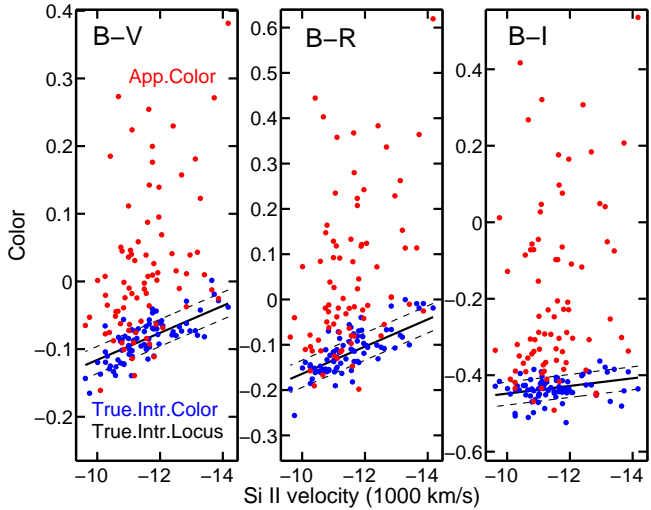}
\caption{\label{fig:sim_gamma_lin_BVBRBI} The joint distribution of intrinsic colors, observed colors and velocities along with the intrinsic color locus for each color for the gamma-linear simulation (\S \ref{sec:gamma_linear}).  This simulation assumes a gamma distribution of velocities, and linear dependence of the mean intrinsic colors on velocities (black solid lines) with some intrinsic scatter (black dashed lines). The observed colors (red points) are generated from the intrinsic colors (blue points) by adding reddening due to a random dust extinction drawn from an exponential distribution, $A_V \sim \text{Expon}( \tau = 0.3\text{ mag})$, and by adding random measurement error.}
\end{figure}

We generated a sample of ejecta velocities from the same gamma distribution, Eq. \ref{eqn:gamma_distr}, that fits the actual velocity data (Fig. \ref{fig:siII_vel_distr_n79}).   We assumed a linear form (\S \ref{sec:linear}) for the intrinsic-color velocity relation.  For the true mean intercepts at $v_0 = -11,800 \text{ km s}^{-1}$, we assumed $\bm{c}_0 = (-0.08, -0.11, -0.43)$ mag for each of the intrinsic colors.  The assumed true slopes were $\bm{b} = (-0.02, -0.03, -0.01)$ mag per $1000 \text{ km s}^{-1}$.  These values are the same as those we will find when fitting the linear model to the actual color-velocity data (\S \ref{sec:app_linear}).
The joint distribution of the sample of intrinsic colors, observed colors and velocities is shown in Fig. \ref{fig:sim_gamma_lin_BVBRBI} along with the intrinsic color locus for each color.

Using only the observable data $\{ \bm{O}_s, v_s\}$ for the simulated SN Ia sample, and assuming $R_V = 2.5$, we  ran the \textsc{MCRC} sampler to estimate the unknown parameters and compute the information criteria.  Table \ref{table:dic_sim_gam_lin} shows the resulting DIC for the various models.   The constant-Gaussian model with no intrinsic color-velocity trend has the highest DIC value and is clearly disfavored.  The linear model has the lowest DIC value by a large margin, and also has a clearly better DIC than the most complex model (quadratic).  Once again, the model with the lowest DIC value coincides with the true model that generated the data.

\begin{deluxetable}{lrrrrrr}
\tabletypesize{\small}
\tablecaption{Information Criteria for Gamma-Linear Simulation}
\tablewidth{0pt}
\tablehead{ \colhead{Model} & \colhead{$\hat{D}$} & \colhead{$\langle D \rangle$} &\colhead{$p_D$} & \colhead{DIC} & \colhead{$\Delta$DIC\tablenotemark{a}} & \colhead{$\overline{\Delta\text{DIC}}$\tablenotemark{b}}}
\startdata
Constant & -484.2 & -476.2 & 8.1 & -468.1& 0.0 & 0.0 \\ 
Linear &  -501.4 & -490.2 	& 11.2 &  -479.0  & -10.9 & -13.5 \\ 
Step & -496.4 & -485.5 & 10.9 & -474.5  &  -6.4 & -8.2 \\
Quadratic & -503.1 & -489.2 & 13.9 & -475.3 & -7.2 & -9.8  
 \enddata
\tablenotetext{a}{Difference in DIC relative to constant-Gaussian.}
\tablenotetext{b}{$\Delta$DIC averaged over 10 simulations generated from the same model.}
\tablecomments{\label{table:dic_sim_gam_lin} See \S \ref{sec:modelcomp} for details.  The first five numerical columns refer to the simulated dataset in Fig. \ref{fig:sim_gamma_lin_BVBRBI}, described in \S \ref{sec:gamma_linear}.}
\end{deluxetable}

Within the linear model, we check that the true values of the hyperparameters $\tau, \bm{\theta} = (\bm{c}_0, \bm{b})$, and $\bm{\Sigma}_C$ are recovered within the uncertainties of the posterior.    In particular, the intercepts $\bm{c}_0$ and slope of intrinsic colors versus velocity $\bm{b}$, are recovered.  The posterior mean and standard deviations computed from the Markov chains were $\bm{\hat{b}} = (-0.026 \pm 0.008, -0.038 \pm 0.011, -0.019 \pm 0.014)$ mag per $10^3 \text{ km s}^{-1}$ for the slopes, and $\bm{\hat{c}}_0 = (-0.073 \pm 0.008, -0.090 \pm 0.010, = -0.405 \pm 0.014)$ mag for the intercepts.  The inferred  population average dust extinction was $\hat{\tau} = 0.27 \pm 0.03$ mag.

\section{Application to Data}\label{sec:application}

We apply our statistical method to the observed colors and velocity data set of 79 nearby SNe Ia described in \S \ref{sec:data}.   We analyze this data set using our statistical model and Gibbs sampler to estimate the unknown parameters and hyperparameters.  The inputs to the \textsc{MCRC} code (Appendix \ref{sec:mcrc}) were the velocities $\{v_s\}$, the observed peak optical colors $\{\bm{O}_s = (B-V, B-R, B-I) \}$, and their estimation uncertainties, $\{ \bm{W}_s\}$.  For the dust reddening law we assumed a CCM law \citep{ccm89} with the coefficients from \citet{jha07}.  We adopted the value $R_V = 2.5$, as found by \citet{foleykasen11}.  Although changing $R_V$ modifies the dust extinction estimates and the average dust extinction of the population, the results for the intrinsic properties were not very sensitive to this.  This is because the intrinsic color locus is mainly anchored by SNe Ia with the lowest dust extinction, for which the dust reddening corrections are small and insensitive to $R_V$.  We fit each model by running the Gibbs sampler for $2 \times 10^4$ cycles, recording every 10th sample.  We used the MCMC samples to compute the DIC (as described in Appendix \ref{sec:modelcomp}) for model comparison between different models.

\subsection{Model Comparison using DIC}\label{sec:apply_dic}

We fit the data set with the \textsc{constant}-Gaussian, \textsc{Linear}, \textsc{Step}, \textsc{Quadratic}, and \textsc{Cubic} (polynomials of order $p  = 3$, c.f. \S \ref{sec:poly}) models for the mean intrinsic colors vs. velocity function $\bm{\mu}_C(v)$. We examined the deviance information criteria computed from the model fits to the data (Table \ref{table:dic_data}).    The DIC values are compared against the baseline \textsc{constant}-Gaussian model with a constant mean intrinsic color versus velocity.   Information criterion differences greater than 2 represent positive support for the model with the lower numerical value, and differences greater than 6 represent strong support.  The more complex models have lower deviance $\hat{D}$ values, indicating that the observed data have a higher probability under these models.  However, after penalizing by the effective number of parameters, the DIC reaches a minimum and then increases with model complexity.  Models with non-constant trends are strongly favored over the \textsc{constant}-Gaussian model.  The most favored model under DIC is $\textsc{Linear}$ with $\Delta \text{DIC} = -11.5$, but it is only marginally better than $\textsc{Step}$.  The DIC increases for \textsc{Quadratic} and \textsc{Cubic}, suggesting that these more complex models are not supported by the current data.  We describe the fits for the \textsc{constant}-Gaussian model, and our two best models (\textsc{Linear} and \textsc{Step}) under DIC.

\begin{deluxetable}{lrrrrr}
\tabletypesize{\small}
\tablecaption{Information Criteria for Nearby Color-Velocity Sample}
\tablewidth{0pt}
\tablehead{ \colhead{Model} & \colhead{$\hat{D}$} & \colhead{$\langle D \rangle$} &\colhead{$p_D$} & \colhead{DIC} & \colhead{$\Delta$DIC\tablenotemark{a} } }
\startdata
Constant & -530.0 & -522.4 & 7.7 & -514.7 & 0.0 \\ 
Linear & -546.8 & -536.5 & 10.3 &-526.2 & -11.5 \\ 
Step & -546.3 & -535.8 &  10.5 & -525.2 & -10.5 \\
Quadratic & -550.9 & -537.3 & 13.6 & -523.7 & -9.0 \\
Cubic & -552.1 & -536.0 & 16.1 & -519.9 & -5.2 
 \enddata
\tablenotetext{a}{Difference in DIC relative to constant-Gaussian.}
\tablecomments{\label{table:dic_data} See \S \ref{sec:modelcomp} for details.}
\end{deluxetable}


\subsection{Application of Gaussian Intrinsic Color Model}\label{sec:app_gaussian}

\begin{figure}[t]
\centering
\includegraphics[angle=0,scale=0.4]{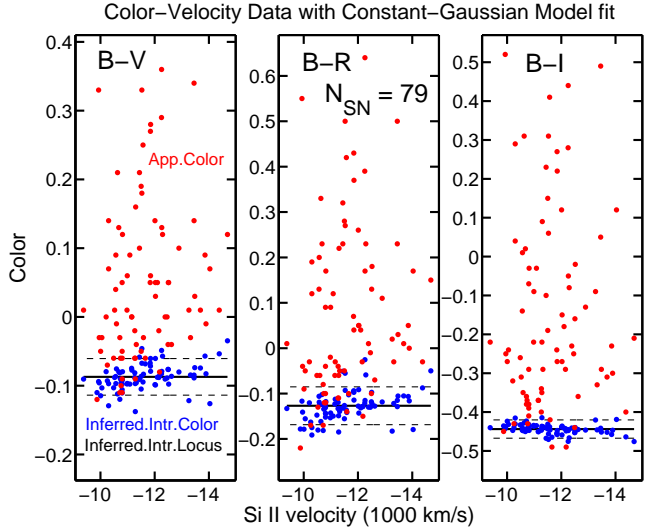}
\caption{\label{fig:plot_const_BVBRBI_n79_233404}  Fit of the constant-Gaussian intrinsic color model to the apparent color data.  The red points are the measured apparent colors and Si II velocities.  The black solid lines indicate the inferred mean intrinsic colors for the population, and the dashed lines indicate plus or minus one standard deviation of the population.  The blue points indicate the inferred intrinsic colors for each individual SN Ia.}
\end{figure}

We fit the \textsc{constant}-Gaussian model for the intrinsic color distribution that ignores the Si II velocity information.  This model assumes that the joint population distribution of the intrinsic colors is multivariate Gaussian (also implying that the marginal population distribution of each color is univariate Gaussian).   The mean function $\bm{\mu}_C(v; \bm{\theta}) = \bm{c}_0$ assumes no trend with velocity $v$.  The result of this fit is depicted in Fig. \ref{fig:plot_const_BVBRBI_n79_233404}.    From the posterior density, we estimate the hyperparamters governing the SN Ia and dust population distributions.  The population mean extinction was estimated: $\hat{\tau} = 0.35 \pm 0.04$ mag.  Table \ref{table:gaussian} lists the posterior estimates of the population mean intrinsic colors and standard deviations.

By comparing the apparent color measurements (red dots) with the inferred intrinsic color distribution (black lines), one can see that for $B-V$ and $B-R$, at normal absolute velocities $|v| < 11,800 \text{ km s}^{-1}$, there are SNe with peak apparent colors both below and above the inferred mean intrinsic value.  However, at high absolute velocities $|v| > 11,800 \text{ km s}^{-1}$, there are only SNe with peak apparent colors at or redder than the inferred mean intrinsic value. Similarly, at low absolute velocities, there are more SNe with inferred intrinsic $B-V$ and $B-R$ colors (blue) less than the population mean (black solid line), while at high absolute velocities, there are more with intrinsic colors greater than the population mean.  These are clues that a model with an intrinsic color-velocity trend would describe the data better.

\begin{deluxetable}{lrrrrr}
\tabletypesize{\small}
\tablecaption{Estimates of constant-Gaussian Model Hyperparameters}
\tablewidth{0pt}
\tablehead{ \colhead{ } & \colhead{$B-V$} & \colhead{$B-R$} &\colhead{$B-I$} }
\startdata
$c_0$  & $-0.086 \pm 0.008$ & $-0.12 \pm 0.01$ & $-0.44 \pm 0.02$ \\ 
 $\sigma_C$ & $0.027 \pm 0.007$ & $0.04 \pm 0.01$ & $0.02 \pm 0.01$ 
 \enddata
\tablecomments{\label{table:gaussian} The mean intrinsic color at all velocities is $c_0$.  The  intrinsic color scatter is  $\sigma_C$.  Numbers are the posterior means and standard deviations of each parameter in units of magnitude.}
\end{deluxetable}

\subsection{Application of Linear Model}\label{sec:app_linear}

\begin{figure}[t]
\centering
\includegraphics[angle=0,scale=0.375]{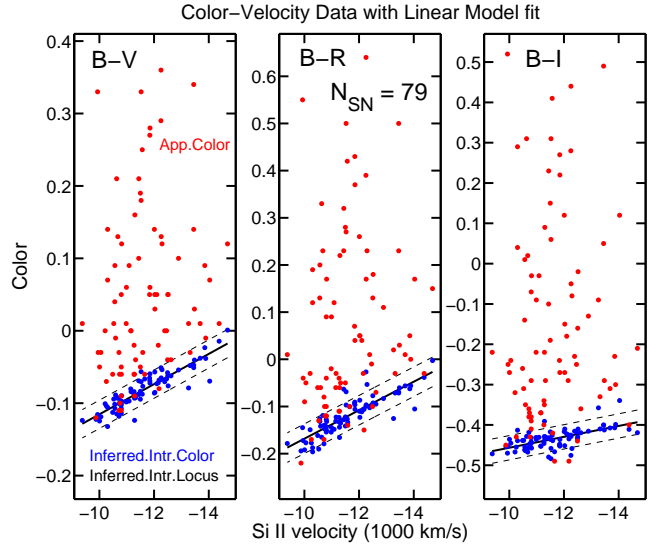}
\caption{\label{fig:plot_linear_BVBRBI_n79} The apparent colors and velocity data are shown (red points), along with posterior inferences from the application of the linear model.  The blue points are the marginal posterior means for each of the intrinsic colors of each SN Ia.  The black solid line represents the conditional mean intrinsic color-velocity function $\bm{\mu}_C(v; \bm{\theta})$ using the posterior mean estimates of the hyperparameters $\bm{\theta} = \{\bm{c}_0, \bm{b}\}$.  The black dashed lines indicate the residual intrinsic scatter about the mean relation.}
\end{figure}

\begin{figure}[h]
\centering
\includegraphics[angle=0,scale=0.39]{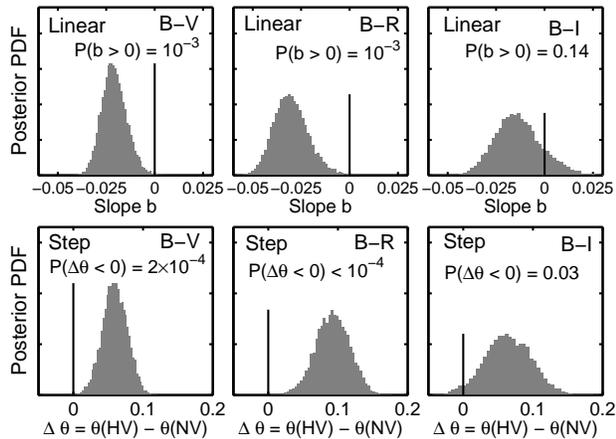}
\caption{\label{fig:post_steplinear_BVBRBI_n79} (Top row) Marginal posterior probability densities of the slopes $\bm{b}$ (in units of mag $(10^3 \text{ km s}^{-1})^{-1}$) of the \textsc{linear} intrinsic colors-velocity model applied to the data.  Each posterior density is normalized to integrate to one, so that more precise, narrow posterior pdfs are taller.  The vertical lines indicate zero slope.  We also indicate the posterior probability in the right tail $p_\text{tail} = P( b > 0)$.  The slopes of the intrinsic $B-V$ and $B-R$ colors versus velocity are most significantly different from zero.  The slope of intrinsic $B-I$ color versus velocity is not statistically significant.  (Bottom row) Marginal posterior probability of the mean intrinsic color difference between HV and NV SNe Ia, $\Delta \theta = \theta_\text{HV} - \theta_\text{NV}$ (in units of mag), under the \textsc{Step} intrinsic colors-velocity model applied to the data.  The vertical lines indicate zero intrinsic color offset.  We also indicate the posterior probability in the left tail $p_\text{tail} = P(  \Delta \theta < 0)$.  The intrinsic color offsets are most significant in $B-V$ and $B-R$, whereas the offset in $B-I$ is marginal.}
\end{figure}

We fit the \textsc{Linear} model of \S \ref{sec:linear} assuming $R_V = 2.5$.  The data and posterior inferred intrinsic colors and color loci are shown in Fig. \ref{fig:plot_linear_BVBRBI_n79}.  Posterior estimates of the hyperparameters are summarized in Table \ref{table:linear}.
The posterior estimate of the population average dust extinction was $\hat{\tau} = 0.35 \pm 0.04$ mag.  The marginal posterior densities of the slopes in each color are shown in the top row of Fig. \ref{fig:post_steplinear_BVBRBI_n79} as histograms of the MCMC samples.  The slopes are generally negative, so that SNe Ia with more negative velocities (and higher absolute velocities) tend to be intrinsically redder (positive color).  For each slope, we show the posterior mean and standard deviation, as well as the tail probability that the slope is greater than zero.  The most significant results are that the slopes of intrinsic color versus velocity are $\hat{b} = -0.021 \pm 0.006$ for $B-V$, and  $\hat{b} = -0.030 \pm 0.009$ mag per $10^3 \text{ km s}^{-1}$ for $B-R$.   The $B-I$ slope is not significantly different from zero, and its posterior variance is the largest of the three colors.  A larger data set may help determine if there is a real velocity effect in $B-I$.

\begin{deluxetable}{lrrrrr}
\tabletypesize{\small}
\tablecaption{Estimates of Linear Model Hyperparameters}
\tablewidth{0pt}
\tablehead{ \colhead{ } & \colhead{$B-V$} & \colhead{$B-R$} &\colhead{$B-I$} }
\startdata
$c_0$  & $-0.078 \pm 0.008$ & $-0.11 \pm 0.01$ & $-0.43 \pm 0.02$ \\ 
$b$  & $-0.021 \pm 0.006$ & $-0.030 \pm 0.009$ & $-0.013 \pm 0.012$  \\ 
$p_\text{tail}$ & 0.001 &  0.001 & 0.14 \\
 $\sigma_C$ & $0.020 \pm 0.006$ & $0.03 \pm 0.01$ & $0.03 \pm 0.01$ 
 \enddata
\tablecomments{\label{table:linear} The mean intrinsic color at $v_0 = -11,800 \text{ km s}^{-1}$ is $c_0$ in units of mag.  The slope is $b$ in units of mag per $10^3 \text{ km s}^{-1}$.  The residual intrinsic color scatter is  $\sigma_C$.  Numbers are the posterior means and standard deviations of each parameter, except for the tail probability $p_\text{tail} = P(b > 0)$.}
\end{deluxetable}

The \textsc{MCRC} code also computes the posterior estimates of the residual intrinsic color correlation matrix.  The mean and standard deviations of each residual correlation were
\begin{equation}
\bm{\hat{R}}_C = \begin{pmatrix} 1 & 0.5 \pm 0.3 & -0.5 \pm 0.4 \\ 0.5 \pm 0.3 & 1 & -0.5 \pm 0.4 \\  -0.5 \pm 0.4 & -0.5 \pm 0.4 & 1 \end{pmatrix}.
\end{equation}
The residual intrinsic correlations were not strongly constrained, and together have a complex joint uncertainty, owing to the positive-definiteness of correlation matrices.

To test the sensitivity to the dust reddening law, we alternatively fitted the data assuming $R_V = 1.7$.  The posterior results for the intrinsic color locus versus velocity were not substantially changed.  The intercepts $c_0$ changed by less than $1\sigma$, while the slopes changed by $\approx 0.001$.  The estimate of the average dust extinction changed to $\hat{\tau} = 0.25 \pm 0.03$.  We also fitted the data using other color combinations, specifically, ($B-V, V-R, V-I$) and ($B-V, V-R, R-I$), to examine possible velocity trends with other colors.  We did not find slopes significantly different from zero for $V-R$, $V-I$ or $R-I$ with either $R_V = 1.7$ or $R_V = 2.5$.

\subsection{Application of Step Function Model}\label{sec:app_step}

\begin{figure}[t]
\centering
\includegraphics[angle=0,scale=0.41]{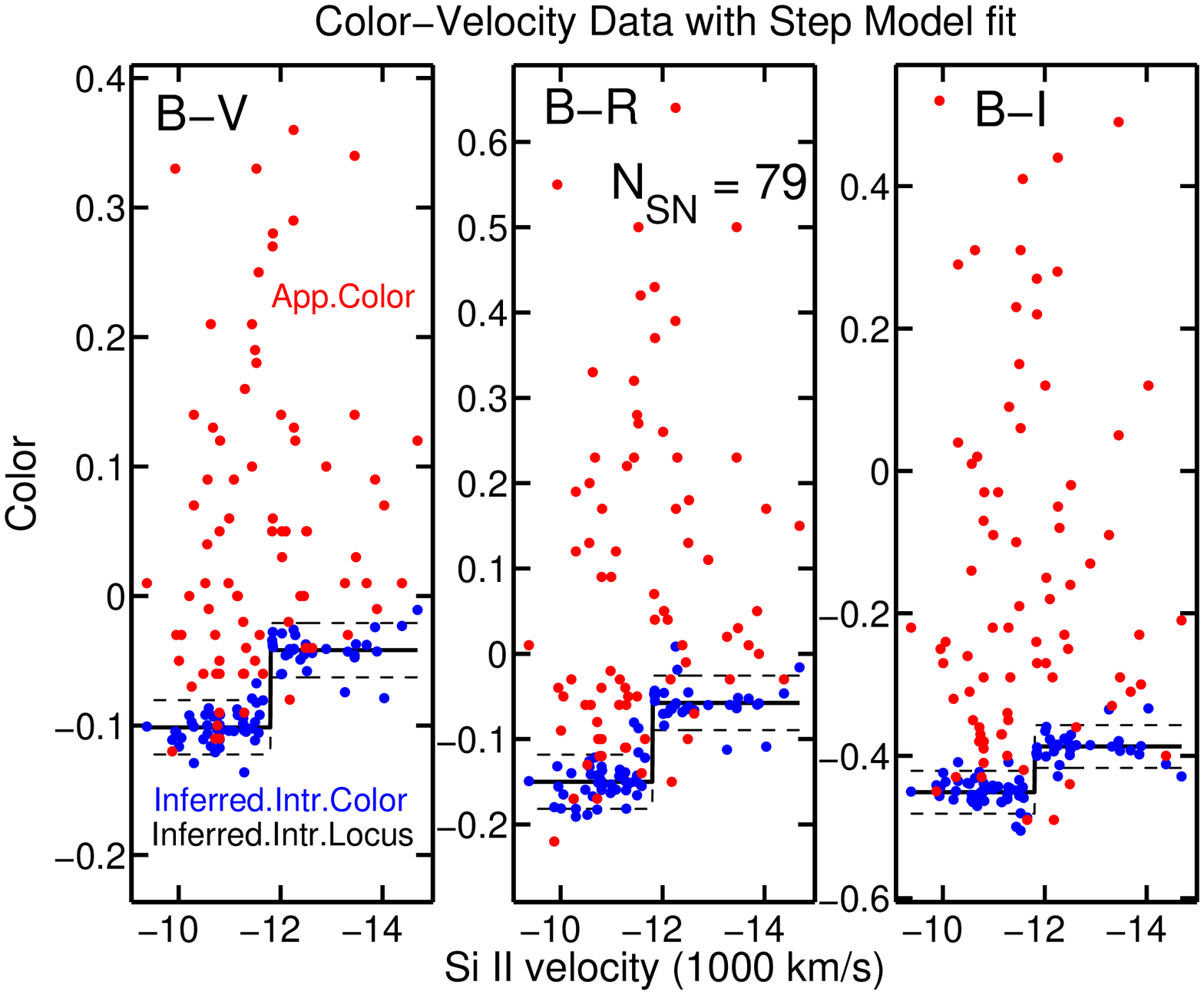}
\caption{\label{fig:plot_step_BVBRBI_n79_232952} The apparent colors and velocity data are shown (red points), along with posterior inferences from the application of the step model.  The blue points are the marginal posterior means for each of the intrinsic colors of each SN Ia.  The black solid line represents the conditional mean intrinsic color-velocity function $\bm{\mu}_C(v; \bm{\theta})$ using the posterior mean estimates of the hyperparameters $\bm{\theta} = \{\bm{\theta}_\text{HV}, \bm{\theta}_\text{LV}  \}$.}
\end{figure}

Next, we fit the \textsc{Step} model of \S \ref{sec:step} to the data, assuming $R_V = 2.5$ and a break at $v_0 = -11,800 \text{ km s}^{-1}$, between the NV and HV groups.  The data and posterior estimates of intrinsic colors and color loci are shown in Fig. \ref{fig:plot_step_BVBRBI_n79_232952}.  The posterior estimate of the population average dust extinction was $\hat{\tau} = 0.33 \pm 0.04$ mag.  The hyperparameters estimates  are summarized in Table f\ref{table:step}.  The marginal posterior densities of the mean intrinsic color offsets between the HV and NV SNe Ia, $\Delta \theta = \theta_{HV} - \theta_{NV}$ are shown in the bottom row of Fig. \ref{fig:post_steplinear_BVBRBI_n79}.

The most significant results are that the mean intrinsic color differences between the two velocity groups are $\Delta \theta = 0.06 \pm 0.02$ mag for $B-V$ and $0.09 \pm 0.02$ mag for $B-R$, such that the intrinsic colors of the HV group are redder (more positive).  The intrinsic color difference in $B-I$ is intriguing but of lower significance.  The uncertainty of $\Delta \theta$ is the largest in $B-I$, so more data may help ascertain if the velocity effect in this color is real.

The inferred values of the mean intrinsic colors at normal velocities $\bm{\theta}_{NV}$ in Table \ref{table:step} are bluer (more negative) than mean intrinsic colors $\bm{c}_0$ found by fitting the \textsc{constant}-Gaussian model (\S \ref{sec:app_gaussian}) at all velocities.  The inferred values of $\bm{\theta}_{HV}$ are redder (more positive) than the mean intrinsic colors in the \textsc{constant}-Gaussian model.  However, the $\bm{\theta}_{NV}$ under the \textsc{Step} model are much closer to the $\bm{c}_0$ under the \textsc{constant}-Gaussian model.  This is because the majority of SNe are in the NV group, and hence the estimation of the global mean intrinsic colors $\bm{c}_0$ is weighted more towards the intrinsic colors of the NV SNe Ia.  Thus, relative to the \textsc{Step} model, the global mean intrinsic colors $\bm{c}_0$ of the \textsc{constant}-Gaussian model will tend to underestimate the intrinsic colors (too blue) for HV objects much more than they overestimate the intrinsic colors (too red) for NV objects.

Fig. \ref{fig:plot_step_BVBRBI_n79_232952} shows that, in just the NV group, there appears to be more SNe with  inferred intrinsic $B-V$ and $B-R$ colors below the NV mean at low velocities, and more SNe with intrinsic colors greater than the NV mean at moderate velocities.  This is suggestive of a trend within just the NV velocity group.

We also fitted the data using other color combinations, specifically, ($B-V, V-R, V-I$) and ($B-V, V-R, R-I$), to examine possible velocity trends with other colors.  For $V-R$, we find a small mean intrinsic color difference of $0.03 \pm 0.01$ between the HV and NV groups.  The mean intrinsic color differences in $V-I$ and $R-I$ were consistent with zero.

\begin{deluxetable}{lrrrrr}
\tabletypesize{\small}
\tablecaption{Estimates of Step Model Hyperparameters}
\tablewidth{0pt}
\tablehead{ \colhead{ } & \colhead{$B-V$} & \colhead{$B-R$} &\colhead{$B-I$} }
\startdata
$\theta_\text{NV}$  & $-0.10 \pm 0.01$ & $-0.15 \pm 0.01$ & $-0.45 \pm 0.02$ \\ 
$\theta_\text{HV}$  & $-0.04 \pm 0.01$ & $-0.06 \pm 0.02$ & $-0.39 \pm 0.03$  \\ 
$\Delta \theta$  & $0.06 \pm 0.02$ & $0.09 \pm 0.02$ &  $0.06 \pm 0.03$ \\
$p_\text{tail}$  & $2 \times 10^{-4}$ & $< 10^{-4}$ & 0.03 \\
 $\sigma_C$ & $0.021 \pm 0.006$ & $0.03 \pm 0.01$ & $0.03 \pm 0.01$ 
 \enddata
\tablecomments{\label{table:step} The mean intrinsic color of normal velocity SN Ia is $\theta_\text{NV}$.   The mean intrinsic color of high velocity SN Ia is $\theta_\text{HV}$.  The intrinsic color offset is $\Delta \theta \equiv \theta_\text{HV} - \theta_\text{NV}$.  The residual intrinsic color scatter is  $\sigma_C$.  Numbers are the posterior means and standard deviations of each parameter in units of magnitude, except for the tail probability $p_\text{tail} = P(\Delta \theta < 0)$.}
\end{deluxetable}

\subsection{Implied Population Distributions of Intrinsic Colors}\label{sec:implied_color_distr}

If it is assumed that there is no trend with ejecta velocity, then the intrinsic color distribution implied by the fitted model is Gaussian by default and is simply described by the estimated hyperparameters, the population means $\bm{c}_0$ and standard deviations $\bm{\sigma}_C$, as given in Table \ref{table:gaussian}.  For example, in $B-V$, the implied intrinsic color distribution is a Gaussian with a mean color of $-0.09$ mag and a standard deviation of 0.03 mag.  In $B-R$, the population  mean intrinsic color is $-0.12$ mag and the population standard deviation is 0.04 mag.   

\begin{figure}[t]
\centering
\includegraphics[angle=0,scale=0.36]{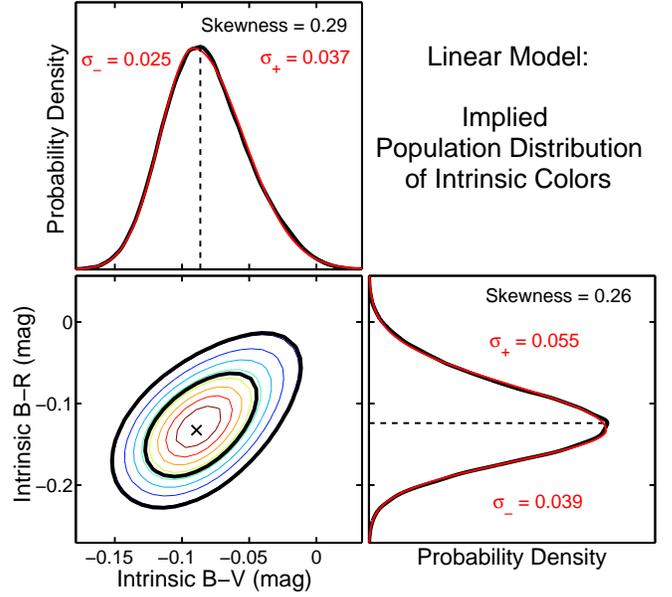}
\caption{\label{fig:implied_intrcol_distr_linear_n79} The intrinsic color population distributions implied by the fitted \textsc{Linear} model.   These are computed using Eq. \ref{eqn:implied_intrcolor_distr} and the empirical distribution of the ejecta velocities (Fig. \ref{fig:siII_vel_distr_n79}). (bottom left) The joint population distribution of intrinsic $B-V$ and $B-R$ colors, showing the mode, and the 68\% and 95\% contours in black.  (top left) The marginal intrinsic color population distribution in $B-V$ (black).  (bottom right) The marginal intrinsic color population distribution of $B-R$ (black). The long tails towards redder values of intrinsic color result in positive skewness.    The vertical dashed lines indicate the marginal modes.  The red curves indicate the best-fit split-Normal approximation to the marginal distributions (Eq. \ref{eqn:splitgaussian}), and the widths of the left and right half-Gaussians are $\sigma_-$ and $\sigma_+$.}
\end{figure}

\begin{figure}[t]
\centering
\includegraphics[angle=0,scale=0.36]{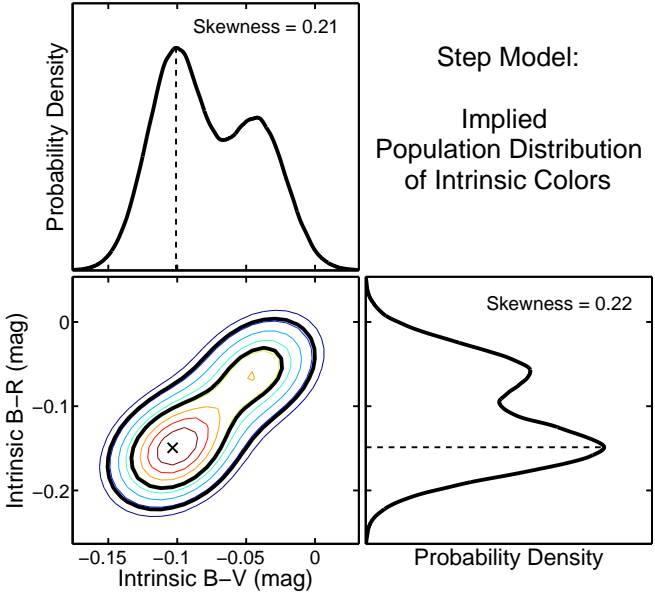}
\caption{\label{fig:implied_intrcol_distr_step_n79_232952} The intrinsic color population distributions implied by the fitted \textsc{Step} function model.   These are computed using Eq. \ref{eqn:implied_intrcolor_distr} and the empirical distribution of the ejecta velocities (Fig. \ref{fig:siII_vel_distr_n79}). (bottom left) The joint population distribution of intrinsic $B-V$ and $B-R$ colors, showing the mode, and the 68\% and 95\%  contours in black.  (top left) The marginal intrinsic color population distribution in $B-V$ (black).  (bottom right) The marginal intrinsic color population distribution of $B-R$ (black). The asymmetries in the intrinsic color distributions result in positive skewness.    The vertical dashed lines indicate the marginal modes.}
\end{figure}

However, we have found statistically significant non-constant trends of SN Ia intrinsic colors with ejecta velocity, particularly for $B-V$ and $B-R$.   As illustrated in Figures \ref{fig:pv_pcv_pc_linear} and \ref{fig:pv_pcv_pc_step}, the non-Gaussian Si II velocity distribution (Fig. \ref{fig:siII_vel_distr_n79}), with a long tail towards higher absolute velocities, together with a significant non-constant mean intrinsic color-velocity relation will generically imply a non-Gaussian marginal intrinsic color distribution.  With the posterior estimates of the hyperparameters $\bm{\theta}, \bm{\Sigma}_C$ governing the intrinsic color-velocity relation, we can compute the implied marginal intrinsic color distribution using Eq. \ref{eqn:implied_intrcolor_distr} to marginalize over the empirical distribution of the ejecta velocities, shown in Fig. \ref{fig:siII_vel_distr_n79}.   This is easily accomplished by Monte Carlo sampling:  sample a random velocity $v_i$ from the empirical velocity distribution, and generate a random set of intrinsic colors $\bm{C}_i \sim N[ \bm{\mu}_C(v_i; \bm{\theta}), \bm{\Sigma}_C]$, using the posterior mean estimates of the hyperparameters $\bm{\theta}, \bm{\Sigma}_C$.  
In this way, we generate a large number of samples  $\{ \bm{C}_i\}$ from the intrinsic color population distribution $P( \bm{C} | \, \bm{\theta}, \bm{\Sigma}_C )$ in Eq. \ref{eqn:implied_intrcolor_distr}.   Kernel density estimation on the samples computes the probability densities shown in Figs. \ref{fig:implied_intrcol_distr_linear_n79} and  \ref{fig:implied_intrcol_distr_step_n79_232952}.

For the \textsc{linear} model, the implied intrinsic distributions, both joint and marginal, for $B-V$ and $B-R$ are shown in Fig. \ref{fig:implied_intrcol_distr_linear_n79}.    The (mean, mode, standard deviation) for each marginal distribution are $(-0.082, -0.090, 0.031)$ mag for $B-V$ and $(-0.12, -0.126, 0.047)$ mag for $B-R$.  The non-Gaussianity of the velocity distribution, together with the linear intrinsic color-velocity relation, give rise to a longer tail at redder (positive) values of each color.  Whereas a Gaussian distribution is symmetric and has zero skewness, the $B-V$ and $B-R$ intrinsic distributions have skewness of 0.29 and 0.26, respectively.   
The skewness of the implied intrinsic color distribution is a function of the hyperparameters $s = s(\bm{\theta}, \bm{\Sigma}_C)$.  We can compute the posterior distribution of the skewness, $P( s(\bm{\theta}, \bm{\Sigma}_C) | \{ \bm{O}_s, v_s \})$, by using the MCMC samples of the hyperparameters, $\bm{\theta}^i, \bm{\Sigma}_C^i$ as draws from the posterior distribution $P(\bm{\theta}, \bm{\Sigma}_C | \{ \bm{O}_s, v_s \})$.   Using $10^3$ random MCMC draws, we compute the posterior probability of positive skewness, $P( s > 0 ) = (0.988, 0.998, 0.811)$ for $B-V$, $B-R$, and $B-I$, respectively.  

Under the \textsc{linear} model, the marginal intrinsic color population distributions have no simple analytic form and we have computed them numerically.  We can approximate each marginal pdf by fitting a split-Normal density, $\mathcal{SN}(\tilde{\mu}, \sigma_{-}, \sigma_{+})$ (Eq. \ref{eqn:splitgaussian}) to our Monte Carlo samples $\{ \bm{C}_i\} \sim P( \bm{C} | \, \bm{\theta}, \bm{\Sigma}_C )$.  For $B-V$,  the best-fitting split-Normal approximation has parameters $\tilde{\mu} = -0.0916$ mag, $\sigma_{-} = 0.0247$ mag, and $\sigma_{+} = 0.0369$ mag.  For $B-R$, we find $\tilde{\mu} = -0.1323$ mag, $\sigma_{-} = 0.0387$ mag, and  $\sigma_{+} = 0.0546$ mag.   These approximations accurately capture the skewness of the marginal intrinsic distributions computed via Eq. \ref{eqn:implied_intrcolor_distr}.  These analytic approximations are shown as the red curves in Fig. \ref{fig:implied_intrcol_distr_linear_n79}, and may be useful for simulating these intrinsic color distributions.

We repeat these calculations for the \textsc{step} function model.  The implied joint and marginal $B-V$ and $B-R$ intrinsic color distributions under the posterior mean estimates of $\bm{\theta}, \bm{\Sigma}_C$ are shown in Fig. \ref{fig:implied_intrcol_distr_step_n79_232952}.  The (mean, mode, standard deviation) for each marginal distribution are $(-0.077, -0.101, 0.036)$ mag for $B-V$ and $(-0.113, -0.149, 0.055)$ mag for $B-R$.  The marginal intrinsic color distributions have skewness of 0.22 and 0.21 for $B-V$ and $B-R$, respectively.  The posterior probabilities of positive skewness $P( s > 0) = (0.996, 0.998, 0.939)$  for $B-V$, $B-R$, and $B-I$, respectively. 

Under the \textsc{step} model, the marginal implied intrinsic color population distributions have a simple analytic form as the mixture of two Gaussians:
\begin{equation}
\begin{split}
P(C | \, \hat{\theta}_\text{NV}, \hat{\theta}_\text{HV}, \hat{\sigma}_C &, \hat{\pi}_\text{NV} ) = \hat{\pi}_\text{NV}  N( C | \, \hat{\theta}_\text{NV}, \hat{\sigma}^2_C ) \\ &+ (1-\hat{\pi}_\text{NV} ) \, N( C | \, \hat{\theta}_\text{HV}, \hat{\sigma}^2_C ),
\end{split}
\end{equation}
where the  estimates $(\hat{\theta}_\text{NV}, \hat{\theta}_\text{HV}, \hat{\sigma}_C)$ are given in Table \ref{table:step}, and $\hat{\pi}_\text{NV}$ is the fraction of SNe Ia with normal velocities $|v| < 11,800 \text{ km s}^{-1}$.  For this sample, $\hat{\pi}_\text{NV} = 0.595$.

With both linear and step function models, the implied intrinsic distributions of $B-V$ and $B-R$ have statistically significant skewness of about 0.2 to 0.3.  This numerically captures the non-Gaussianity implied in the intrinsic colors by the skewed velocity distribution and the fitted intrinsic color-velocity relations.   Relative to both of these models, the fitted constant-Gaussian model with zero skewness underestimates the probability of intrinsically red, high velocity objects.  It also slightly underestimates the width (standard deviation) of the intrinsic color distributions, but not significantly so.

\subsection{Effect on Estimation of Intrinsic Colors and Host Galaxy Dust Extinction of Individual SNe Ia}\label{sec:effect}

In this section, we illustrate the effect of the intrinsic color-velocity trend and the use of velocity measurements to improve estimates of intrinsic colors and dust extinction to individual SNe Ia.    Using the SNe Ia with color and velocity measurements as a training set, we have trained statistical models that capture the intrinsic colors-velocity relations by estimating their population hyperparameters.  For a new SN with apparent color and velocity measurements, we can use the trained model to infer its velocity-dependent intrinsic colors and thus the dust reddening.  For a new SN with only apparent color measurements and no velocity data, the population color-velocity information is still useful.  We can still use the trained model's non-Gaussian implied intrinsic colors distribution (\S \ref{sec:implied_color_distr}) to obtain a skewed posterior probability for its dust extinction that marginalizes over the unknown specific velocity of the SN.

For each model $\mathcal{M}$ (e.g. \textsc{constant}-Gaussian, \textsc{Linear}, \textsc{Step}), we obtain estimates of the hyperparameters of the dust population ($\hat{\tau}$) and the intrinsic color population ($\bm{\hat{\theta}}, \bm{\hat{\Sigma}}_C$) within that model by training on the full data set of colors and velocity measurements.  We calculate the posterior mean estimates of these hyperparameters using the MCMC samples from the global posterior, Eq. \ref{eqn:globalposterior}.  Using these estimates, we can estimate the intrinsic colors and dust extinction for any new supernova $s$ with measured colors and velocity by computing the posterior $P( \bm{C}_s, A_V^s  | \, \bm{\hat{\theta}}, \bm{\hat{\Sigma}}_C, \hat{\tau}; \bm{O}_s, v_s, \mathcal{M}  )$ using Eq. \ref{eqn:posterior_CAv}\footnote{For a more fully Bayesian approach, one could also average over the posterior of the hyperparameters conditional on the training set.}.   If the specific velocity $v_s$ of the individual SN is unobserved, then we marginalize over the the unknown velocity using the empirical distribution of velocities $P_V(v)$ (Fig. \ref{fig:siII_vel_distr_n79}) to get the posterior of $\bm{C}_s, A_V^s$ using only the apparent color measurements:
\begin{equation}\label{eqn:post_CAv_O}
\begin{split}
P&( \bm{C}_s, A_V^s  | \, \bm{\hat{\theta}}, \bm{\hat{\Sigma}}_C, \hat{\tau}; \bm{O}_s, \mathcal{M}  ) \\ &= \int dv  \,P( \bm{C}_s, A_V^s  | \, \bm{\hat{\theta}}, \bm{\hat{\Sigma}}_C, \hat{\tau}; \bm{O}_s, v, \mathcal{M}  ) \, P_V(v).
\end{split}
\end{equation}

\begin{figure}[b]
\centering
\includegraphics[angle=0,scale=0.39]{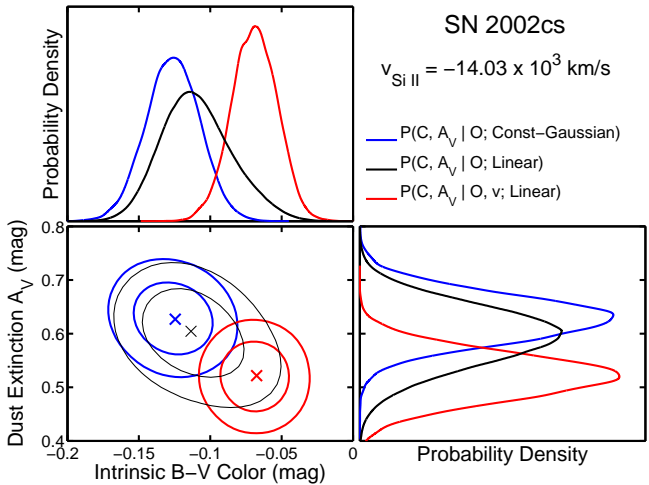}
\caption{\label{fig:sn2002cs_condmarg_linear_vs_const} The posterior density of the inferred intrinsic color $B-V$ and dust extinction given the data for high-velocity SN 2002cs under the trained \textsc{constant}-Gaussian or \textsc{Linear} models.  (bottom left) The mode, 68\%, and 95\% highest posterior density contours of the joint inference, (bottom right) The normalized marginal posterior density for $A_V$ for this SN. (top left) The normalized marginal posterior density of intrinsic $B-V$ for this SN.  (blue)  The \textsc{constant}-Gaussian model does not incorporate a trend of intrinsic colors versus ejecta velocity, and tends to underestimate the intrinsic color and overestimate the host galaxy dust extinction $A_V$ for this high-velocity SN.  (black) The trained \textsc{Linear} model incorporates a trend between intrinsic colors and versus ejecta velocity.   If the specific velocity of this SN is not measured, then conditioning only on the SN apparent colors $\bm{O}_s$ and marginalizing over its unknown velocity, the posterior probability density of the intrinsic color and dust extinction is broad and skewed towards redder (more positive) intrinsic color and lower dust extinction, to account for population velocity distribution. (red) Conditioning on both the measured ejecta velocity $v_s$ and the apparent colors $\bm{O}_s$ for this SN, using the trained \textsc{Linear} model, the posterior density focuses on a solution with a red (more positive) intrinsic $B-V$ color and a lower extinction $A_V$.  The red and black pdfs are related through Eq. \ref{eqn:post_CAv_O}.  The red pdf is the posterior evaluated at the specific velocity of this SN.  The black pdf is the result of marginalizing the red pdf evaluated over the empirical velocity distribution of the SN Ia population, shown in Fig. \ref{fig:siII_vel_distr_n79}.  The measurement of the ejecta velocity adds valuable information that increases the precision of the intrinsic color and dust estimate when used with the trained \textsc{Linear} model, and corrects the error that would be incurred from using the \textsc{constant}-Gaussian model that ignores the intrinsic color-velocity trends.}
\end{figure}

In Figure \ref{fig:sn2002cs_condmarg_linear_vs_const}, we demonstrate these inferences with a particular object, SN 2002cs, with a high velocity $v = -14,030 \text{ km s}^{-1}$ and moderate reddening.   Using the hyperparameters learned from the training set with the \textsc{constant}-Gaussian model, we computed $P( \bm{C}_s, A_V^s  | \, \bm{\hat{\theta}}, \bm{\hat{\Sigma}}_C, \hat{\tau}; \bm{O}_s, \textsc{Constant} )$.  This is the same as what one would obtain by also conditioning on the velocity measurement, since the mean intrinsic colors are independent of velocity in this model, so the specific velocity of the SN is ignored.  The posterior estimates of the peak intrinsic $B-V$ color and the host galaxy dust extinction $A_V$ of this SN are shown in blue.  Since this model does not account for the intrinsic colors-velocity trend, the inferred intrinsic colors are too blue (negative) and the inferred $A_V$ is overestimated for this SN.  For a normal velocity SN, the estimate of the intrinsic colors will tend to be too red (positive).

Next, we used the trained \textsc{Linear} model and computed Eq. \ref{eqn:post_CAv_O}, conditioning on the specific apparent color measurements, but not the velocity measurement of this SN.  This result is shown in black.  In this case, the relevant intrinsic color population distribution is that shown in Fig. \ref{fig:implied_intrcol_distr_linear_n79}, obtained by integrating over the population distribution of ejecta velocities.      Because the \textsc{linear} model incorporates the population intrinsic colors-velocity trend, the integration over the population distribution of velocities results in a broad posterior probability for $(\bm{C}_s, A_V^s)$, given the apparent color measurement.  The posterior density of the specific intrinsic $B-V$ color of the SN is skewed towards redder (positive) values, to account for the chance that the SN has a high velocity.  Consequently, the posterior density of the  dust extinction $A_V$ of the SN is skewed towards lower values.  The joint posterior of intrinsic color and dust extinction exhibits an expected anti-correlation, reflecting the trade-off between the two effects.  Under the assumption that the \textsc{Linear} model is true, these posterior estimates of dust extinction and intrinsic color, given the specific apparent colors but without the specific velocity measurement, will be correct on the average, with respect to the population velocity distribution.  However, this requires that the specific velocity of the new SN can be considered a random draw from the same population velocity distribution as the training set.  Relative to these estimates, the \textsc{constant}-Gaussian  estimates are biased to bluer (more negative) intrinsic colors and more dust extinction.

Using the \textsc{Linear} model, we next conditioned on both the specific apparent colors and ejecta velocity of this SN.   We calculated $P( \bm{C}_s, A_V^s  | \, \bm{\hat{\theta}}, \bm{\hat{\Sigma}}_C, \hat{\tau}; \bm{O}_s, v_s, \textsc{Linear} )$, shown in red.  The impact of the velocity information is to make the posterior probability of $(\bm{C}_s, A_V^s)$ more sharply peaked upon values of the intrinsic color redder (more positive) than those of the overall population.  This is accounting for the intrinsic color-velocity trend captured by the \textsc{Linear} model, in combination with the actual velocity measurement for this SN.  Furthermore, the dust extinction $A_V$ estimate is consequently smaller and more precise.  Within the \textsc{Linear} model, the use of both apparent color and velocity measurements for this SN improves the precision (inverse variance) of the intrinsic $B-V$ color estimate by a factor of 2.4 and improves the precision of the $A_V$ estimate by a factor of 1.7, compared to using apparent colors alone (black).   The accuracies are also improved by using the velocity information to adjust the estimate of the intrinsic colors and, thus, the dust extinction.  Relative to these estimates, the \textsc{constant}-Gaussian model incurs an error of $-0.06$ mag (too blue) in estimating the intrinsic $B-V$ color, and an error of $+0.11$ mag in the $A_V$ extinction estimate.  Ignoring velocity information, both of the population and of the specific SN, results in a velocity-dependent systematic error in dust and intrinsic color estimates.

\begin{figure}[t]
\centering
\includegraphics[angle=0,scale=0.39]{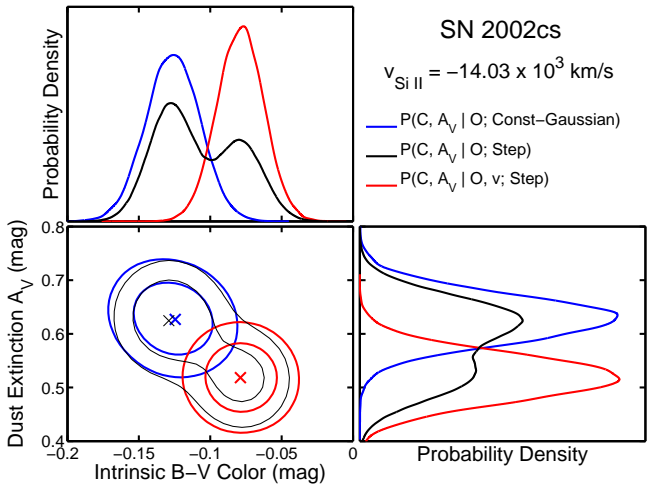}
\caption{\label{fig:sn2002cs_condmarg_step_vs_const} The posterior density of the inferred intrinsic color $B-V$ and dust extinction given the data for high-velocity SN 2002cs under the trained \textsc{constant}-Gaussian versus \textsc{Step} models.  (bottom left) The mode, 68\%, and 95\% highest posterior density contours of the joint inference. (bottom right) The normalized marginal posterior density for $A_V$ for this SN, (top left) The normalized marginal posterior density of intrinsic $B-V$ for this SN.  (blue)  The \textsc{constant}-Gaussian model does not incorporate a trend of intrinsic colors versus ejecta velocity, and tends to underestimate the intrinsic color and overestimate the host galaxy dust extinction $A_V$ for this SN.  (black) The trained \textsc{Step} model incorporates a trend between intrinsic colors and versus ejecta velocity.  If the specific velocity of this SN is not measured, then conditioning only on the SN apparent colors $\bm{O}_s$ and marginalizing over its unknown  velocity, the posterior probability density of the intrinsic color and dust extinction is broad and double peaked, to account for the possibilities that the SN has low ejecta velocity and is intrinsically red with low extinction, or has high ejecta velocity and is intrinsically blue with a large dust extinction. (red) Conditioning on both the measured ejecta velocity $v_s$ and the apparent colors $\bm{O}_s$ for this SN, using the trained Step model, the posterior density focuses on a solution with a red (more positive) intrinsic $B-V$ color and a lower extinction $A_V$.}
\end{figure}

In Figure \ref{fig:sn2002cs_condmarg_step_vs_const} we show the same calculations with the same SN, but using the  \textsc{Step} model.   The posterior probability of $(\bm{C}_s, A_V^s)$ for this SN under the \textsc{constant}-Gaussian model is shown in blue.   The estimate of intrinsic color is too blue (negative), and the dust extinction $A_V$ is too large.  In black, we compute Eq. \ref{eqn:post_CAv_O}, using the trained \textsc{Step} model and conditioning on apparent color measurements, but not the specific velocity, of this SN.  In the case that the specific velocity is unknown, the relevant implied intrinsic color population distribution is the bimodal distribution shown in Fig. \ref{fig:implied_intrcol_distr_step_n79_232952}.  This is reflected in the posterior density for the individual supernova parameters $(\bm{C}_s, A_V^s)$.  The posterior probability of the intrinsic colors and dust extinction of this SN is spread over two modes, corresponding to the possibilities that the unknown specific velocity is high or normal.  If the \textsc{Step} model is true, these posterior estimates, given the apparent colors but no velocity measurement, will be correct on the average with respect to the population velocity distribution.  Relative to this posterior, the estimates obtained with the \textsc{constant}-Gaussian model are again biased to bluer (more negative) intrinsic colors and more dust extinction.

The posterior density conditioning on both the apparent colors and the specific velocity measurement, $P( \bm{C}_s, A_V^s  | \, \bm{\hat{\theta}}, \bm{\hat{\Sigma}}_C, \hat{\tau}; \bm{O}_s, v_s, \textsc{Step} )$ is shown in red.  The velocity information improves the accuracy of the inferences by concentrating the posterior probability on the redder intrinsic color/lower dust extinction solution.  Within the \textsc{Step} model, the use of both apparent color and velocity measurements for this SN improves the precision (inverse variance) of the intrinsic $B-V$ color by a factor of 3.2 and improves the precision of the $A_V$ estimate by a factor of 2.6, compared to using apparent colors alone.   Relative to these estimates, the \textsc{constant}-Gaussian model incurs a velocity-dependent systematic error of $-0.05$ mag (too blue) in estimating the intrinsic $B-V$ color, and an error of $+0.12$ mag in the $A_V$ extinction estimate.


\subsection{Impact on Dust Extinction Estimates}\label{sec:dust_impact}

We find statistically significant non-constant trends of intrinsic color versus Si II ejecta velocity.  Hence, a model accounting for the color-velocity effect changes the estimate of host galaxy dust extinction, relative to a model that assumes zero trend.   We examine and quantify these corrections to the extinction by comparing the $A_V$ estimates from the \textsc{linear} and \textsc{step} models to those of the \textsc{constant}-Gaussian model for the 79 SNe Ia in the data set.  Current analysis methods typically explicitly or implicitly assume a Gaussian intrinsic color distribution with a single mean color that is constant with respect to expansion velocity.  Comparing the $A_V$ estimates to those obtained from this baseline model quantifies the expected improvement from incorporating velocity information.  The $A_V$ estimates are obtained from the global posterior density, Eq. \ref{eqn:globalposterior}, as part of the Gibbs sampling routine (Appendix \ref{sec:mcrc}).
 
\begin{figure}[t]
\centering
\includegraphics[angle=0,scale=0.395]{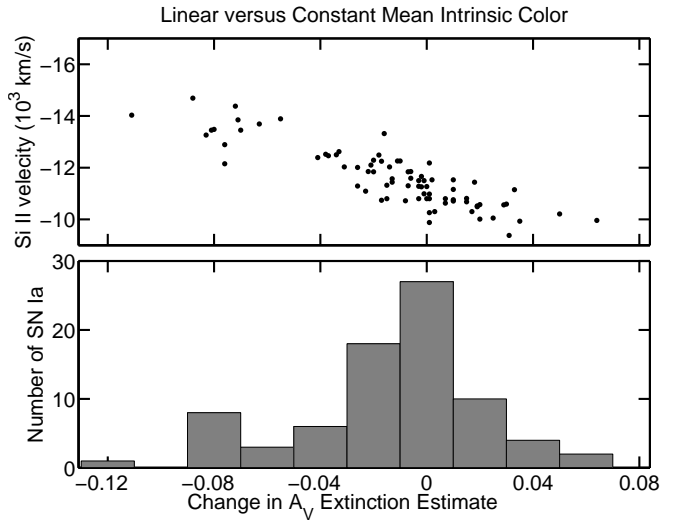}
\caption{\label{fig:deltaAv_linear_vs_const_n79_232314} The change in the individual extinction estimates, $\Delta A_V$, as a function of ejecta velocity, from the \textsc{constant}-Gaussian  (constant mean intrinsic color) model to the \textsc{Linear} model.  This quantifies the error in $A_V$ incurred by ignoring velocity information.  At high absolute ejecta velocities, the \textsc{constant}-Gaussian model underestimates the intrinsic colors (too blue), and overestimates $A_V$.  The correction to $A_V$ is as negative as $-0.11$ mag.  At low absolute velocities, the \textsc{constant}-Gaussian  model overestimates the intrinsic colors (too red), leading to corrections as positive as $+0.06$ mag.}
\end{figure}

 In Figure \ref{fig:deltaAv_linear_vs_const_n79_232314}, we compare the extinction estimates obtained from the \textsc{linear} model (\S \ref{sec:app_linear}) to those from the \textsc{constant}-Gaussian  model (\S \ref{sec:app_gaussian}).   For each model, we compute the posterior marginal mode of $P(A_V^s | \{ \bm{O}_s, v_s \}, \textsc{Model} )$ for each SN $s$, and show differences  $\Delta A_V^s$ in the modes between  \textsc{Linear} and \textsc{constant}-Gaussian.   At high velocities, $|v| > 11,800 \text{ km s}^{-1}$, the extinction estimate under \textsc{Linear} is a smaller positive number than under \textsc{constant}-Gaussian, because the intrinsic colors  are inferred to be redder under the \textsc{Linear} model.    The $\Delta A_V^s$ estimates quantify the systematic error in the extinction estimate incurred by ignoring velocity information.  While most of the $\Delta A_V$ estimates cluster around zero, at high ejecta velocities, the extinction correction is as negative as $\Delta A_V \approx -0.11$ mag, and at low ejecta velocities it is as positive as $\Delta A_V \approx 0.06$ mag.  The (mean, median) $\Delta A_V$ of the sample are $(-0.013, -0.006)$ mag.

\begin{figure}[t]
\centering
\includegraphics[angle=0,scale=0.395]{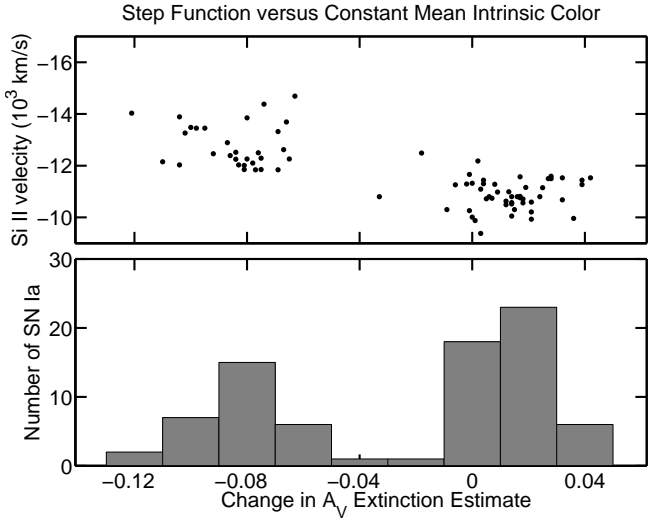}
\caption{\label{fig:deltaAv_step_vs_const_BVBRBI_n79_232952} The change in the individual extinction estimates, $\Delta A_V$, as a function of velocity, from the \textsc{constant}-Gaussian  (constant mean intrinsic color) model to the step function model. This quantifies the error in $A_V$ incurred by ignoring velocity information.  At high absolute ejecta velocities, the \textsc{constant}-Gaussian  model underestimates the intrinsic colors (too blue), and overestimates $A_V$.  The correction to $A_V$ is as negative as $-0.12$ mag.  At low absolute velocities, the \textsc{constant}-Gaussian  model overestimates the intrinsic colors (too red), leading to corrections as positive as $+0.04$ mag.}
\end{figure}

In Figure \ref{fig:deltaAv_step_vs_const_BVBRBI_n79_232952}, we show the change in the extinction estimates, $\Delta A_V$, from switching from the \textsc{constant}-Gaussian  intrinsic color model to the \textsc{Step} model.  Because the step function model infers a redder mean intrinsic color for high velocity events, the extinction estimates for high velocity events are shifted to smaller positive numbers.   The mean intrinsic color for normal velocity events ($|v| < 11,800 \text{ km s}^{-1}$), however, is bluer than the global mean intrinsic color using the \textsc{constant}-Gaussian  model, so the extinction estimates are larger positive numbers. At high Si II ejecta velocities, the extinction corrections have a distribution peaked near $\Delta A_V \sim -0.08$ mag, whereas at normal velocities, they have a distribution clustered around $+0.02$ mag.    The extinction correction is as negative as $-0.11$ mag or as positive as $0.04$ mag.  The (mean, median) $\Delta A_V$ of the sample are $(-0.024, 0.001)$ mag.

The extinction $A_V$ estimates within each model are sensitive to the assumed value of $R_V$ controlling the dust reddening law.  Regardless of the model (\textsc{constant}-Gaussian , \textsc{Linear}, or \textsc{Step}), the average $A_V$ extinction was found to be $\hat{\tau}= (0.25 \pm 03, 0.35 \pm 0.04, 0.40 \pm 0.05)$ mag for $R_V = (1.7, 2.5, 3.1)$, respectively.  Hence, a typical  extinction value will increase by a factor $\approx 1.14$ going from $R_V = 2.5$ to $R_V = 3.1$, or decrease by a factor of $\approx 0.71$ going from $R_V = 2.5$ to $R_V =1.7$.   These factors are not exactly equal to the ratios of the assumed $R_V$ values because SN Ia spectra differ from stellar spectra and the $A_V$ estimates rely not just on $B-V$ colors but also use information from $B-R$ and $B-I$.

\section{Conclusion}\label{sec:conclusion}

We have constructed a hierarchical Bayesian model to estimate the relation between multiple peak intrinsic colors of SNe Ia and their photospheric expansion velocities measured from the Si II $\lambda$6355 absorption feature.     We model the distribution of the observed apparent colors, conditional on the velocity measurement, as a probabilistic combination of the intrinsic color locus, a dust reddening distribution and measurement error scatter.   The hyperparameters of the underlying distributions are determined from the posterior density conditional on all the SN Ia data.  Bayesian inference with the hierarchical model can be thought of as a probabilistic deconvolution of the data into the different sources of randomness generating it.  We developed and implemented a Gibbs sampling code (Appendix \ref{sec:mcrc}) to generate MCMC samples from the global posterior density of the unknowns conditional on the data.  The deviance information criterion (DIC) (Appendix \ref{sec:modelcomp}) is computed to evaluate the relative fits of models with different levels of complexity.  We used this model to analyze a dataset of $79$ nearby SNe Ia with $BVRI$ light curves and Si II spectroscopic  ejecta velocity measurements.  

The empirical distribution of Si II ejecta velocities (Fig. \ref{fig:siII_vel_distr_n79}) is well described by a gamma distribution with a long tail towards high absolute velocities, and a skewness of $0.6 \pm 0.2$.  A simple analysis of the distributions of apparent $B-V$ and $B-R$ colors of SNe Ia in the HV and NV groups (Fig.  \ref{fig:plot_appcol_ecdfs_all_n79}) reveals significant differences in the ``blue edge'' (left tail) of each distribution, where one would expect objects with the least dust reddening.  However, no similar significant differences are found in other colors, as one would expect if the effects were due to host galaxy dust (i.e. overall more host galaxy dust reddening for high velocity objects), since dust reddening affects all colors simultaneously.  This strongly suggests that the velocity dependence of the apparent color distribution in $B-R$ and $B-V$ is due to spectroscopic physics intrinsic to the supernova explosion, rather than to extrinsic host galaxy dust.  We applied our hierarchical regression method to directly model the relations between intrinsic colors and velocity, treated as a continuous variable, using the apparent color and velocity data.

We confirm previously published findings with our new method, using only the apparent color and velocity data, and a larger number of SNe.  \citet{foleykasen11}, analyzing the sample of \citet{wangx09b}, found that the $B_\text{max}-V_\text{max}$ pseudocolors of high-velocity SNe Ia were offset by $\sim 0.06$ mag to the red of those of normal velocity SNe Ia.  \citet{foleysanderskirshner11} analyzed the scatter around the mean relation between SN Ia absolute magnitudes, controlling for light curve shape, versus  $B_\text{max}-V_\text{max}$ pseudocolors.   Fitting the residuals, which they interpret as being due to variations of the intrinsic colors, versus velocities, they estimated a linear relation with slope of $-0.033$ mag $(10^3 \text{ km s}^{-1} )^{-1}$.   \citet{blondin12} regressed the intrinsic $B-V$ colors, inferred from fitting the apparent light curves with \textsc{BayeSN} \citep{mandel11}, against velocity measurements.  They estimated a linear slope of $-0.013 \pm 0.005$ mag $(10^3 \text{ km s}^{-1} )^{-1}$, and a mean intrinsic $B-V$ color difference between high- and normal-velocity groups of $0.030 \pm 0.013$ mag.

In this work, by jointly modeling the multiple peak colors in $BVRI$ versus velocity, we find a $B-V$ mean intrinsic offset of $0.06 \pm 0.02$ mag between the high velocity and normal velocity SNe Ia in $B-V$ using a step-function model.  With a linear model, we estimate the slope to be $-0.021 \pm 0.006$ mag $(10^3 \text{ km s}^{-1} )^{-1}$.  While the mean intrinsic color difference is closer to that found by \citet{foleysanderskirshner11}, the slope under a linear fit is closer to that of \citet{blondin12}.   For the first time, the velocity effect on intrinsic $B-R$ colors is examined, and we find that it is more pronounced than in $B-V$.  The mean intrinsic color of the high velocity group is $0.09 \pm 0.02$ mag redder than that of the normal velocity SNe Ia.  The slope under a linear model is $-0.030 \pm 0.009$ mag $(10^3 \text{ km s}^{-1} )^{-1}$.  The effects in $B-I$ are intriguing but less statistically significant; further data will help ascertain the intrinsic $B-I$ color vs. velocity relation.  The intrinsic color-velocity effects in longer wavelength colors, i.e. $V-R$, $V-I$, and $R-I$ are not significant.

We have focused in this work on the optical colors in the $BVRI$ bands, wavelengths at which there is the most nearby SN Ia data.  Although there is currently comparatively less data in the near infrared and ultraviolet (UV), future empirical work can explore the correlations between velocities and other spectroscopic properties and intrinsic colors at those wavelengths.  Although our statistical model allows for both intrinsic variations correlated with velocity, and uncorrelated intrinsic scatter, whether or not there are additional intrinsic color variations that are not captured by the model in practice is an open question.   Expanding the analysis to include NIR and UV colors, and their dependence on velocity, and potentially other spectroscopic properties may help us get a better handle on both the intrinsic variations of SNe Ia and the properties of host galaxy dust.

Our results are broadly consistent with the heuristic physical explanation for the effect described by \citet{foleykasen11}.
The higher ejecta velocities correlate with broader absorption lines. In the $B$-band, the line opacity is important, so this has the effect of
reducing the flux in $B$.  In the $V$-band, electron scattering is more important, so the effect is minimal.   Therefore, one would expect
redder intrinsic $B-V$ colors with higher ejecta velocities.   This is seen in the asymmetric, detonating failed deflagration explosion models of \citet{kasenplewa07}.  The same reasoning would explain the strong effect in $B-R$ colors, and lack of a clear effect in $V-R$, $V-I$, and $R-I$, colors, as electron scattering generally dominates at wavelengths longer than $\sim 4300$ \AA.  However, \citet{blondin11b} found that a strong correlation between the $B-V$ intrinsic color and ejecta velocity was not a generic feature of the 2D delayed detonation models of \citet{krw09}.  Hence, the reality of these effects in the observed data place important constraints on theoretical explosion mechanisms.  Thus, we urge supernova theorists to present the intrinsic color-velocity trends predicted by their explosion models, in $B-V$, as well as other colors.
 
Using the measurement of the ejecta velocity increases the precision and accuracy of the dust extinction and intrinsic color estimates for an individual event, particularly for high ejecta velocity events, as we demonstrated in \S \ref{sec:effect}.  However, there is useful information about the velocity-color relations captured in the implied intrinsic color population distribution even when applied to inference for a SN without a velocity measurement.  In \S \ref{sec:implied_color_distr}, for the first time, we have modeled the non-Gaussianity of the population distribution of colors intrinsic to the SNe Ia by marginalizing over the  ejecta velocity distribution, using the training set of SNe Ia with both color and velocity data.    In particular, regardless of the choice of step or linear model, we estimated a skewness of $+0.2$ to $+0.3$ towards redder colors. This skewness in the implied intrinsic colors distribution  also alters the probability distribution of the inferred extinction for a particular SN Ia with no ejecta velocity measurement.  Specifically, as redder intrinsic colors are more likely due to the long positive tail, this will make the posterior probability density of extinction to a particular SN Ia have a skewed tail towards lower values (less dust), to account for the chance that it has a high ejecta velocity.  The implied intrinsic color distribution leads to dust and color inferences that are correct on the average with respect to the population's ejecta velocity distribution.  However, the cost of not using the specific velocity measurement of a SN Ia is lower precision in the dust and color estimates.  In the absence of a specific velocity measurement for a SN, the skewed implied intrinsic color distribution should be used instead of the constant-Gaussian model, which ignores the population velocity-color relation and underestimates the probability of intrinsically red, high velocity events, leading to inferences that tend to be biased toward bluer intrinsic colors and higher dust extinction.  The applicability of the intrinsic colors-velocity trends we find and the marginal intrinsic color distributions they imply to other SNe Ia depends upon the assumption that they belongs to the same effective population of events used in the training set.  This caveat is important to consider when comparing SN Ia  samples from different surveys.  

The presence of a significant intrinsic color-velocity relation impacts the estimation of host galaxy dust reddening and extinction using color information.  As the intrinsic colors of SNe Ia depend on the ejecta velocity, the amount of dust  inferred from the difference between the apparent colors and intrinsic colors will also depend on it (\S \ref{sec:dust_impact}).   Relative to the constant-Gaussian intrinsic color model with no velocity-color trend, the velocity-dependent correction in $A_V$ could be as large as $-0.12$ mag at high velocities and $+0.06$ mag at low velocities.  Hence, ignoring velocity information and its relation to the intrinsic color distribution will result in errors in host galaxy extinction estimates that will propagate to errors in the distance modulus.   The particular impact on distance estimates will depend on the method used to infer the distance, how it models SN Ia color, and what other information or wavelengths are used.  We would expect these errors to be largest for methods that depend solely on optical light curves to derive extinction and distance estimates;  methods utilizing near-infrared light curves may be less susceptible to these errors, as dust extinction is a much smaller factor in those wavelengths.  These errors will be incurred in a \emph{systematic} direction depending on the ejecta velocity.  This would lead to systematic errors in cosmological inference if, for example, the observed distribution of ejecta velocities changes with redshift.  \citet{foley12a} find that the low-$z$ and high-$z$ SN Ia samples have similar Si II velocity distributions; this should be monitored for future surveys.

A challenge to fully utilizing intrinsic color-velocity information for cosmological SN Ia samples is obtaining accurate velocity measurements from high-$z$  spectra near maximum light .  \citet{foley12a} examined the published spectra and light curve parameters of the high-$z$ SNLS and SDSS-II SN Ia samples.   Of the 282 SNe Ia with both spectra and photometry, 154 objects pass the cuts on light curve shape and the phase of the spectrum, while only 40 SNe Ia survive a further cut based on spectrum quality.  At $z > 0.4$, the Si II $\lambda 6455$ feature redshifts out of the observer-frame optical range.  Alternatively, the Ca II H\&K feature may be used  to measure expansion velocity at higher redshifts up to $z \approx 1.2$.  Near-infrared spectroscopy would be useful for measuring these spectroscopic features to the highest redshifts.   Improved modeling of their velocity evolution over a broad range of decline rates may help broaden the fraction of useable SNe Ia.  Optimizing future survey strategies for obtaining adequate maximum-light spectra would increase the yield of SNe Ia with useful color-velocity data.  Even if the expansion velocity cannot be measured for all SNe Ia because of observational limitations, measurements for a representative training subset can be used to estimate the intrinsic color distribution, which could then be used as a prior for estimating dust and distances for the SNe Ia lacking velocity measurements, as demonstrated in \S \ref{sec:effect}.

In this paper, we have focused on modeling the apparent color and ejecta velocity data, inferring the intrinsic colors-velocity trends for multiple optical colors simultaneously, and assessing their significance and impact on dust extinction estimates.   By focusing on the velocity effects on colors, our analysis is insensitive to errors in distance or luminosity.  Properly assessing the impact of these effects on distance estimates will require embedding these effects in a statistical model incorporating SN Ia magnitudes, colors, ejecta velocities, and distance moduli.  This will be the subject of future research.  The initial analyses of \citet{foleykasen11}  assessed this effect by simply splitting a nearby sample into  HV and NV groups, and accounting for a relative intrinsic color difference between the two.  This led to promising, significant improvements in the Hubble diagram scatter (from 0.19 mag to 0.13 mag) for that sample.  Hence, a more sophisticated analysis that treats the ejecta velocity as a continuous parameter with a skewed population distribution, and incorporates its effects on the inference of intrinsic colors and host galaxy dust extinction, SN Ia magnitudes and distances, will have the potential to improve the accuracy and precision of SN Ia distances. Leveraging the intrinsic color-velocity effect to improve cosmological distance estimates can potentially lead to more reliable SN Ia constraints on the cosmological parameters and the properties of dark energy that are less prone to systematic errors from host galaxy dust extinction.  Our statistical modeling of intrinsic colors, dust and ejecta velocities in this work is a first step towards that goal.

\acknowledgements

K.M. thanks R. Kessler, B. Schmidt, R. Trotta and W.M. Wood-Vasey for useful discussions, and K. Krisciunas for a careful reading of the manuscript.  We thank the anonymous referee for useful suggestions that led to an improved manuscript. Supernova research at Harvard University is supported in part by NSF grant AST-1211196.

 \appendix

\section{Mathematical Formulae for the Marginal Likelihood}\label{sec:math_lkhd}

The marginal likelihood for a single SN $s$ with a given velocity $v_s$ is the predictive probability density of its $n_C$ observable, apparent colors $\bm{O}_s$, under a set of population hyperparameters.    Given the modeling assumptions (\S \ref{sec:statmodel}), we derive this analytically by integrating out the latent variables $\bm{C}_s, A_V^s$ from  $P(\bm{O}_s, \bm{C}_s, A_V^s | \, v_s ; \bm{\theta}, \bm{\Sigma_C}, \tau)$:
\begin{equation}\label{eqn:marginal_lkhd}
\begin{split}
P(\bm{O}_s | \, v_s ; \bm{\theta}, \bm{\Sigma_C}, \tau) &= \int dA_V^s \, d\bm{C}_s   \Big( N[\bm{O}_s | \, \bm{C}_s + A_V^s \bm{\gamma}(R_V), \bm{W}_s ]\, N[ \bm{C}_s | \, \bm{\mu}_C(v_s; \bm{\theta}), \bm{\Sigma}_C]\,  \text{Expon}( A_V^s | \tau) \Big)  \\
& = \sqrt{2 \pi} \left(\frac{\sigma_{A,s}}{\tau}\right) \exp\left[ \frac{1}{2} \left( \frac{\sigma_{A,s}}{\tau}\right)^2 - \frac{\hat{A}_s(v_s)}{\tau} \right]  \Phi\left( \frac{\hat{A}_s(v_s)}{\sigma_{A,s}} - \frac{\sigma_{A,s}}{\tau} \right) \\
& \times N\left( \bm{O}_s \Big| \, \bm{\mu}_C(v_s; \bm{\theta})+ \hat{A}_s(v_s) \bm{\gamma}(R_V), \bm{\Sigma}_C + \bm{W}_s \right),
\end{split}
\end{equation}
where $N( \bm{x} | \, \bm{\mu}, \bm{\Sigma})$ is a placeholder for $| 2 \pi \bm{\Sigma} |^{-1/2} \exp[-(\bm{x}-\bm{\mu})^T \bm{\Sigma}^{-1} (\bm{x}-\bm{\mu})/2]$, and 
$\Phi(x)  = \int^x_{-\infty} dx \, N(x | 0,1) $ is the cumulative distribution function for the unit Gaussian density, and we define
\begin{equation}
\sigma_{A,s}^{-2} \equiv \bm{\gamma}^T \left( \bm{\Sigma}_C + \bm{W}_s \right)^{-1} \bm{\gamma}
\end{equation}
\begin{equation}
\hat{A}_s(v_s) \equiv \sigma_{A,s}^2 \bm{\gamma}^T  \left( \bm{\Sigma}_C + \bm{W}_s \right)^{-1}  \left[ \bm{O}_s - \bm{\mu}_C(v_s; \bm{\theta}) \right].
\end{equation}
The coefficients $\bm{\gamma}(R_V) = ( \Delta \bm{\alpha} + \Delta \bm{\beta}/R_V )$ model the effect of dust reddening, e.g. for  $B-V$, $\gamma = R_V^{-1}$.   The marginal likelihood for the full sample is the product of $N_{SN}$ individual marginal likelihoods.  This marginal likelihood can be maximized with respect to the hyperparameters ($ \bm{\theta}, \bm{\Sigma_C}, \tau$) to yield point estimates.  It is needed to compute the deviance information criterion for model comparison (Appendix \ref{sec:modelcomp}).

If the population hyperparameters are known or estimated, then the posterior density of the extinction for an individual SN $s$ has the simple form of a  Gaussian, truncated to $A_V \ge 0$ by a Heaviside step function $H(x)$.
\begin{equation}
P(A_V^s | \, \bm{O}_s, v_s; \bm{\theta}, \bm{\Sigma}_C, \tau) \propto N( A_V^s | \, \hat{A}_s(v_s) - \sigma^2_{A,s}/\tau, \sigma^2_{A,s}) \, H(A_V^s)
\end{equation}

In the case of $n_C = 1$ color, the shape of the likelihood function depends critically on the relative values of $\tau_E \equiv \gamma \tau$ (i.e. the average color excess due to dust) and $\sigma_s^2 \equiv \sigma_C^2 + w_s^2$, the sum of the intrinsic variance ($\bm{\Sigma}_C = \sigma^2_C$) plus measurement variance ($\bm{W}_s = w_s^2$).   The marginal likelihood for one color simplifies to
\begin{equation}\label{eqn:marginal_lkhd_1c}
P( O_s | v_s; \bm{\theta}, \sigma_C^2, \tau_E) = \tau_E^{-1} \exp\left[\frac{1}{2}\left(\frac{\sigma_s }{\tau_E} \right)^2 - \frac{O_s - \mu_c( v_s; \bm{\theta})}{\tau_E} \right] \Phi\left( \frac{O_s - \mu_c(v_s; \bm{\theta})}{\sigma_s} - \frac{\sigma_s }{\tau_E} \right).
\end{equation}
Given values of the hyperparameters, the posterior of the color excess due to dust $E_s = A_V^s \gamma$ for SN $s$ is simply
\begin{equation}
P( E_s | \, O_s, v_s; \bm{\theta},\sigma_C^2, \tau_E) \propto N( E_s | \, O_s - \mu_C(v_s; \bm{\theta}) - \sigma^2_s / \tau_E, \sigma^2_s ) \,H(E_s).
\end{equation}

\citet*{jha07} modeled the SN Ia apparent color distribution at +35 days after maximum light as the combination of a Gaussian distribution of intrinsic colors  and an exponential distribution of dust reddening. In our new framework, that model corresponds to the special case of $n_C = 1$ and constant mean intrinsic colors, $\mu_C(v_s; \bm{\theta}) = c_0$, with intrinsic color scatter $\sigma_C$.  However, we have also explicitly included heteroscedastic measurement errors $w_s$ in both our marginal likelihood and posterior, and derived the analytic form of the marginal likelihood, Eq, \ref{eqn:marginal_lkhd_1c}.

\section{Sampling the Posterior: Markov Chain for Regressing Colors}\label{sec:mcrc}

We sketch an MCMC Gibbs sampling algorithm to generate samples from the global posterior, Eq. \ref{eqn:globalposterior}, of the hierararchical model.   The goal of an MCMC algorithm is to generate a sequence of random parameter vectors with a long-run distribution equal to the global posterior.
Our algorithm, Markov Chain for Regressing Colors (\textsc{MCRC}), proceeds by sequentially drawing new parameter values from a full set of conditional posterior densities derived from Eq. \ref{eqn:globalposterior}.     The algorithm works for models in which the conditional mean intrinsic colors can be written as linear function of the hyperparameters $\mathbb{E}[\bm{C} | \, v] = \bm{\mu}_C(v; \bm{\theta}) = \bm{M}(v) \bm{\theta}$, and the color likelihood function is in the form of  Eq. \ref{eqn:intcol_lkhd}.  This includes the linear and nonlinear dependencies on velocity described in \S \ref{sec:linear}, \ref{sec:nonlinear}.  We begin with randomized guesses for the intrinsic colors and dust extinctions $\{\bm{C}_s, A_V^s\}$, as well as the residual covariance matrix $\bm{\Sigma}_C$.   We alternate between updating the population hyperparameters ($ \bm{\theta}, \bm{\Sigma_C}, \tau$) conditional on the current values of the set of individual SN parameters, and updating the individual SN parameters $\{\bm{C}_s, A_V^s\}$ conditional on the population hyperparameters.

\begin{enumerate}
\item Sample the hyperparameters $\bm{\theta}$ describing the conditional mean intrinsic colors function $\bm{\mu}_C(v; \bm{\theta})$ from the conditional posterior $P(\bm{\theta} | \, \{\bm{C}_s, v_s \}, \bm{\Sigma}_C)$, which is a Gaussian density.  
We sample a Gaussian random vector $\bm{\theta} \sim N( \bm{\hat{\theta}}, \bm{V_\theta})$ to update the hyperparameters $\bm{\theta}$, where the mean $\bm{\hat{\theta}}$ and covariance $\bm{V_\theta}$ can be derived in terms of the covariate matrices $\{\bm{M}_s = \bm{M}(v_s)\}$, the colors $\{\bm{C}_s\}$ and the residual covariance matrix $\bm{\Sigma}_C$.  

\item Sample the hyperparameters of the residual covariance matrix $\bm{\Sigma}_C$ from $P( \bm{\Sigma}_C | \, \bm{\theta}, \{ \bm{C}_s, v_s\} )$.  The covariance matrix conditional on $\bm{\theta}$, the intrinsic colors and velocities has  an inverse Wishart distribution with a scale matrix  equal to the sum of squared deviations from the current fit of the mean intrinsic colors vs. velocity and the prior matrix $\bm{\Lambda}_0$, and degrees of freedom equal to the prior degrees $\nu_0$ plus the number of objects, $N_\text{SN}$.

\item Draw a new extinction scale $\tau$ from $P(\tau| \, \{ A_V^s\}) = \text{Inv-Gamma}(\tau | \, N_\text{SN}, \sum_s {A_V^s} )$, an inverse-Gamma distribution.

\item Each step below updates the parameters for single SN $s$ given the current values of the hyperparameters.  We cycle through these steps for every SN.

\begin{enumerate}

\item Sample new intrinsic colors $\bm{C}_s$ from the conditional posterior, $P( \bm{C}_s |\, A_V^s; \bm{\theta}, \bm{\Sigma}_C, \bm{O}_s, v_s)$, which is a multivariate Gaussian with mean and covariance that can be derived from Eq. \ref{eqn:posterior_CAv}.

\item Sample a new dust extinction $A_V^s$ from the conditional posterior $P(A_V^s | \, \bm{C}_s; \tau, \bm{O}_s)$.  This has the shape of a Gaussian, with a mean and variance that can be derived from Eq. \ref{eqn:posterior_CAv}, but truncated below $A_V^s = 0$.  We can sample a new $A_V^s$ from this truncated Gaussian distribution using the algorithm of \citet{robert95}.

\end{enumerate}

\end{enumerate}
We repeat these steps until convergence.   To monitor convergence, we typically run 4-8 parallel chains starting from different initial guesses, and compute the Gelman-Rubin statistic \citep{gelman92}.  We run the MCMC until the maximum G-R statistic is less than 1.02.  We find that for moderate supernova samples ($N_\text{SN} \sim 100$) and $\dim(\bm{C}_s) = 3$, the chains converge within a few minutes for $n = 10^4$ full cycles of steps 1-4.    We discard the initial 20\% of each chain as burn-in, and concatenate the remaining chains for posterior analysis.

\section{Model Comparison using the Deviance Information Criterion}\label{sec:modelcomp}

In \S \ref{sec:statmodel}, we built a general model for the apparent color and velocity data that could be used with any parameterized function for the mean relation between intrinsic colors and velocity $\bm{\mu}_C( v; \bm{\theta})$.   The mean function can be as simple as a constant, i.e. $\bm{\mu}_C(v; \bm{c}_0) = \bm{c}_0$, or more complex, e.g. $p$-degree polynomial functions of $v$ (\S \ref{sec:poly}) or multiple covariates (\S \ref{sec:mult_covariates}).   More complex models will typically have more parameters, degrees of freedom, and flexibility to fit the data, but also are more vulnerable to overfitting the noise in the data.   The more complex model will usually fit the data better than the simpler models, especially if the simpler models are special cases of the complex model (i.e. they are nested models).   However, the improvement in fit ought to be weighed against the increase in complexity to avoid overfitting the data.  Information criteria are useful numerical summaries for model selection that balance the goodness-of-fit of the data and model versus the additional parameters of model complexity.   \citet{liddle07} reviewed various information criteria for astrophysical and cosmological model selection.  Information criteria such as AIC, DIC, and WAIC are measures of predictive accuracy that include a correction for the bias introduced by evaluating the model fit on the data used to fit the model \citep{gelman14}.  We employ the Deviance Information Criterion, which is well suited for model comparison of hierarchical Bayesian models fit using MCMC \citep{spiegelhalter02}.

A familiar summary of model fit is the $\chi^2$ statistic, the weighted sum of squared deviations of the data from the model. In this paper, we work with the \emph{deviance}, defined as twice the negative (natural) log-likelihood function (or log predictive density), as a summary of model fit.   It is a generalized summary of model fit that reduces to $\chi^2$ in the case of a Gaussian likelihood function with known variances.  The expected deviance is proportional to the Kullback-Leibler information of the model.  Under standard conditions, with large samples sizes, the model with the lowest Kullback-Leibler information will have the highest posterior probability \citep{gelman_bda}.
We use the marginal likelihood function, integrating out the individual latent variables (i.e. the intrinsic colors and extinctions of supernovae), as given in Eq. \ref{eqn:marginal_lkhd}. The deviance is a function of the data and the hyperparameters.
\begin{equation}\label{eqn:deviance}
D( \mathcal{D},  \bm{H} ) = -2 \sum_{s=1}^{N_{SN}} \log P( \bm{O}_s | \, v_s, \bm{\theta}, \bm{\Sigma_C}, \tau)
\end{equation}
where the set of hyperparameters is denoted $\bm{H} = \{ \bm{\theta}, \bm{\Sigma_C}, \tau \}$, and the total data are $\mathcal{D} = \{\bm{O}_s, v_s\}$.

In simple cases, the deviance is related to the familiar $\chi^2$ goodness-of-fit statistic.  For example, suppose we could measure the scalar \emph{intrinsic} colors $\{ C_s \}$, with known variances $\sigma_s^2$, and covariates $\{v_s \}$ for a set of SNe.    We could run a regression of $C$ versus $v$, using a mean function $\mu_C(v, \bm{\theta})$.  The probability model is then $C_s  | v_s, \bm{\theta} \sim N[ \mu_C(v_s, \bm{\theta}), \sigma_s^2]$, with a Gaussian likelihood function. In this case, the deviance, as $-2 \times$ log-likelihood, is the $\chi^2$, up to a  constant:
\begin{equation}
D = \chi^2 = \sum_{s=1}^N \frac{[C_s - \mu_C(v_s; \bm{\theta})]^2}{\sigma_s^2}.
\end{equation}
However, in our case, we do not measure the intrinsic colors; those are latent variables.  We measure the apparent colors $\{ \bm{O}_s\}$, which differ from the intrinsic colors by a random, positive amount of dust reddening.  Furthermore, the residual intrinsic variances are also unknown.  In this more realistic situation, the likelihood function (Eq. \ref{eqn:marginal_lkhd}) takes into account these effects.   The deviance (Eq. \ref{eqn:deviance}) is an appropriate generalization of the $\chi^2$ as a summary of model fit for non-Gaussian likelihoods.

Using the MCMC samples from the posterior density, $\bm{H}_i \sim P(\bm{H} | \, \mathcal{D})$, we can compute the deviance evaluated at the posterior mean of the hyperparameters, $\hat{D} \equiv D(\mathcal{D}, \bm{\hat{H}})$, where
\begin{equation}
\bm{\hat{H}} = \int \bm{H} P(\bm{H} | \, \mathcal{D}) \, d\bm{H} \approx \frac{1}{L}\sum_{i=1}^L \bm{H}_i.
\end{equation}
and $L$  is the number of MCMC samples.   We can also compute the posterior average of the deviance:
\begin{equation}
\langle D \rangle \equiv \int D( \mathcal{D},  \bm{H} ) P(\bm{H} | \, \mathcal{D}) \, d\bm{H} \approx \frac{1}{L} \sum_{i=1}^L D( \mathcal{D},  \bm{H}_i ).
\end{equation}
An estimate of the \emph{effective number of parameters} is the difference between the mean deviance and the deviance at the mean: $p_D = \langle D \rangle - \hat{D}$.  This estimate need not be an integer; for Bayesian models, $p_D$ can be less than the naive parameter count if prior information is important or parameter estimates are degenerate.   A scalar parameter will contribute 1 if its prior constraints are unimportant, and 0 if its estimate is entirely determined by the prior.   It will contribute an intermediate value if both the data and prior constraints are important.  The deviance information criterion (DIC) is formed by penalizing the deviance measure of model fit by the effective number of parameters: 
\begin{equation}
\text{DIC} \equiv \hat{D} + 2 p_D = \langle D \rangle + p_D
\end{equation} 
\citep{spiegelhalter02}. This gives an estimate of the expected predictive deviance.  The DIC weighs the improvement in the model fit (summarised by $\langle D \rangle$) against the complexity of the model (summarised by $p_D$), with lower values of DIC being more favorable.  The DIC is well-suited for model comparison with hierarchical Bayesian models with an analytic expression for the likelihood (Eq. \ref{eqn:marginal_lkhd}), as it is straightforward to compute with the MCMC samples already in hand, and it accounts for parameter degeneracies and the influence of the prior in computing the effective degrees of freedom in the model.    Under the conventional interpretation of information criteria, differences greater than 2 represent positive support for the model with the lower numerical value, and differences greater than 6 represent strong support \citep{jeffreys61,kassraferty95, spiegelhalter02}).

\bibliographystyle{apj}
\bibliography{apj-jour,sn,stat}{}

\begin{thebibliography}{102}
\expandafter\ifx\csname natexlab\endcsname\relax\def\natexlab#1{#1}\fi

\bibitem[{{Altavilla} {et~al.}(2007){Altavilla}, {Stehle}, {Ruiz-Lapuente},
  {Mazzali}, {Pignata}, {Balastegui}, {Benetti}, {Blanc}, {Canal},
  {Elias-Rosa}, {Goobar}, {Harutyunyan}, {Pastorello}, {Patat}, {Rich},
  {Salvo}, {Schmidt}, {Stanishev}, {Taubenberger}, {Turatto}, \&
  {Hillebrandt}}]{altavilla07}
{Altavilla}, G., et~al.\ 2007, \aap, 475, 585

\bibitem[{{Amanullah} {et~al.}(2010){Amanullah}, {Lidman}, {Rubin}, {Aldering},
  {Astier}, {Barbary}, {Burns}, {Conley}, {Dawson}, {Deustua}, {Doi}, {Fabbro},
  {Faccioli}, {Fakhouri}, {Folatelli}, {Fruchter}, {Furusawa}, {Garavini},
  {Goldhaber}, {Goobar}, {Groom}, {Hook}, {Howell}, {Kashikawa}, {Kim}, {Knop},
  {Kowalski}, {Linder}, {Meyers}, {Morokuma}, {Nobili}, {Nordin}, {Nugent},
  {{\"O}stman}, {Pain}, {Panagia}, {Perlmutter}, {Raux}, {Ruiz-Lapuente},
  {Spadafora}, {Strovink}, {Suzuki}, {Wang}, {Wood-Vasey}, {Yasuda}, \&
  {Supernova Cosmology Project}}]{amanullah10}
{Amanullah}, R., et~al.\ 2010, \apj, 716, 712

\bibitem[{{Astier} {et~al.}(2006){Astier}, {Guy}, {Regnault}, {Pain},
  {Aubourg}, {Balam}, {Basa}, {Carlberg}, {Fabbro}, {Fouchez}, {Hook},
  {Howell}, {Lafoux}, {Neill}, {Palanque-Delabrouille}, {Perrett}, {Pritchet},
  {Rich}, {Sullivan}, {Taillet}, {Aldering}, {Antilogus}, {Arsenijevic},
  {Balland}, {Baumont}, {Bronder}, {Courtois}, {Ellis}, {Filiol}, {Gon{\c
  c}alves}, {Goobar}, {Guide}, {Hardin}, {Lusset}, {Lidman}, {McMahon},
  {Mouchet}, {Mourao}, {Perlmutter}, {Ripoche}, {Tao}, \& {Walton}}]{astier06}
{Astier}, P., et~al.\ 2006, \aap, 447, 31

\bibitem[{{Bailey} {et~al.}(2009){Bailey}, {Aldering}, {Antilogus}, {Aragon},
  {Baltay}, {Bongard}, {Buton}, {Childress}, {Chotard}, {Copin}, {Gangler},
  {Loken}, {Nugent}, {Pain}, {Pecontal}, {Pereira}, {Perlmutter}, {Rabinowitz},
  {Rigaudier}, {Runge}, {Scalzo}, {Smadja}, {Swift}, {Tao}, {Thomas}, {Wu}, \&
  {The Nearby Supernova Factory}}]{bailey09}
{Bailey}, S., et~al.\ 2009, \aap, 500, L17

\bibitem[{Barnard {et~al.}(2000)Barnard, McCulloch, \& Meng}]{barnard00}
Barnard, J., McCulloch, R., \& Meng, X.-L. 2000, Statistica Sinica, 10, 1281

\bibitem[{{Barone-Nugent} {et~al.}(2012){Barone-Nugent}, {Lidman}, {Wyithe},
  {Mould}, {Howell}, {Hook}, {Sullivan}, {Nugent}, {Arcavi}, {Cenko}, {Cooke},
  {Gal-Yam}, {Hsiao}, {Kasliwal}, {Maguire}, {Ofek}, {Poznanski}, \&
  {Xu}}]{barone-nugent12}
{Barone-Nugent}, R.~L., et~al.\ 2012, \mnras,  425, 1007

\bibitem[{{Benetti} {et~al.}(2005){Benetti}, {Cappellaro}, {Mazzali},
  {Turatto}, {Altavilla}, {Bufano}, {Elias-Rosa}, {Kotak}, {Pignata}, {Salvo},
  \& {Stanishev}}]{benetti05}
{Benetti}, S., et~al.\ 2005, \apj, 623, 1011

\bibitem[{{Blondin} {et~al.}(2011{\natexlab{a}}){Blondin}, {Kasen},
  {R{\"o}pke}, {Kirshner}, \& {Mandel}}]{blondin11b}
{Blondin}, S., {Kasen}, D., {R{\"o}pke}, F.~K., {Kirshner}, R.~P., \& {Mandel},
  K.~S. 2011{\natexlab{a}}, \mnras, 417, 1280

\bibitem[{{Blondin} {et~al.}(2011{\natexlab{b}}){Blondin}, {Mandel}, \&
  {Kirshner}}]{blondin11}
{Blondin}, S., {Mandel}, K.~S., \& {Kirshner}, R.~P. 2011{\natexlab{b}}, \aap,
  526, A81+

\bibitem[{{Blondin} {et~al.}(2012){Blondin}, {Matheson}, {Kirshner}, {Mandel},
  {Berlind}, {Calkins}, {Challis}, {Garnavich}, {Jha}, {Modjaz}, {Riess}, \&
  {Schmidt}}]{blondin12}
{Blondin}, S., et~al.\ 2012, \aj, 143, 126

\bibitem[{{Bongard} {et~al.}(2006){Bongard}, {Baron}, {Smadja}, {Branch}, \&
  {Hauschildt}}]{bongard06}
{Bongard}, S., {Baron}, E., {Smadja}, G., {Branch}, D., \& {Hauschildt}, P.~H.
  2006, \apj, 647, 513

\bibitem[{{Branch} {et~al.}(2006){Branch}, {Dang}, {Hall}, {Ketchum},
  {Melakayil}, {Parrent}, {Troxel}, {Casebeer}, {Jeffery}, \&
  {Baron}}]{branch06}
{Branch}, D., et~al.\ 2006, \pasp, 118, 560

\bibitem[{{Branch} \& {Tammann}(1992)}]{branch92}
{Branch}, D. \& {Tammann}, G.~A. 1992, \araa, 30, 359

\bibitem[{{Brewer} \& {Elliott}(2014)}]{brewer14}
{Brewer}, B.~J. \& {Elliott}, T.~M. 2014, \mnras, 439, L31

\bibitem[{{Burns} {et~al.}(2014){Burns}, {Stritzinger}, {Phillips}, {Hsiao},
  {Contreras}, {Persson}, {Folatelli}, {Boldt}, {Campillay}, {Castell{\'o}n},
  {Freedman}, {Madore}, {Morrell}, {Salgado}, \& {Suntzeff}}]{burns14}
{Burns}, C.~R., et~al.\ 2014, \apj, 789, 32

\bibitem[{{Cardelli} {et~al.}(1989){Cardelli}, {Clayton}, \& {Mathis}}]{ccm89}
{Cardelli}, J.~A., {Clayton}, G.~C., \& {Mathis}, J.~S. 1989, \apj, 345, 245

\bibitem[{{Cartier} {et~al.}(2011){Cartier}, {F{\"o}rster}, {Coppi}, {Hamuy},
  {Maeda}, {Pignata}, \& {Folatelli}}]{cartier11}
{Cartier}, R., {F{\"o}rster}, F., {Coppi}, P., {Hamuy}, M., {Maeda}, K.,
  {Pignata}, G., \& {Folatelli}, G. 2011, \aap, 534, L15

\bibitem[{{Chotard} {et~al.}(2011){Chotard}, {Gangler}, {Aldering},
  {Antilogus}, {Aragon}, {Bailey}, {Baltay}, {Bongard}, {Buton}, {Canto},
  {Childress}, {Copin}, {Fakhouri}, {Hsiao}, {Kerschhaggl}, {Kowalski},
  {Loken}, {Nugent}, {Paech}, {Pain}, {Pecontal}, {Pereira}, {Perlmutter},
  {Rabinowitz}, {Runge}, {Scalzo}, {Smadja}, {Tao}, {Thomas}, {Weaver}, \&
  {Wu}}]{chotard11}
{Chotard}, N., et~al.\ 2011, \aap, 529, L4+

\bibitem[{{Conley} {et~al.}(2007){Conley}, {Carlberg}, {Guy}, {Howell}, {Jha},
  {Riess}, \& {Sullivan}}]{conley07}
{Conley}, A., {Carlberg}, R.~G., {Guy}, J., {Howell}, D.~A., {Jha}, S.,
  {Riess}, A.~G., \& {Sullivan}, M. 2007, \apjl, 664, L13

\bibitem[{{Conley} {et~al.}(2011){Conley}, {Guy}, {Sullivan}, {Regnault},
  {Astier}, {Balland}, {Basa}, {Carlberg}, {Fouchez}, {Hardin}, {Hook},
  {Howell}, {Pain}, {Palanque-Delabrouille}, {Perrett}, {Pritchet}, {Rich},
  {Ruhlmann-Kleider}, {Balam}, {Baumont}, {Ellis}, {Fabbro}, {Fakhouri},
  {Fourmanoit}, {Gonz{\'a}lez-Gait{\'a}n}, {Graham}, {Hudson}, {Hsiao},
  {Kronborg}, {Lidman}, {Mourao}, {Neill}, {Perlmutter}, {Ripoche}, {Suzuki},
  \& {Walker}}]{conley11}
{Conley}, A., et~al.\ 2011, \apjs, 192, 1

\bibitem[{{Contreras} {et~al.}(2010){Contreras}, {Hamuy}, {Phillips},
  {Folatelli}, {Suntzeff}, {Persson}, {Stritzinger}, {Boldt}, {Gonz{\'a}lez},
  {Krzeminski}, {Morrell}, {Roth}, {Salgado}, {Jos{\'e} Maureira}, {Burns},
  {Freedman}, {Madore}, {Murphy}, {Wyatt}, {Li}, \& {Filippenko}}]{contreras10}
{Contreras}, C., et~al.\ 2010, \aj, 139, 519

\bibitem[{{Draine}(2003)}]{draine03}
{Draine}, B.~T. 2003, \araa, 41, 241

\bibitem[{{Elias} {et~al.}(1985){Elias}, {Matthews}, {Neugebauer}, \&
  {Persson}}]{elias85}
{Elias}, J.~H., {Matthews}, K., {Neugebauer}, G., \& {Persson}, S.~E. 1985,
  \apj, 296, 379

\bibitem[{{Elias-Rosa} {et~al.}(2006){Elias-Rosa}, {Benetti}, {Cappellaro},
  {Turatto}, {Mazzali}, {Patat}, {Meikle}, {Stehle}, {Pastorello}, {Pignata},
  {Kotak}, {Harutyunyan}, {Altavilla}, {Navasardyan}, {Qiu}, {Salvo}, \&
  {Hillebrandt}}]{elias-rosa06}
{Elias-Rosa}, N., et~al.\ 2006, \mnras,  369, 1880

\bibitem[{{Elias-Rosa} {et~al.}(2008){Elias-Rosa}, {Benetti}, {Turatto},
  {Cappellaro}, {Valenti}, {Arkharov}, {Beckman}, {di Paola}, {Dolci},
  {Filippenko}, {Foley}, {Krisciunas}, {Larionov}, {Li}, {Meikle},
  {Pastorello}, {Valentini}, \& {Hillebrandt}}]{elias-rosa07}
{Elias-Rosa}, N., et~al.\ 2008, \mnras, 384, 107

\bibitem[{{Filippenko} {et~al.}(1992){Filippenko}, {Richmond}, {Branch},
  {Gaskell}, {Herbst}, {Ford}, {Treffers}, {Matheson}, {Ho}, {Dey}, {Sargent},
  {Small}, \& {van Breugel}}]{filippenko92}
{Filippenko}, A.~V., et~al.\ 1992, \aj,  104, 1543

\bibitem[{{Finkelman} {et~al.}(2008){Finkelman}, {Brosch}, {Kniazev},
  {Buckley}, {O'Donoghue}, {Hashimoto}, {Loaring}, {Romero-Colmenero}, {Still},
  {Sefako}, \& {V{\"a}is{\"a}nen}}]{finkelman08}
{Finkelman}, I., et~al.\ 2008, \mnras, 390, 969

\bibitem[{{Finkelman} {et~al.}(2010){Finkelman}, {Brosch}, {Kniazev},
  {V{\"a}is{\"a}nen}, {Buckley}, {O'Donoghue}, {Gulbis}, {Hashimoto},
  {Loaring}, {Romero-Colmenero}, \& {Sefako}}]{finkelman10}
{Finkelman}, I., et~al.\ 2010, \mnras, 409, 727

\bibitem[{{Folatelli} {et~al.}(2013){Folatelli}, {Morrell}, {Phillips},
  {Hsiao}, {Campillay}, {Contreras}, {Castell{\'o}n}, {Hamuy}, {Krzeminski},
  {Roth}, {Stritzinger}, {Burns}, {Freedman}, {Madore}, {Murphy}, {Persson},
  {Prieto}, {Suntzeff}, {Krisciunas}, {Anderson}, {F{\"o}rster}, {Maza},
  {Pignata}, {Rojas}, {Boldt}, {Salgado}, {Wyatt}, {Olivares E.}, {Gal-Yam}, \&
  {Sako}}]{folatelli13}
{Folatelli}, G., et~al.\ 2013, \apj, 773, 53

\bibitem[{{Folatelli} {et~al.}(2010){Folatelli}, {Phillips}, {Burns},
  {Contreras}, {Hamuy}, {Freedman}, {Persson}, {Stritzinger}, {Suntzeff},
  {Krisciunas}, {Boldt}, {Gonz{\'a}lez}, {Krzeminski}, {Morrell}, {Roth},
  {Salgado}, {Madore}, {Murphy}, {Wyatt}, {Li}, {Filippenko}, \&
  {Miller}}]{folatelli10}
{Folatelli}, G., et~al.\ 2010, \aj, 139,  120

\bibitem[{{Foley}(2012)}]{foley12a}
{Foley}, R.~J. 2012, \apj, 748, 127

\bibitem[{{Foley} {et~al.}(2008){Foley}, {Filippenko}, \& {Jha}}]{foley08}
{Foley}, R.~J., {Filippenko}, A.~V., \& {Jha}, S.~W. 2008, \apj, 686, 117

\bibitem[{{Foley} \& {Kasen}(2011)}]{foleykasen11}
{Foley}, R.~J. \& {Kasen}, D. 2011, \apj, 729, 55

\bibitem[{{Foley} {et~al.}(2011){Foley}, {Sanders}, \&
  {Kirshner}}]{foleysanderskirshner11}
{Foley}, R.~J., {Sanders}, N.~E., \& {Kirshner}, R.~P. 2011, \apj, 742, 89

\bibitem[{{Foley} {et~al.}(2012){Foley}, {Simon}, {Burns}, {Gal-Yam}, {Hamuy},
  {Kirshner}, {Morrell}, {Phillips}, {Shields}, \& {Sternberg}}]{foley12b}
{Foley}, R.~J., et~al.\ 2012, \apj, 752, 101

\bibitem[{{Foster} {et~al.}(2013){Foster}, {Mandel}, {Pineda}, {Covey}, {Arce},
  \& {Goodman}}]{foster13}
{Foster}, J.~B., {Mandel}, K.~S., {Pineda}, J.~E., {Covey}, K.~R., {Arce},
  H.~G., \& {Goodman}, A.~A. 2013, \mnras, 428, 1606

\bibitem[{{Freedman} {et~al.}(2009){Freedman}, {Burns}, {Phillips}, {Wyatt},
  {Persson}, {Madore}, {Contreras}, {Folatelli}, {Gonzalez}, {Hamuy}, {Hsiao},
  {Kelson}, {Morrell}, {Murphy}, {Roth}, {Stritzinger}, {Sturch}, {Suntzeff},
  {Astier}, {Balland}, {Bassett}, {Boldt}, {Carlberg}, {Conley}, {Frieman},
  {Garnavich}, {Guy}, {Hardin}, {Howell}, {Kessler}, {Lampeitl}, {Marriner},
  {Pain}, {Perrett}, {Regnault}, {Riess}, {Sako}, {Schneider}, {Sullivan}, \&
  {Wood-Vasey}}]{freedman09}
{Freedman}, W.~L., et~al.\ 2009, \apj, 704, 1036

\bibitem[{{Friedman} {et~al.}(2014){Friedman}, {Wood-Vasey}, {Marion},
  {Challis}, {Mandel}, {Bloom}, {Modjaz}, {Narayan}, {Hicken}, {Foley},
  {Klein}, {Starr}, {Morgan}, {Rest}, {Blake}, {Miller}, {Falco}, {Wyatt},
  {Mink}, {Skrutskie}, \& {Kirshner}}]{friedman14}
{Friedman}, A.~S., et~al.\ 2014, submitted to \apj, ArXiv  e-prints, arXiv:1408.0465

\bibitem[{{Ganeshalingam} {et~al.}(2010){Ganeshalingam}, {Li}, {Filippenko},
  {Anderson}, {Foster}, {Gates}, {Griffith}, {Grigsby}, {Joubert}, {Leja},
  {Lowe}, {Macomber}, {Pritchard}, {Thrasher}, \& {Winslow}}]{ganeshalingam10}
{Ganeshalingam}, M., et~al.\ 2010, \apjs, 190, 418

\bibitem[{{Garnavich} {et~al.}(1998){Garnavich}, {Jha}, {Challis},
  {Clocchiatti}, {Diercks}, {Filippenko}, {Gilliland}, {Hogan}, {Kirshner},
  {Leibundgut}, {Phillips}, {Reiss}, {Riess}, {Schmidt}, {Schommer}, {Smith},
  {Spyromilio}, {Stubbs}, {Suntzeff}, {Tonry}, \& {Carroll}}]{garnavich98b}
{Garnavich}, P.~M., et~al.\ 1998, \apj, 509, 74

\bibitem[{Gelman {et~al.}(2003)Gelman, Carlin, Stern, \& Rubin}]{gelman_bda}
Gelman, A., Carlin, J.~B., Stern, H.~S., \& Rubin, D.~B. 2003, Bayesian Data
  Analysis, Second Edition (Boca Raton, Fla.: {Chapman \& Hall/CRC})

\bibitem[{Gelman {et~al.}(2014)Gelman, Hwang, \& Vehtari}]{gelman14}
Gelman, A., Hwang, J., \& Vehtari, A. 2014, Statistics and Computing, 24, 997

\bibitem[{Gelman \& Rubin(1992)}]{gelman92}
Gelman, A. \& Rubin, D.~B. 1992, Statistical Science, 7, 457

\bibitem[{{Guy} {et~al.}(2007){Guy}, {Astier}, {Baumont}, {Hardin}, {Pain},
  {Regnault}, {Basa}, {Carlberg}, {Conley}, {Fabbro}, {Fouchez}, {Hook},
  {Howell}, {Perrett}, {Pritchet}, {Rich}, {Sullivan}, {Antilogus}, {Aubourg},
  {Bazin}, {Bronder}, {Filiol}, {Palanque-Delabrouille}, {Ripoche}, \&
  {Ruhlmann-Kleider}}]{guy07}
{Guy}, J., et~al.\ 2007, \aap, 466, 11

\bibitem[{{Hachinger} {et~al.}(2008){Hachinger}, {Mazzali}, {Tanaka},
  {Hillebrandt}, \& {Benetti}}]{hachinger08}
{Hachinger}, S., {Mazzali}, P.~A., {Tanaka}, M., {Hillebrandt}, W., \&
  {Benetti}, S. 2008, \mnras, 389, 1087

\bibitem[{{Hamuy} {et~al.}(1996){Hamuy}, {Phillips}, {Suntzeff}, {Schommer},
  {Maza}, {Antezan}, {Wischnjewsky}, {Valladares}, {Muena}, {Gonzales},
  {Aviles}, {Wells}, {Smith}, {Navarrete}, {Covarrubias}, {Williger}, {Walker},
  {Layden}, {Elias}, {Baldwin}, {Hernandez}, {Tirado}, {Ugarte}, {Elston},
  {Saavedra}, {Barrientos}, {Costa}, {Lira}, {Ruiz}, {Anguita}, {Gomez},
  {Ortiz}, {della Valle}, {Danziger}, {Storm}, {Kim}, {Bailyn}, {Rubenstein},
  {Tucker}, {Cersosimo}, {Mendez}, {Siciliano}, {Sherry}, {Chaboyer},
  {Koopmann}, {Geisler}, {Sarajedini}, {Dey}, {Tyson}, {Rich}, {Gal},
  {Lamontagne}, {Caldwell}, {Guhathakurta}, {Phillips}, {Szkody}, {Prosser},
  {Ho}, {McMahan}, {Baggley}, {Cheng}, {Havlen}, {Wakamatsu}, {Janes},
  {Malkan}, {Baganoff}, {Seitzer}, {Shara}, {Sturch}, {Hesser}, {Hartig},
  {Hughes}, {Welch}, {Williams}, {Ferguson}, {Francis}, {French}, {Bolte},
  {Roth}, {Odewahn}, {Howell}, \& {Krzeminski}}]{hamuy96_29sne}
{Hamuy}, M., et~al.\ 1996, \aj, 112, 2408

\bibitem[{{Hicken} {et~al.}(2009{\natexlab{a}}){Hicken}, {Challis}, {Jha},
  {Kirshner}, {Matheson}, {Modjaz}, {Rest}, {Michael Wood-Vasey}, {Bakos},
  {Barton}, {Berlind}, {Bragg}, {Brice{\~n}o}, {Brown}, {Caldwell}, {Calkins},
  {Cho}, {Ciupik}, {Contreras}, {Dendy}, {Dosaj}, {Durham}, {Eriksen},
  {Esquerdo}, {Everett}, {Falco}, {Fernandez}, {Gaba}, {Garnavich}, {Graves},
  {Green}, {Groner}, {Hergenrother}, {Holman}, {Hradecky}, {Huchra},
  {Hutchison}, {Jerius}, {Jordan}, {Kilgard}, {Krauss}, {Luhman}, {Macri},
  {Marrone}, {McDowell}, {McIntosh}, {McNamara}, {Megeath}, {Mochejska},
  {Munoz}, {Muzerolle}, {Naranjo}, {Narayan}, {Pahre}, {Peters}, {Peterson},
  {Rines}, {Ripman}, {Roussanova}, {Schild}, {Sicilia-Aguilar}, {Sokoloski},
  {Smalley}, {Smith}, {Spahr}, {Stanek}, {Barmby}, {Blondin}, {Stubbs},
  {Szentgyorgyi}, {Torres}, {Vaz}, {Vikhlinin}, {Wang}, {Westover}, {Woods}, \&
  {Zhao}}]{hicken09a}
{Hicken}, M., et~al.\ 2009{\natexlab{a}},  \apj, 700, 331

\bibitem[{{Hicken} {et~al.}(2012){Hicken}, {Challis}, {Kirshner}, {Rest},
  {Cramer}, {Wood-Vasey}, {Bakos}, {Berlind}, {Brown}, {Caldwell}, {Calkins},
  {Currie}, {de Kleer}, {Esquerdo}, {Everett}, {Falco}, {Fernandez},
  {Friedman}, {Groner}, {Hartman}, {Holman}, {Hutchins}, {Keys}, {Kipping},
  {Latham}, {Marion}, {Narayan}, {Pahre}, {Pal}, {Peters}, {Perumpilly},
  {Ripman}, {Sipocz}, {Szentgyorgyi}, {Tang}, {Torres}, {Vaz}, {Wolk}, \&
  {Zezas}}]{hicken12}
{Hicken}, M., et~al.\ 2012, \apjs, 200,  12

\bibitem[{{Hicken} {et~al.}(2009{\natexlab{b}}){Hicken}, {Wood-Vasey},
  {Blondin}, {Challis}, {Jha}, {Kelly}, {Rest}, \& {Kirshner}}]{hicken09b}
{Hicken}, M., et~al.\ 2009{\natexlab{b}}, \apj,  700, 1097

\bibitem[{{Hogg} {et~al.}(2010){Hogg}, {Myers}, \& {Bovy}}]{hogg10}
{Hogg}, D.~W., {Myers}, A.~D., \& {Bovy}, J. 2010, \apj, 725, 2166

\bibitem[{Jeffreys(1961)}]{jeffreys61}
Jeffreys, H. 1961, Theory of Probability, 3rd Ed. (Oxford: Clarendon Press)

\bibitem[{{Jha} {et~al.}(2006){Jha}, {Kirshner}, {Challis}, {Garnavich},
  {Matheson}, {Soderberg}, {Graves}, {Hicken}, {Alves}, {Arce}, {Balog},
  {Barmby}, {Barton}, {Berlind}, {Bragg}, {Brice{\~n}o}, {Brown}, {Buckley},
  {Caldwell}, {Calkins}, {Carter}, {Concannon}, {Donnelly}, {Eriksen},
  {Fabricant}, {Falco}, {Fiore}, {Garcia}, {G{\'o}mez}, {Grogin}, {Groner},
  {Groot}, {Haisch}, {Hartmann}, {Hergenrother}, {Holman}, {Huchra},
  {Jayawardhana}, {Jerius}, {Kannappan}, {Kim}, {Kleyna}, {Kochanek},
  {Koranyi}, {Krockenberger}, {Lada}, {Luhman}, {Luu}, {Macri}, {Mader},
  {Mahdavi}, {Marengo}, {Marsden}, {McLeod}, {McNamara}, {Megeath}, {Moraru},
  {Mossman}, {Muench}, {Mu{\~n}oz}, {Muzerolle}, {Naranjo}, {Nelson-Patel},
  {Pahre}, {Patten}, {Peters}, {Peters}, {Raymond}, {Rines}, {Schild},
  {Sobczak}, {Spahr}, {Stauffer}, {Stefanik}, {Szentgyorgyi}, {Tollestrup},
  {V{\"a}is{\"a}nen}, {Vikhlinin}, {Wang}, {Willner}, {Wolk}, {Zajac}, {Zhao},
  \& {Stanek}}]{jha06}
{Jha}, S., et~al.\ 2006, \aj, 131, 527

\bibitem[{{Jha} {et~al.}(2007){Jha}, {Riess}, \& {Kirshner}}]{jha07}
{Jha}, S., {Riess}, A.~G., \& {Kirshner}, R.~P. 2007, \apj, 659, 122

\bibitem[{{Kasen} \& {Plewa}(2007)}]{kasenplewa07}
{Kasen}, D. \& {Plewa}, T. 2007, \apj, 662, 459

\bibitem[{{Kasen} {et~al.}(2009){Kasen}, {R{\"o}pke}, \& {Woosley}}]{krw09}
{Kasen}, D., {R{\"o}pke}, F.~K., \& {Woosley}, S.~E. 2009, \nat, 460, 869

\bibitem[{Kass \& Raftery(1995)}]{kassraferty95}
Kass, R.~E. \& Raftery, A.~E. 1995, Journal of the American Statistical
  Association, 90, pp. 773

\bibitem[{{Kelly}(2007)}]{bkelly07}
{Kelly}, B.~C. 2007, \apj, 665, 1489

\bibitem[{{Kelly} {et~al.}(2012){Kelly}, {Shetty}, {Stutz}, {Kauffmann},
  {Goodman}, \& {Launhardt}}]{kelly12}
{Kelly}, B.~C., {Shetty}, R., {Stutz}, A.~M., {Kauffmann}, J., {Goodman},
  A.~A., \& {Launhardt}, R. 2012, \apj, 752, 55

\bibitem[{{Kessler} {et~al.}(2009){Kessler}, {Becker}, {Cinabro}, {Vanderplas},
  {Frieman}, {Marriner}, {Davis}, {Dilday}, {Holtzman}, {Jha}, {Lampeitl},
  {Sako}, {Smith}, {Zheng}, {Nichol}, {Bassett}, {Bender}, {Depoy}, {Doi},
  {Elson}, {Filippenko}, {Foley}, {Garnavich}, {Hopp}, {Ihara}, {Ketzeback},
  {Kollatschny}, {Konishi}, {Marshall}, {Mc Millan}, {Miknaitis}, {Morokuma},
  {M{\"o}rtsell}, {Pan}, {Prieto}, {Richmond}, {Riess}, {Romani}, {Schneider},
  {Sollerman}, {Takanashi}, {Tokita}, {van der Heyden}, {Wheeler}, {Yasuda}, \&
  {York}}]{kessler09}
{Kessler}, R., et~al.\ 2009, \apjs, 185,  32

\bibitem[{{Kowalski} {et~al.}(2008){Kowalski}, {Rubin}, {Aldering},
  {Agostinho}, {Amadon}, {Amanullah}, {Balland}, {Barbary}, {Blanc}, {Challis},
  {Conley}, {Connolly}, {Covarrubias}, {Dawson}, {Deustua}, {Ellis}, {Fabbro},
  {Fadeyev}, {Fan}, {Farris}, {Folatelli}, {Frye}, {Garavini}, {Gates},
  {Germany}, {Goldhaber}, {Goldman}, {Goobar}, {Groom}, {Haissinski}, {Hardin},
  {Hook}, {Kent}, {Kim}, {Knop}, {Lidman}, {Linder}, {Mendez}, {Meyers},
  {Miller}, {Moniez}, {Mour{\~a}o}, {Newberg}, {Nobili}, {Nugent}, {Pain},
  {Perdereau}, {Perlmutter}, {Phillips}, {Prasad}, {Quimby}, {Regnault},
  {Rich}, {Rubenstein}, {Ruiz-Lapuente}, {Santos}, {Schaefer}, {Schommer},
  {Smith}, {Soderberg}, {Spadafora}, {Strolger}, {Strovink}, {Suntzeff},
  {Suzuki}, {Thomas}, {Walton}, {Wang}, {Wood-Vasey}, \& {Yun}}]{kowalski08}
{Kowalski}, M., et~al.\ 2008, \apj,  686, 749

\bibitem[{{Krisciunas} {et~al.}(2007){Krisciunas}, {Garnavich}, {Stanishev},
  {Suntzeff}, {Prieto}, {Espinoza}, {Gonzalez}, {Salvo}, {Elias de la Rosa},
  {Smartt}, {Maund}, \& {Kudritzki}}]{krisciunas07}
{Krisciunas}, K., et~al.\ 2007, \aj,  133, 58

\bibitem[{{Krisciunas} {et~al.}(2004{\natexlab{a}}){Krisciunas}, {Phillips}, \&
  {Suntzeff}}]{krisciunas04a}
{Krisciunas}, K., {Phillips}, M.~M., \& {Suntzeff}, N.~B. 2004{\natexlab{a}},
  \apjl, 602, L81

\bibitem[{{Krisciunas} {et~al.}(2004{\natexlab{b}}){Krisciunas}, {Suntzeff},
  {Phillips}, {Candia}, {Prieto}, {Antezana}, {Chassagne}, {Chen}, {Dickinson},
  {Eisenhardt}, {Espinoza}, {Garnavich}, {Gonz{\' a}lez}, {Harrison}, {Hamuy},
  {Ivanov}, {Krzemi{\' n}ski}, {Kulesa}, {McCarthy}, {Moro-Mart{\'{\i}}n},
  {Muena}, {Noriega-Crespo}, {Persson}, {Pinto}, {Roth}, {Rubenstein},
  {Stanford}, {Stringfellow}, {Zapata}, {Porter}, \&
  {Wischnjewsky}}]{krisciunas04c}
{Krisciunas}, K., et~al.\ 2004{\natexlab{b}}, \aj, 128, 3034

\bibitem[{{Leibundgut} {et~al.}(1993){Leibundgut}, {Kirshner}, {Phillips},
  {Wells}, {Suntzeff}, {Hamuy}, {Schommer}, {Walker}, {Gonzalez}, {Ugarte},
  {Williams}, {Williger}, {Gomez}, {Marzke}, {Schmidt}, {Whitney}, {Coldwell},
  {Peters}, {Chaffee}, {Foltz}, {Rehner}, {Siciliano}, {Barnes}, {Cheng},
  {Hintzen}, {Kim}, {Maza}, {Parker}, {Porter}, {Schmidtke}, \&
  {Sonneborn}}]{leibundgut93}
{Leibundgut}, B., et~al.\ 1993, \aj, 105, 301

\bibitem[{{Leonard} {et~al.}(2005){Leonard}, {Li}, {Filippenko}, {Foley}, \&
  {Chornock}}]{leonard05}
{Leonard}, D.~C., {Li}, W., {Filippenko}, A.~V., {Foley}, R.~J., \& {Chornock},
  R. 2005, \apj, 632, 450

\bibitem[{{Liddle}(2007)}]{liddle07}
{Liddle}, A.~R. 2007, \mnras, 377, L74

\bibitem[{{Loredo}(2012)}]{loredo12}
{Loredo}, T.~J. 2012, ArXiv e-prints, arXiv:1208.3036

\bibitem[{{Loredo} \& {Hendry}(2010)}]{loredohendry10}
{Loredo}, T.~J. \& {Hendry}, M.~A. 2010, in Bayesian Methods in Cosmology, ed.
  M. Hobson et al. (Cambridge: Cambridge University Press), 245

\bibitem[{{Maeda} {et~al.}(2010){Maeda}, {Benetti}, {Stritzinger}, {R{\"o}pke},
  {Folatelli}, {Sollerman}, {Taubenberger}, {Nomoto}, {Leloudas}, {Hamuy},
  {Tanaka}, {Mazzali}, \& {Elias-Rosa}}]{maeda10}
{Maeda}, K., et~al.\ 2010, \nat,  466, 82

\bibitem[{{Maeda} {et~al.}(2011){Maeda}, {Leloudas}, {Taubenberger},
  {Stritzinger}, {Sollerman}, {Elias-Rosa}, {Benetti}, {Hamuy}, {Folatelli}, \&
  {Mazzali}}]{maeda11}
{Maeda}, K., et~al.\ 2011, \mnras, 413, 3075

\bibitem[{Mandel(2012)}]{mandel_scmav}
Mandel, K. 2012, in Statistical Challenges in Modern Astronomy V, ed. E.~D.
  Feigelson \& G.~J. Babu, Lecture Notes in Statistics (Springer New York),
  209--218

\bibitem[{{Mandel}(2011)}]{mandelthesis}
{Mandel}, K.~S. 2011, Ph.D.~Thesis, Harvard University

\bibitem[{{Mandel} {et~al.}(2011){Mandel}, {Narayan}, \& {Kirshner}}]{mandel11}
{Mandel}, K.~S., {Narayan}, G., \& {Kirshner}, R.~P. 2011, \apj, 731, 120

\bibitem[{{Mandel} {et~al.}(2009){Mandel}, {Wood-Vasey}, {Friedman}, \&
  {Kirshner}}]{mandel09}
{Mandel}, K.~S., {Wood-Vasey}, W.~M., {Friedman}, A.~S., \& {Kirshner}, R.~P.
  2009, \apj, 704, 629

\bibitem[{{March} {et~al.}(2011){March}, {Trotta}, {Berkes}, {Starkman}, \&
  {Vaudrevange}}]{march11}
{March}, M.~C., {Trotta}, R., {Berkes}, P., {Starkman}, G.~D., \&
  {Vaudrevange}, P.~M. 2011, \mnras, 418, 2308

\bibitem[{{Matheson} {et~al.}(2008){Matheson}, {Kirshner}, {Challis}, {Jha},
  {Garnavich}, {Berlind}, {Calkins}, {Blondin}, {Balog}, {Bragg}, {Caldwell},
  {Dendy Concannon}, {Falco}, {Graves}, {Huchra}, {Kuraszkiewicz}, {Mader},
  {Mahdavi}, {Phelps}, {Rines}, {Song}, \& {Wilkes}}]{matheson08}
{Matheson}, T., et~al.\ 2008, \aj, 135,  1598

\bibitem[{{Meikle}(2000)}]{meikle00}
{Meikle}, W.~P.~S. 2000, \mnras, 314, 782

\bibitem[{{Nobili} \& {Goobar}(2008)}]{nobili08}
{Nobili}, S. \& {Goobar}, A. 2008, \aap, 487, 19

\bibitem[{{Nugent} {et~al.}(1995){Nugent}, {Phillips}, {Baron}, {Branch}, \&
  {Hauschildt}}]{nugent95}
{Nugent}, P., {Phillips}, M., {Baron}, E., {Branch}, D., \& {Hauschildt}, P.
  1995, \apjl, 455, L147+

\bibitem[{{Perlmutter} {et~al.}(1999){Perlmutter}, {Aldering}, {Goldhaber},
  {Knop}, {Nugent}, {Castro}, {Deustua}, {Fabbro}, {Goobar}, {Groom}, {Hook},
  {Kim}, {Kim}, {Lee}, {Nunes}, {Pain}, {Pennypacker}, {Quimby}, {Lidman},
  {Ellis}, {Irwin}, {McMahon}, {Ruiz-Lapuente}, {Walton}, {Schaefer}, {Boyle},
  {Filippenko}, {Matheson}, {Fruchter}, {Panagia}, {Newberg}, {Couch}, \& {The
  Supernova Cosmology Project}}]{perlmutter99}
{Perlmutter}, S., et~al.\ 1999, \apj, 517, 565

\bibitem[{{Rest} {et~al.}(2014){Rest}, {Scolnic}, {Foley}, {Huber}, {Chornock},
  {Narayan}, {Tonry}, {Berger}, {Soderberg}, {Stubbs}, {Riess}, {Kirshner},
  {Smartt}, {Schlafly}, {Rodney}, {Botticella}, {Brout}, {Challis}, {Czekala},
  {Drout}, {Hudson}, {Kotak}, {Leibler}, {Lunnan}, {Marion}, {McCrum},
  {Milisavljevic}, {Pastorello}, {Sanders}, {Smith}, {Stafford}, {Thilker},
  {Valenti}, {Wood-Vasey}, {Zheng}, {Burgett}, {Chambers}, {Denneau}, {Draper},
  {Flewelling}, {Hodapp}, {Kaiser}, {Kudritzki}, {Magnier}, {Metcalfe},
  {Price}, {Sweeney}, {Wainscoat}, \& {Waters}}]{rest14}
{Rest}, A., et~al.\ 2014, \apj, 795, 44

\bibitem[{{Riess} {et~al.}(1998){Riess}, {Filippenko}, {Challis},
  {Clocchiatti}, {Diercks}, {Garnavich}, {Gilliland}, {Hogan}, {Jha},
  {Kirshner}, {Leibundgut}, {Phillips}, {Reiss}, {Schmidt}, {Schommer},
  {Smith}, {Spyromilio}, {Stubbs}, {Suntzeff}, \& {Tonry}}]{riess98}
{Riess}, A.~G., et~al.\ 1998, \aj, 116, 1009

\bibitem[{{Riess} {et~al.}(1999){Riess}, {Kirshner}, {Schmidt}, {Jha},
  {Challis}, {Garnavich}, {Esin}, {Carpenter}, {Grashius}, {Schild}, {Berlind},
  {Huchra}, {Prosser}, {Falco}, {Benson}, {Brice{\~n}o}, {Brown}, {Caldwell},
  {dell'Antonio}, {Filippenko}, {Goodman}, {Grogin}, {Groner}, {Hughes},
  {Green}, {Jansen}, {Kleyna}, {Luu}, {Macri}, {McLeod}, {McLeod}, {McNamara},
  {McLean}, {Milone}, {Mohr}, {Moraru}, {Peng}, {Peters}, {Prestwich},
  {Stanek}, {Szentgyorgyi}, \& {Zhao}}]{riess99}
{Riess}, A.~G., et~al.\ 1999, \aj, 117, 707

\bibitem[{Robert(1995)}]{robert95}
Robert, C.~P. 1995, Statistics and Computing, 5, 121

\bibitem[{{Sanders} {et~al.}(2014){Sanders}, {Betancourt}, \&
  {Soderberg}}]{sanders14}
{Sanders}, N., {Betancourt}, M., \& {Soderberg}, A. 2014, ArXiv e-prints,
  arXiv:1404.3619

\bibitem[{Scholz \& Stephens(1987)}]{scholzstephens87}
Scholz, F.~W. \& Stephens, M.~A. 1987, Journal of the American Statistical
  Association, 82, pp. 918

\bibitem[{{Scolnic} {et~al.}(2014{\natexlab{a}}){Scolnic}, {Rest}, {Riess},
  {Huber}, {Foley}, {Brout}, {Chornock}, {Narayan}, {Tonry}, {Berger},
  {Soderberg}, {Stubbs}, {Kirshner}, {Rodney}, {Smartt}, {Schlafly},
  {Botticella}, {Challis}, {Czekala}, {Drout}, {Hudson}, {Kotak}, {Leibler},
  {Lunnan}, {Marion}, {McCrum}, {Milisavljevic}, {Pastorello}, {Sanders},
  {Smith}, {Stafford}, {Thilker}, {Valenti}, {Wood-Vasey}, {Zheng}, {Burgett},
  {Chambers}, {Denneau}, {Draper}, {Flewelling}, {Hodapp}, {Kaiser},
  {Kudritzki}, {Magnier}, {Metcalfe}, {Price}, {Sweeney}, {Wainscoat}, \&
  {Waters}}]{scolnic14b}
{Scolnic}, D., et~al.\ 2014{\natexlab{a}}, \apj, 795, 45

\bibitem[{{Scolnic} {et~al.}(2014{\natexlab{b}}){Scolnic}, {Riess}, {Foley},
  {Rest}, {Rodney}, {Brout}, \& {Jones}}]{scolnic14}
{Scolnic}, D.~M., {Riess}, A.~G., {Foley}, R.~J., {Rest}, A., {Rodney}, S.~A.,
  {Brout}, D.~J., \& {Jones}, D.~O. 2014{\natexlab{b}}, \apj, 780, 37

\bibitem[{{Shetty} {et~al.}(2013){Shetty}, {Kelly}, \& {Bigiel}}]{shetty13}
{Shetty}, R., {Kelly}, B.~C., \& {Bigiel}, F. 2013, \mnras, 430, 288

\bibitem[{{Silverman} {et~al.}(2012){Silverman}, {Foley}, {Filippenko},
  {Ganeshalingam}, {Barth}, {Chornock}, {Griffith}, {Kong}, {Lee}, {Leonard},
  {Matheson}, {Miller}, {Steele}, {Barris}, {Bloom}, {Cobb}, {Coil},
  {Desroches}, {Gates}, {Ho}, {Jha}, {Kandrashoff}, {Li}, {Mandel}, {Modjaz},
  {Moore}, {Mostardi}, {Papenkova}, {Park}, {Perley}, {Poznanski}, {Reuter},
  {Scala}, {Serduke}, {Shields}, {Swift}, {Tonry}, {Van Dyk}, {Wang}, \&
  {Wong}}]{silverman12}
{Silverman}, J.~M., et~al.\ 2012, \mnras,  425, 1789

\bibitem[{Spiegelhalter {et~al.}(2002)Spiegelhalter, Best, Carlin, \& Van
  Der~Linde}]{spiegelhalter02}
Spiegelhalter, D.~J., Best, N.~G., Carlin, B.~P., \& Van Der~Linde, A. 2002,
  Journal of the Royal Statistical Society: Series B (Statistical Methodology),
  64, 583

\bibitem[{{Stritzinger} {et~al.}(2011){Stritzinger}, {Phillips}, {Boldt},
  {Burns}, {Campillay}, {Contreras}, {Gonzalez}, {Folatelli}, {Morrell},
  {Krzeminski}, {Roth}, {Salgado}, {DePoy}, {Hamuy}, {Freedman}, {Madore},
  {Marshall}, {Persson}, {Rheault}, {Suntzeff}, {Villanueva}, {Li}, \&
  {Filippenko}}]{stritzinger11}
{Stritzinger}, M.~D., et~al.\ 2011, \aj, 142, 156

\bibitem[{{Sullivan} {et~al.}(2011){Sullivan}, {Guy}, {Conley}, {Regnault},
  {Astier}, {Balland}, {Basa}, {Carlberg}, {Fouchez}, {Hardin}, {Hook},
  {Howell}, {Pain}, {Palanque-Delabrouille}, {Perrett}, {Pritchet}, {Rich},
  {Ruhlmann-Kleider}, {Balam}, {Baumont}, {Ellis}, {Fabbro}, {Fakhouri},
  {Fourmanoit}, {Gonz{\'a}lez-Gait{\'a}n}, {Graham}, {Hudson}, {Hsiao},
  {Kronborg}, {Lidman}, {Mourao}, {Neill}, {Perlmutter}, {Ripoche}, {Suzuki},
  \& {Walker}}]{sullivan11}
{Sullivan}, M., et~al.\ 2011, \apj, 737, 102

\bibitem[{{Tripp}(1998)}]{tripp98}
{Tripp}, R. 1998, \aap, 331, 815

\bibitem[{{Tripp} \& {Branch}(1999)}]{trippbranch99}
{Tripp}, R. \& {Branch}, D. 1999, \apj, 525, 209

\bibitem[{{Wang} {et~al.}(2006){Wang}, {Baade}, {H{\"o}flich}, {Wheeler},
  {Kawabata}, {Khokhlov}, {Nomoto}, \& {Patat}}]{lwang06-04dt}
{Wang}, L., {Baade}, D., {H{\"o}flich}, P., {Wheeler}, J.~C., {Kawabata}, K.,
  {Khokhlov}, A., {Nomoto}, K., \& {Patat}, F. 2006, \apj, 653, 490

\bibitem[{{Wang} {et~al.}(2009){Wang}, {Filippenko}, {Ganeshalingam}, {Li},
  {Silverman}, {Wang}, {Chornock}, {Foley}, {Gates}, {Macomber}, {Serduke},
  {Steele}, \& {Wong}}]{wangx09b}
{Wang}, X., et~al.\ 2009,  \apjl, 699, L139

\bibitem[{{Wang} {et~al.}(2008){Wang}, {Li}, {Filippenko}, {Krisciunas},
  {Suntzeff}, {Li}, {Zhang}, {Deng}, {Foley}, {Ganeshalingam}, {Li}, {Lou},
  {Qiu}, {Shang}, {Silverman}, {Zhang}, \& {Zhang}}]{wangx08}
{Wang}, X., et~al.\ 2008, \apj, 675, 626

\bibitem[{{Wang} {et~al.}(2013){Wang}, {Wang}, {Filippenko}, {Zhang}, \&
  {Zhao}}]{wangx13}
{Wang}, X., {Wang}, L., {Filippenko}, A.~V., {Zhang}, T., \& {Zhao}, X. 2013,
  Science, 340, 170

\bibitem[{{Weyant} {et~al.}(2014){Weyant}, {Wood-Vasey}, {Allen}, {Garnavich},
  {Jha}, {Joyce}, \& {Matheson}}]{weyant13}
{Weyant}, A., {Wood-Vasey}, W.~M., {Allen}, L., {Garnavich}, P.~M., {Jha},
  S.~W., {Joyce}, R., \& {Matheson}, T. 2014, \apj, 784, 105

\bibitem[{{Wood-Vasey} {et~al.}(2008){Wood-Vasey}, {Friedman}, {Bloom},
  {Hicken}, {Modjaz}, {Kirshner}, {Starr}, {Blake}, {Falco}, {Szentgyorgyi},
  {Challis}, {Blondin}, {Mandel}, \& {Rest}}]{wood-vasey08}
{Wood-Vasey}, W.~M., et~al.\ 2008, \apj, 689, 377

\bibitem[{{Wood-Vasey} {et~al.}(2007){Wood-Vasey}, {Miknaitis}, {Stubbs},
  {Jha}, {Riess}, {Garnavich}, {Kirshner}, {Aguilera}, {Becker}, {Blackman},
  {Blondin}, {Challis}, {Clocchiatti}, {Conley}, {Covarrubias}, {Davis},
  {Filippenko}, {Foley}, {Garg}, {Hicken}, {Krisciunas}, {Leibundgut}, {Li},
  {Matheson}, {Miceli}, {Narayan}, {Pignata}, {Prieto}, {Rest}, {Salvo},
  {Schmidt}, {Smith}, {Sollerman}, {Spyromilio}, {Tonry}, {Suntzeff}, \&
  {Zenteno}}]{wood-vasey07}
{Wood-Vasey}, W.~M., et~al.\ 2007, \apj, 666, 694


\end{thebibliography}

\end{document}